
%
%
%
\font\tenbf=cmbx10

\font\eightrm=cmr8
\font\eightit=cmti8

\font\germ=eufm10
\def\g{\hbox{\germ g}}
\def\sectiontitle#1\par{\vskip0pt plus.1\vsize\penalty-250
\vskip0pt plus-.1\vsize\bigskip\vskip\parskip
\message{#1}\leftline{\tenbf#1}\nobreak\vglue 5pt}
\def\ds{\displaystyle}
\def\wt{\widetilde}
\def\wh{\widehat}
\def\eno{\eqalignno}
\def\ld{\lambda}
\def\ol{\overline}
\def\al{\alpha}
\def\vp{\varphi}
\def\tt{\theta}

\def\mod{\hbox{\rm mod~}}

\def\frac#1#2{{#1\over#2}}

\def\m@th{\mathsurround=0pt}

\def\Fsquare(#1,#2){
\hbox{\vrule$\hskip-0.4pt\vcenter to #1{\normalbaselines\m@th
\hrule\vfil\hbox to #1{\hfill$#2$\hfill}\vfil\hrule}$\hskip-0.4pt
\vrule}}

\def\Addsquare(#1,#2){\hbox{$
	\dimen1=#1 \advance\dimen1 by -0.8pt
	\vcenter to #1{\hrule height0.4pt depth0.0pt\vss%
	\hbox to #1{\hss{%
	\vbox to \dimen1{\vss%
	\hbox to \dimen1{\hss$~#2~$\hss}%
	\vss}\hss}%
	\vrule width0.4pt}\vss%
	\hrule height0.4pt depth0.0pt}$}}

\def\Hthreebox(#1,#2,#3){%
	\Fsquare(0.3cm,#1)\Addsquare(0.3cm,#2)\Addsquare(0.3cm,#3)}

\def\Htwobox(#1,#2){%
	\Fsquare(0.3cm,#1)\Addsquare(0.3cm,#2)}

\def\Threeone(#1,#2,#3,#4){%
	\normalbaselines\m@th\offinterlineskip
	\vtop{\hbox{\Hthreebox({#1},{#2},{#3})~0}
	      \vskip-0.4pt
	      \hbox{\Fsquare(0.3cm,#4)$~_{\ds 0}$}}}

\def\Twotwo(#1,#2,#3,#4){%
	\normalbaselines\m@th\offinterlineskip
	\vtop{\hbox{\Htwobox({#1},{#2})}
	      \vskip-0.4pt
	      \hbox{\Htwobox({#3},{#4})}}}

\magnification=\magstep1

\parindent=15pt
\nopagenumbers
\baselineskip=10pt
\vglue 4pc
\baselineskip=13pt
\headline{\ifnum\pageno=1\hfil\else%
{\ifodd\pageno\rightheadline \else \leftheadline\fi}\fi}
\def\rightheadline{\hfil\eightit
Dilogarithm identities
\quad\eightrm\folio}
\def\leftheadline{\eightrm\folio\quad
\eightit
Anatol N. Kirillov
\hfil}
\voffset=2\baselineskip
\centerline{\tenbf
  DILOGARITHM   \hskip 0.1cm  IDENTITIES  
}
\vglue 24pt
\centerline{\eightrm
ANATOL N. KIRILLOV
}
\baselineskip=12pt
\centerline{\eightit
Department of Mathematical Sciences, University of Tokyo,
}
\baselineskip=10pt
\centerline{\eightit
Komaba, Meguro-ku, Tokyo 153, Japan
}
\baselineskip=12pt
\centerline{\eightit
and }
\baselineskip=12pt
\centerline{\eightit
Steklov Mathematical Institute,
}
\baselineskip=10pt
\centerline{\eightit
Fontanka 27, St.Petersburg, 191011, Russia
}
\vglue 20pt
\centerline{\eightrm ABSTRACT}
{\rightskip=1.5pc
\leftskip=1.5pc
\eightrm\parindent=1pc
We study the dilogarithm identities from algebraic, analytic,
asymptotic, $K$-theoretic, combinatorial and representation-theoretic points
of view. We prove that a lot of dilogarithm identities (hypothetically
all!) can be obtained by using the five-term relation only. Among those
the Coxeter, Lewin, Loxton and Browkin ones are contained. Accessibility
of Lewin's one variable and Ray's multivariable (here for $n\le 2$ only)
functional equations is given. For odd levels the $\wh{sl_2}$ case of
Kuniba-Nakanishi's dilogarithm conjecture is proven and additional
results about remainder term are obtained. The connections between
dilogarithm identities and Rogers-Ramanujan-Andrews-Gordon type partition
identities via their asymptotic behavior are discussed. Some new results
about the string functions for level $k$ vacuum representation of the
affine Lie algebra $\wh{sl_n}$ are obtained. Connection between
dilogarithm identities and algebraic $K$-theory (torsion in $K_3({\bf
R})$) is discussed. Relations between crystal basis, branching functions
$b_{\ld}^{k\Lambda_0}(q)$ and Kostka-Foulkes polynomials (Lusztig's
$q$-analog of weight multiplicity) are considered. The Melzer and Milne
conjectures are proven. In some special cases we are proving that the
branching functions $b_{\ld}^{k\Lambda_0}(q)$ are equal to an appropriate
limit of Kostka polynomials (the so-called Thermodynamic Bethe
Ansatz limit). Connection between "finite-dimensional part of crystal base"
and Robinson-Schensted-Knuth correspondence is considered.}
\vglue12pt
\baselineskip=13pt
\overfullrule=0pt
\def\qed{\hfill$\vrule height 2.5mm width 2.5mm depth 0mm$}
\vskip 0.3cm

{\bf 0. Introduction.}
\medbreak

In this paper we consider mainly the properties of
dilogarithm related  to the so-called accessibility problem,
representation theory of Virasoro and
Kac-Moody algebras, and Conformal Field Theory (CFT). The remarkable and
mysterious dilogarithm [Za2], [Za3] will be the main hero of our
paper, but we start with definition and list some applications of more
general and may be even more amazing functions:

\centerline{polylogarithm \ and \ hyperlogarithm.}

\medbreak
{\bf Definition 1} (G. Leibnitz, 1696; L. Euler, 1768). {\it The
polylogarithm function Li$_k(x)$ is defined for \hbox{$0<x<1$} by
$$Li_k(x):=\sum_{n=1}^{\infty}{x^n\over n^k}\eqno (0.1)
$$
and admits analytic continuation to the complex plane as a multivalued
analytical function on $x$.}

\medbreak
{\bf Definition 2} (E. Kummer, 1840; I. Lappo-Danilevsky, 1910; K.-T.
Chen, 1977). {\it For given vector $n\in{\bf Z}_+^l$, the hyperlogarithm of
type $n$ is defined for a point $x$ lying in the unit cube
$I_l=\{x=(x_1,\ldots ,x_l)~|~|x_i|<1,~1\le i\le l\}$ by}
$$\Phi_n(x_1,\ldots ,x_l)=\sum_{0<k_1<k_2<\cdots
<k_l}{x_1^{k_1}x_2^{k_2}\ldots x_l^{k_l}\over k_1^{n_1}k_2^{n_2}\ldots
k_l^{n_l}}.\eqno (0.2)
$$

The function $\Phi_n(x)$ admits an analytic continuation to the complex
space ${\bf C}^l$.

\bigbreak
{\bf 0.1. Ubiquity of polylogarithms.}
\medbreak

The polylogarithm and hyperlogarithm have many intriguing properties and
have appeared
in various branches of mathematics and physics. We mention the
following ones:

\item{---} volumes of polytopes in hyperbolic and spheric geometry
(N. Lobachevsky (1836),
\break L. Schl\"{a}fli (1852), H. Coxeter (1935), G. Kneser (1936),
$\ldots$)
\item{---} volumes of 3-dim hyperbolic manifolds (S. Kojima, J. Milnor,
W. Neu\-mann, T. Yoshida, D. Zagier, ...)
\item{---} combinatorial formula for characteristic classes (I.M. Gel'fand,
R. MacPherson,\break A. Gabrielov,  M. Losik, ...)
\item{---} cohomology of $GL_n({\bf C})$ (A. Borel, J. Dupont, D. Quillen,
A. Suslin, $ \ldots$)
\item{---} special values of zeta functions for number fields/elliptic
curves
(S. Bloch, A. Beilinson, P. Deligne, A. Goncharov, D. Zagier, $\ldots$)
\item{---} algebraic $K$-theory (A. Borel, S. Bloch, A. Goncharov, A. Suslin,
D. Zagier, $\ldots$)
\item{---} number theory : Rogers-Ramanujan's type identities, asymptotic
behavior of partitions (S. Ramanujan, G. Hardy, D. Littlewood, B. Richmond,
G. Meinardus, $\ldots$)
\item{---} representation theory of infinite dimensional algebras
(A.~Feingold, J.~Lepowsky,\break V.~Kac, D. Peterson, M. Wakimoto, B. Feigin,
D. Fuchs, E. Frenkel, $\ldots$)
\item{---} Exactly Solvable Models (R. Baxter, V. Bazhanov, A.N.
Kirillov, N.Yu. Reshetikhin, W. Nahm, A. Varchenko, ...)
\item{---} Conformal  Field Theory (Al. Zamolodchikov, T. Klassen, E.
Melzer, F. Ravanini, A.N.~Kirillov, A. Kuniba, ...)
\item{---} Low dimensional topology; Vassiliev-Kontsevich's knot
invariants; Drinfeld's associator, ...(V. Drinfeld, M. Kontsevich, D.
Bar-Natan, T. Kohno, X.-S. Lin, T.Q.T. Le, J.~Murakami, ...)
\item{---} Quantum dilogarithm (L.D. Faddeev, R. Kashaev, A.N. Kirillov,
...)

See also [Ca], [BGSV], [DS], [Go], [JKS], [Le5], [Mi2], [Mi3], [O], [V],
[Za1]-[Za4].

\vfil\eject
\bigbreak
{\bf 0.2. Dilogarithm.}
\medbreak

The main subject of this paper is the dilogarithm identities. More
exactly, we are study the relations of the following type
$$\sum_{i=1}^NL(x_i)=c{\pi^2\over 6},\eqno (0.3)
$$
where $L(x)$ is the Rogers dilogarithm, all $x_i$ are assumed to be
the algebraic and $c$ is a rational number. The main problems we are
interested in are the following:

$i)$ how to find/classify the relations of type (0.3),

$ii)$ how one can prove a dilogarithm identity.

Both problems have the deep connections with algebraic $K$-theory,
representation theory of infinite dimensional algebras, combinatorics and
mathematical physics.

The main purpose of the paper presented is to study the dilogarithm
identities and related topics from the different points of view, namely,
the algebraic, analytic, asymptotic, group-theoretic and combinatorial
ones.

The first non-trivial examples of dilogarithm relations were obtained by
Euler and Landen in the second half of 18-th century. And it happened
only after 150 years that Coxeter and Watson had found the new examples
of the type (0.3) dilogarithm relations (see (1.11) and Section 2.1.2).
The most complicated among the Coxeter relations is the following one
$$L(\rho^{20})-2L(\rho^{10})-15L(\rho^4)+10L(\rho^2)={\pi^2\over
5}.\eqno(0.4)
$$
The methods used by Coxeter and
Watson were entirely different. Coxeter's  relations were developed from
the properties of a certain infinite series, whereas Watson deduced his
relations by "simple elimination" using Euler's and Abel's one-variable
functional equations (i.e. (1.3) and (1.7)). Following L. Lewin, let us
give

\medbreak
{\bf Definition 3.} {\it A dilogarithm relation (0.3) is called to be
accessible if it can be obtained
by applying the five-term relation (1.4) a number of times.}

Although Watson's dilogarithm relations are accessible from the five term
relations only, they are not obviously so. As Watson indicated, "he had
long suspected the existence of certain results, and although his
eventual proof is easy enough to follow, it is clear that it was not all
that easy to come by" (see [Le5], p.5). This comment is applicable to the
Coxeter relations as well. In fact, the last Coxeter relation had long
been believed to be inaccessible (on the basis of many failed tries).
However, it has been shown by Dupont in 1990 (see [Le5], p.52) that the
relation (0.4) can be really deduced from the five-term relation (1.4) and
the so-called factorization (or multiplication) formula
$$L(y^n)=n\sum_{l=1}^nL(\exp\left({2\pi il\over n}\right) y).
$$
There exist also the proofs of Coxeter's relations based on Lewin's
single-variable ([Le4] and Theorem 1) and Ray's multivariable ([Ray] and
Proposition E) functional equations.
In fact, Ray multivariable functional equation is the very powerful means
for proving the dilogarithm identities (see [Ray]). However, the known
proof of Ray's functional equation [Ray] uses the analytical methods
(e.g. the integral representation for ${\rm Li}_2$). Hence, the Ray
functional equation can not be considered {\it ad hoc} as accessible.

Now we came to one of the main problems considered in the Section 1 of our
paper, namely, to that of the dilogarithm identities accessibility.
According to the strong version of Goncharov's conjecture (see (2.3.2)),
any dilogarithm relation (0.3) with real algebraic $x_i$, $1\le i\le N$,
can be obtained as a linear combination with rational coefficients of the
five-term relations:
$$\eno{
&\sum_{i=1}^NL(x_i)-cL(1)\cr
&=\sum_jn_j\left\{L(a_j)+L(b_j)-L(a_jb_j)-L\left({a_j(1-b_j)\over
1-a_jb_j}\right)-L\left({b_j(1-a_j)\over 1-a_jb_j}\right)\right\},}
$$
where all $a_j,b_j$ lie in ${\bf R}\cap\ol{\bf Q}$ and all $n_j\in{\bf Q}$.
Here $L(x)$ is the so-called single-valued dilogarithm (see (1.13)). It
was indicated by Lichtenbaum ([Li], see also Section 1.1.3) that one should
be careful
working with the single-valued dilogarithm, because it does not satisfy
the five-term relation in general (but only modulo $\pi^2$).

The main purpose of the Sections 1.1-1.3 is to prove accessibility of
some well-known dilogarithm identities such as the Coxeter, Lewin,
Loxton, Browkin ones. Let us note, that Browkin's identities have been
found in [Bro2] and [Le3] "numerically", i.e. without, up to-date, any
analytic proof. We give also the proofs of accessibility of the Lewin
single-valued functional equations as well as the Ray multivalued
functional equation (for simplicity, we consider in the paper presented
only the case $n\le 2$, see Appendix).

A remarkable family of dilogarithm identities (see [Kir4], [BR], [Ku])
comes from a consideration of some Mathematical Physics problems. Namely,
in physics, the dilogarithm appeared at first from a calculation of
magnetic susceptibility in the $XXZ$-model at small magnetic field [KR],
[Kir4], [BR]. More recently [Z], the dilogarithm identities (through
Thermodynamic Bethe Ansatz (TBA)) appeared in the context of
investigation of the ultra violet (UV) limit or the critical behavior of
the integrable 2-dimensional quantum field theories and lattice models [Z],
[DR], ... . The meaning of dilogarithm identities in question is to
identify some theories via the central/effective charges and conformal
dimension of primary fields computation. Even more, it was shown (e.g.
[NRT], [DKKMM], [KMM], [KKMM1], [Tr], ...) using the Richmond-Szekeres
method [RS] and the Kac-Wakimoto theorem, that the dilogarithm identities
can be derived from an investigation of the asymptotic behavior of some
characters of 2-dimensional CFT. To be more precise, it is necessary to
have the Rogers-Ramanujan-left-hand-side type formulas for the characters
(and branching functions) of the Virasoro and affine Lie algebras. This
problem is very interesting one and is considered in many papers [LP],
[FeSt]... .

In the Section 1.3 we give a new proof of the $A_n$-type dilogarithm
identity [Kir7], mainly based on the multiple uses of the five-term
relation.

In the Section 1.4 we discuss and partially prove the $\wh{sl_2}$-case of
Kuniba-Nakanishi's conjecture [KN]. This conjecture predicts the
integrality and divisibility by 24 for the remainder term in some
special linear combination of the principal-branch-valued Euler
dilogarithm and logarithms and has many interesting connections with
CFT, TBA and representation theory [KN], [KNS], ... . In principle, using
the $A_n$-dilogarithm identity (1.28), one can reduce a proof of
Kuniba-Nakanishi's identity for $\wh{sl_n}$ to a proving that between the
values of logarithms only. We intend to consider this interesting
subject, including a generalization of Kuniba-Nakanishi's identity to a
"fractional level", in a separate publication.

Summarizing, in the Section 1 we present mainly the algebraic method
for proving the single-valued-dilogarithm identities by means of the
multiple uses of the five-term relation (1.4). Presumably, any identity
of type (0.3) can be obtained by this way (strong form of Goncharov's
conjecture). But even if it is so, to have the more direct ways to
produce the dilogarithm identities are still desirable. One of such ways,
the so-called asymptotic method (S. Ramanujan, G. Hardy, G. Meinardus, B.
Richmond, S. Szekeres, W. Nahm, B. MacCoy, M. Terhoeven, ...) and closely
related with Rogers-Ramanujan type identities (L. Rogers, S. Ramanujan,
I. Schur, G. Watson, L. Slater, B. Gordon, G. Andrews, D.~Bressoud,
B.~Feigin, E. Melzer, ...) will be discussed in the Section 2.1. Another
one, related with the algebraic $K$-theory (torsion in $K_3({\bf R})$,
S. Bloch, A. Suslin, D. Zagier, E.~Frenkel, A. Szenes, ...) will be
considered in Section 2.3. Finally, we discuss the so-called
character-asymptotic method due to V. Kac and M. Wakimoto (see also E.
Frenkel, A.~Szenes, ...). This method is closely related with the
following problems:

$i)$ combinatorial (i.e. Rogers-Ramanujan's and Schur's type) formulae for
the branching and string functions in the representation theory of
integrable highest weight modules over affine Kac-Moody algebras [Kac];

$ii)$ finite-dimensional version of crystal basis (see e.g. [FrSz2]) and
its connection with Robinson-Schensted type correspondence;

$iii)$ a natural "finitization" (or polynomial version) of the Weyl-Kac [Kac]
and
Feigin-Fuchs-Rocha-Caridi character formulae;

$iv)$ branching functions as the Thermodynamic-Bethe-Ansatz-like limit of
the Kostka-Foul\-kes polynomials (or Lusztig's $q$-analog of weight
multiplicity).

We are going to discuss these problems in the Sections 2.2 and 2.4.

Finally, in the Section 2.5 we consider some properties of the quantum
dilogarithm (cf. [FK], [FV]) and quantum polylogarithm.

\bigbreak
{\bf Acknowledgments.}
This work was written during my stay at Mathematical Department of
Tokyo University and is based on the series of lectures given at Tokyo,
Nagoya, Osaka, Kobe and Kyushu Universities and at RIMS and Yukawa
Institute, Kyoto.

I would be very glad to express my gratitude to L.D.~Faddeev, K. Aomoto,
E.~Date, B.~Feigin, A.~Kato, A.~Kuniba, B.~Leclerc, N.A.~Liskova, T.~Miwa,
W.~Nahm, M.~Noumi, R.~Sasaki, E.~Sklyanin, M.~Suzuki, V.~Tarasov,
J.-Y. Thibon, A.Yu.~Volkov and M.~Wakimoto for
fruitful discussions and  valuable remarks. I also thank with much
gratitude the colleagues at Tokyo University for their invitation, their
hospitality which made it possible to finish this work.

I would like to
acknowledge my special indebtedness to Dr. N.A. Liskova for the inestimable
help in preparing the manuscript to publication.

\bigbreak
{\bf 1.1.  Dilogarithm }(G. Leibnitz (1696), L. Euler (1768), L. Rogers
(1906)).
\medbreak

{\bf 1.1.1.} Euler's and Rogers' dilogarithm.
\medbreak

The Euler dilogarithm function $Li_2(x)$ defined for $0\le x\le 1$ by
$$
Li_2(x):=\sum_{n\ge 1}{x^n\over n^2}=-\int_0^x{\log (1-t)\over t}dt
\eqno(1.1)
$$
is one of the lesser transcendental function. Nonetheless, it has many
intriguing properties and has appeared in various branches of mathematics
and physics.

It is also convenient to introduce the Rogers dilogarithm function $L(x)$
defined for \hbox{$0<x<1$} by
$$L(x):=\sum^{\infty}_{n=1}{x^n\over n^2}+{1\over 2}\log x\cdot\log (1-x)=
-{1\over 2}\int_0^x\left({\log (1-y)\over y}+{\log y\over 1-y}\right) dy.
\eqno (1.2)
$$
The function $L(x)$ has an analytic continuation to the complex plane cut
analog the real axis from $-\infty$ to 0 and 1 to $+\infty$.

\medbreak
{\bf 1.1.2.} 
Five-term relation and characterization of Rogers' dilogarithm.
\medbreak

{\bf Theorem A} (W. Spence (1809), N. Abel (1830), C. Hill
(1830), E. Kummer (1840)).

{\it The function $L(x)$ belongs to the class
$C^{\infty}((0,1))$ and satisfies the following functional equations
$$\eno{&1.~~L(x)+L(1-x)={\pi^2\over 6},~~0<x<1, & (1.3)\cr
&2.~~L(x)+L(y)=L(xy)+L\left({x(1-y)\over 1-xy}\right) +L\left({y(1-x)
\over 1-xy}\right) ,& (1.4)}
$$
where $0<x,~~y<1$.}
\medbreak

{\bf Theorem B} (Cf. [Ro], Section 4) {\it Let $f(x)$ be a function of class
$C^3((0,1))$ which satisfies the relations (1.3) and (1.4). Then we have}
$$f(x)=L(x).$$

\vfill\eject
\medbreak
{\bf 1.1.3.} Single--valued dilogarithm.
\medbreak

We continue the function $L(x)$ to all real axis ${\bf R}={\bf R}^1\cup
\{\pm\infty\}$ by the following rules
$$\eno{
&L(x)={\pi^2\over 3}-L(x^{-1}),~~{\rm if}~~x>1,\cr
&L(x)=L\left({1\over 1-x}\right)-{\pi^2\over 6},~~{\rm if}~~x<0,& (1.5)\cr
&L(0)=0,~~L(1)={\pi^2\over 6},~~L(+\infty )={\pi^2\over 3},~~L(-\infty )=
-{\pi^2\over 6}.}
$$

{\bf Important remark.} The continuation  of
Rogers' dilogarithm function constructed in this way satisfies the
functional equation (1.3), but does not satisfy in general the five-term
relation (1.4). Namely, one can check ${\rm LHS}(1.4)-{\rm RHS}(1.4)=0$,
except the case $x<0,~~y<0$ and $xy>1$. In the latter case  ${\rm LHS}(1.4)-
{\rm RHS}(1.4)=-\pi^2$.
So, proving accessibility of a dilogarithm identity we must remember
about an existence of the exceptional cases. We leave a checking of
exceptional cases to a reader.

In the last part of this section we are going to show that
(1.3) is a corollary of (1.4). Namely, let us take $x=\displaystyle{a\over
1-b}$ and $y=\displaystyle{b\over 1-a}$ in (1.4). It is clear that
$a=\displaystyle{x(1-y)\over 1-xy}$,\break $b=\displaystyle{y(1-x)\over 1-xy}$
and the five-term relation (1.4) is reduced to the so--called Abel
equation (see e.g. [Le])
$$L\left({a\over 1-b}\right) +L\left({b\over 1-a}\right) =
L\left({a\over 1-b}\cdot{b\over 1-a}\right) +L(a)+L(b).\eqno (1.6)
$$
{}From (1.6), with $1-a$ for $b$, we obtain
$$L(a)+L(1-a)=L(1),
$$
as it was claimed. If in (1.4) we take $x=y$, we get the Abel duplication
formula
$$L(x^2)=2L(x)-2L\left({x\over 1+x}\right).\eqno (1.7)
$$
{}From (1.6), with $-a$ for $b$, and using (1.7), we obtain
$$L(a^2)+L\left({-a^2\over 1-a^2}\right) =0
$$
and consequently,
$$L(a)+L(-a)={1\over 2}L(a^2).\eqno (1.8)
$$
Finally, if in (1.6) we take $\displaystyle{a-1\over a}$ for $b$, we get
$$\eno{
&L(a^2)+L(a^{-2})=2L(1);\cr
&L(-a^2)+L\left({-1\over a^2}\right) =-L(1).& (1.9)}
$$

\vfill\eject
\medbreak
{\bf 1.1.4.} Dilogarithm of complex argument. Wigner-Bloch's
function.
\medbreak

By using the integral representation
$${\rm L}i_2(z)=-\int_0^z{\log (1-t)\over t}dt,
$$
Euler's dilogarithm L$i_2(x)$ can be analytically continued as a
multivalued complex function to the complex plane cut along the real axis
from 1 to $+\infty$.

\medbreak
{\bf Proposition A}  (E. Kummer (1840)). {\it Assume that $r,\theta$ are
the real numbers and $|r|<1$, then
$$\eno{
{\rm L}i_2(re^{i\theta})&=-{1\over 2}\int_0^r{\log (1-2x\cos\theta
+x^2)\over x}dx\cr
&+i\left\{ w\log (r)+{1\over 2}Cl_2(2w)+{1\over
2}Cl_2(2\theta )-{1\over 2}Cl_2(2\theta +2w)\right\},}
$$
where
$$w=\tan^{-1}\left(\ds{r\sin\theta\over 1-r\cos\theta}\right),~~{\rm and}~~
Cl_2(\theta )=\sum_{n=1}^{\infty}{\sin (n\theta )\over
n^2}=-\int_0^{\theta }\log |2\sin (t/2)|dt
$$
is Clausen's function.}

\medbreak
{\bf Proposition B} (F. Newman (1847)).
$${\rm L}i_2(e^{i\theta})=\pi^2{\overline B_2}\left({\theta\over
2\pi}\right) +iCl_2(\theta ),
$$
{\it where ${\overline B_2}(x):=-\ds{1\over\pi^2}\sum_{n=1}^{\infty}{\cos
2n\pi x\over n^2}$ is the second modified Bernoulli polynomial.}

In recent studies in $K$-theory the Wigner-Bloch dilogarithm function
$D(z)$ has played an important role [Bl]. It can be related to Clausen's
function via the formula
$$\eno{
D(z)&:=\arg (1-z)\log |z|-{\rm Im}\int_0^z{\log (1-t)\over t}dt\cr
&={1\over 2}\left\{ Cl_2(2\theta )+Cl_2(2w)-Cl_2(2\theta +2w)\right\},}
$$
where $\theta =\arg (z)$, $w=\arg (1-{\overline z})$ and $D(z)=0$ for
$z\in{\bf R}\cup\{\infty\}$.

The dilogarithm $D$: ${\bf C}\cup\{\infty\}
\to{\bf R}$ is a real-analytic function on ${\bf C}\setminus [0,1]$,
continuous on ${\bf C}\cup\{\infty\}$.
We summarize the basic properties of
Wigner-Bloch's dilogarithm in the next Proposition (see e.g. [Bl], [Bro2]).

\medbreak
{\bf Proposition C.} {\it For $z\in{\bf C}\cup\{\infty\}$, we have}

$i)$ $D(z)+D(1-z)=0$, $D(z)+D(1/z)=0$, $D(z)=D({\overline z})$,

$ii)$ $D(z^n)=n\ds\sum_{k=0}^{n-1}D\left(\exp\left({2\pi ki\over n}\right)
z\right)$,

$iii)$ (Five-term relation) {\it If $z,w\in{\bf C}\setminus [0,1]$, then}
$$D(z)+D(w)=D(zw)+D\left({z(1-w)\over 1-zw}\right)
+D\left({w(1-z)\over 1-zw}\right).
$$

There is a geometric interpretation of the number $D(z)$ and consequently
of $Cl_2(t)=D(e^{it})$. Namely, for a complex number $z$ , the
volume of the asymptotic simplex with vertices $0,1,z$ and $\infty$ in
3-dimensional hyperbolic space is equal to $|D(z)|$. Some formulae for
the Wigner-Bloch dilogarithm can be interpreted as relations between the
volumes of corresponding simplexes. For more details, see [Mi2], [Mi3],
[Za1], [Za4].

\medbreak
{\bf 1.1.5.} Multivariable functional equations (L. Rogers (1906),
G. Ray (1990)).
\medbreak

Using the analytic methods, Rogers [Ro] had proved the following
dilogarithm identity.
\medbreak

{\bf Proposition D} (L. Rogers). {\it Define the polynomial
$$r(x,t):=1-t-\prod^n_{j=1}(1-\alpha_jx)
$$
of degree $n$ in $x$ for nonzero complex numbers $\al_1,\ldots ,\al_n$. Let
$\{ x_l:=x_l(t)\}$, $l=1,\ldots ,n$ be the algebraic functions satisfying
$r(x_l,t)=0$. Then we have}
$$\sum^n_{j=1}\sum^n_{l=1}[L(\al_j/\al_l)-L(\al_jx_l)]=L(1-t).
$$

The special case $n=2$ leads to the form of Hill's equation (see Remark
in Section 1.3) with two independent variables and five dilogarithmic
terms.

{\bf Example.} Let us assume $\al_1=\ldots =\al_n=1$ and $1-t=y^n$. Then
$x_l=\zeta_ly$, where $\zeta_l=\exp\left(\ds{2\pi il\over n}\right)$, and
Rogers' identity is reduced to the so-called factorization formula
$$L(y^n)=n\sum^n_{l=1}L(\zeta_ly).
$$

A remarkable generalization of Rogers' multivariable identity for
dilogarithm was discovered by G. Ray [Ray]. Since the statement of the
most general dilogarithm identity in the paper [Ray] is rather involved,
we present
here only an important particular case of the general Ray's result.
\medbreak

{\bf Proposition E} (G. Ray). {\it Consider the polynomial
$$r(x,t):=(1-t)A\prod^n_{j=1}(1-\beta_jx)-\prod^n_{j=1}(1-\al_jx)
$$
of degree $n$ in $x$ for nonzero complex numbers $\{\al_j,A\}$ and some
complex numbers $\{\beta_j\}$. Let $\{x_l:=x_l(t)\}$, $1\le l\le n$, be
the algebraic functions satisfying $r(x_l,t)=0$. Then we have}
$$\eno{
\sum^n_{j=1}\sum^n_{l=1}&\left[L(\beta_jx_l)-L(\beta_j/\al_l)-
L(\al_jx_l)+L(\al_j/\al_l)\right]\cr
&=L(A(1-t))-L\left({A\beta_1\ldots
\beta_n\over\al_1\ldots\al_n}(1-t)\right).}
$$

In Appendix we are going to prove the accessibility of this identity
for $n=1,2$. General case will be considered elsewhere.

\bigbreak
{\bf 1.2. Accessible dilogarithm relations.}
\medbreak

{\bf 1.2.1.} Accessibility of Coxeter's and Lewin's identities.
\medbreak

It is easy to see from (1.3) and (1.7) that
$$\eno{
&L(1)={\pi^2\over 6},~L(-1)=-{\pi^2\over 12},~L({1\over 2})={\pi^2\over 12},
{}~~\rm{(L. Euler, 1768)};& (1.10)\cr
&L({1\over 2}({\sqrt 5}-1))={\pi^2\over 10},
{}~~L({1\over 2}(3-{\sqrt 5}))={\pi^2\over 15},~~\rm{(J. Landen, 1780)}.}
$$

Proof. Let us put $\rho :=\ds{1\over 2}({\sqrt 5}-1)$. It is clear that
$\rho^2+\rho =1$. So we have $L(\rho^2)=L(1-\rho )=L(1)-L(\rho )$. Now we
use the Abel formula (1.7)
$$
L(\rho^2)=2L(\rho )-2L\left({\rho\over 1+\rho}\right)=2L(\rho )-2L(\rho^2).
$$
Consequently, $3L(\rho^2)=2L(\rho )$. But as we already saw,
$$L(1)=L(\rho )+L(\rho^2)={5\over 3}L(\rho ).
$$
This proves the part of (1.10) due to J. Landen.
\qed

The results of Euler and Landen in equation (1.10) were the only such
expressions until Coxeter [Co] in 1935 paper proved the following formulae
$$\eno{
&L(\rho^6)=4L(\rho^3)+3L(\rho^2)-6L(\rho )+{7\pi^2\over 30};\cr
&L(\rho^{12})=2L(\rho^6)+3L(\rho^4)+4L(\rho^3)-6L(\rho^2)+
{\pi^2\over 10};& (1.11)\cr
&L(\rho^{20})=2L(\rho^{10})+15L(\rho^4)-10L(\rho^2)+{\pi^2\over 5}.}
$$

 Here we rewrote Coxeter's results using Rogers' dilogarithm. H. Coxeter
obtained his results using the properties of a certain infinite series.
Our nearest aim is to show that the Coxeter results (1.11) are accessible
from the functional equation (1.4) only. We prove also the accessibility of
the relation discovered by L. Lewin [Le3]
$$L(\rho^{24})=6L(\rho^8)+8L(\rho^6)-6L(\rho^4)+{\pi^2\over 30}.
\eqno (1.12)
$$
In fact we will prove that the Lewin single-variable functional equations
(see [Le4], or [Le5], Chapter 6, or below) are also accessible from the
five-term relation (1.4) only.

Let us remark, that one can obtain both Coxeter's relations (1.11) and
Lewin's one (1.12) by using the Ray multi-variable functional equation
(Proposition E, see also Exercises to Section 1). It is necessary to
underline, that the accessibility of the last Coxeter's relation among
(1.11) was proven at first by J. Dupont in 1989 (see [Le5], p.52). A
proof of accessibility of Lewin's relation (1.12) and Lewin's
single-variable functional equations (see Theorem 1) seems to be new.

\medbreak
{\bf Theorem 1.} {\it The following single--variable functional equations
(1.13), (1.14) and (1.15) are accessible from the functional equation
(1.4)}
$$\eno{
L\left(-z^7{1-z\over 1+z}\right) &=2L(z^2(1-z))+L\left({-z^3\over 1-z^2}
\right) +2L\left({z^3\over 1+z}\right)& (1.13)\cr
&+L(-z(1-z^2))+{7\over 4}L(z^4)-{9\over 4}L(z^2)+{1\over 2}L\left(
z{1-z\over 1+z}\right) -{1\over 2}L\left( -z{1+z\over 1-z}\right);\cr
L(-z^6(1-z)^3)&=2L(-z(1-z)(1-z^2))+L\left( -{z^4\over (1-z)(1+z)^2}\right)
& (1.14)\cr
&+3L(-z^2(1-z))+4L(z(1-z))+L\left({-z^2\over 1-z}\right)+4L(z^2)-2L(z);\cr
L\left( -{z^9\over (1+z)^3}\right) &=L(-z(1-z)(1-z^2))+2L\left(
-{z^4\over (1-z)(1+z)^2}\right)  &(1.15)\cr
&+3L\left({-z^3\over 1+z}\right)+4L\left({-z^2\over 1+z}\right)
+L(-z(1+z))+3L(z^2)+2L(z).}
$$

Proof. We have the following series of relations coming from functional
equation (1.4):
$$\eno{
\bullet ~L&(1-z^2+z^3)+L(1+z-z^3)-L((1-z^2+z^3)(1+z-z^3))
-L\left({1+z-z^2+z^4\over
1+z}\right)\cr &-L\left({-z(1+z-z^3)\over 1-z^2+z^4}\right) =0;\cr
\bullet ~L&((1-z^2+z^3)(1+z-z^3))+L\left({1+z^2\over 1+z}\right)
-L\left({1+z+z^7-z^8\over 1+z}\right) -L\left({1+z^6\over z^6}\right)\cr
&-L\left( 1-{(1+z)(1-z^2+z^4)\over z^6}\right) =0;\cr
\bullet ~L&\left( -{1+z\over z^5}\right) +L\left({(1+z)^2\over
z(1+z+z^2)}\right) +L\left({1+z-z^3\over 1+z}\right)
-L\left({-z\over 1-z^2+z^4}\right)\cr
& -L\left({(1+z)(1-z^2+z^4)\over z^6}\right) =0;\cr
\bullet ~L&\left({1+z^6\over z^6}\right) +L\left({-z(1+z-z^3)\over
1-z^2+z^4}\right) -L\left( -{1+z+z^2-z^5\over z^5}\right)
-L\left({(1+z)(1+z^2)\over z(1+z+z^2)}\right)\cr
&-L\left({1+z-z^3\over (1+z+z^2)(1-z^2+z^4)}\right) =0;\cr
\bullet ~L&\left({-z^5\over 1+z}\right) +L\left({-z^3\over 1-z^3}
\right) +L\left({z^2\over 1+z+z^2}\right)
-L(z^2(1-z))-L\left({-z^3\over 1-z^2}\right) =0;\cr
\bullet ~L&\left({1+z-z^3\over 1+z}\right) +L\left({-(1+z)z^2\over
1+z-z^3}\right) -L(-z^2)-L\left({1+z+z^2\over (1+z)(1+z^2)}\right)\cr
&-L\left({-z^5\over 1+z+z^2-z^5}\right) =0;\cr
\bullet ~L&\left({1+z\over 1+z+z^2}\right) +L\left({1+z\over z}
\right) -L\left({(1+z)^2\over z(1+z+z^2)}\right)
-L\left({-(1+z)z^2\over 1+z-z^3}\right)\cr
&-L\left({1+z\over 1+z-z^3}\right) =0;\cr
\bullet ~L&\left({z\over 1+z}\right) +L\left({z^2\over 1+z^2}\right)
-L\left({z^3\over (1+z)(1+z^2)}\right) -L\left({z\over 1+z+z^2}
\right)\cr
&-L\left({z^2\over 1+z+z^2}\right) =0;\cr
\bullet ~L&\left({1\over 1+z+z^2}\right) +L\left({-1\over z}\right)
-L\left({-1
\over z(1+z+z^2)}\right) -L\left({1\over 1+z^2}\right)\cr
&-L\left({-z\over 1+z^2}\right) =0;\cr
\bullet ~L&\left( z^3\right) +L\left({z+z^2\over 1+z+z^2}\right)+
L\left({z^2\over 1+z+z^2}\right) -L(z) -L(z^2) =0;\cr
\bullet ~{1\over 2}&L\left({z(1-z)\over 1+z}\right) +
{1\over 2}L\left({-z(1+z)\over
1-z}\right) -{1\over 2}L(-z^2) -{1\over 2}L\left({z\over 1+z}\right) -
{1\over 2}L\left({-z\over 1-z}\right) =0.}
$$
Now let us take a sum of all these equations. Using the identities
(1.3), (1.8) and (1.9) the last sum can be simplified to yield
$$\eno{
{\rm LHS}(1.13)-{\rm RHS}(1.13)&=
{3\over 2}L(-z^2)+L(z)+L(z^2)-{1\over 2}L\left({z\over 1+z}\right)+\cr
&+{1\over 2}L\left({-z\over 1-z}\right)
-2L\left({z^2\over 1+z^2}\right) -
{7\over 4}L(z^4)+{9\over 4}L(z^2)=0.}
$$
The last equality follows from Abel's duplication formula (1.7) and
(1.8). Note that the following identities had played the key role in validity
of the first six equations
$$\eno{
&1+z+z^7-z^8=(1+z^2)(1+z-z^3)(1-z^2+z^3),\cr
&1+z+z^5=(1+z+z^2)(1-z^2+z^3),\cr
&1+z+z^2-z^5=(1+z^2)(1+z-z^3).}
$$

Now let us prove the identity (1.14). We use the following equalities
coming from (1.4)
$$\eno{
\bullet ~L&(1-z+z^2)+L(1+z^2-z^3)-L((1-z+z^2)(1+z^2-z^3))\cr
&-L\left( 1+{z\over (1-z)(1+z^2)}\right) -
L\left( 1-{1\over (1-z)(1+z^2)}\right) =0;\cr
\bullet ~L&(1+z(1-z)(1-z^2))+L((1-z+z^2)((1+z^2-z^3))-L(1+z^6(1-z)^3)\cr
&-L\left( 1-{1+z\over z^5(1-z)}\right) -
L\left( 1+{1+z^2\over z^5(1-z)}\right) =0;\cr
\bullet ~L&\left({-1\over z(1-z)(1-z^2)}\right) +L\left({1+z^2\over 1+z}\right)
+L\left({1\over z^4}\right) -L\left({1+z\over z^5(1-z)}\right)\cr
&-L\left(
{-z^4\over (1-z)(1+z)^2}\right) =0;\cr
\bullet ~L&\left({z\over 1+z^2}\right) +L\left({1\over 1-z}\right)
+L(-z^2)-L(z(1-z))-L\left({1\over (1-z)(1+z^2)}\right)\cr
& -L(-z^2(1-z))=0;\cr
\bullet ~L&\left({z^3\over 1+z^2}\right) +L\left({-z\over 1-z}\right)
+L(-z^2)-
L\left({-z\over (1-z)(1+z^2)}\right) -L\left({z^3\over 1+z^2}\right) =0;\cr
\bullet ~L&\left({-1\over z^2(1-z)}\right) +L\left({-1\over z^3}\right) +
L\left({1+z^3\over 1+z^2}\right) -L\left( -{1+z^2\over z^5(1-z)}\right)
=0;\cr
\bullet ~L&(1-z+z^2)+L\left({1-z+z^2\over 1+z^2}\right) +
L\left({z\over 1+z}\right) -
L\left({1+z^2\over 1+z}\right) -L\left({1+z^3\over 1+z^2}\right) =0;\cr
\bullet ~L&(-z^3)+L\left({-z^2\over 1-z^2}\right) +
L\left({z\over 1+z}\right) -
L\left({-z^2\over 1-z}\right) -L(z(1-z))=0.}
$$
Now let us take a sum of all these equations. Using the classical
identities
(1.3), (1.8) and (1.9) this sum can be simplified to yield
$$LHS(1.14)-RHS(1.14)=2L(-z^2)+2L\left({z\over 1+z}\right) +
L\left({-z^2\over 1-z^2}\right) +4L(z^2)-2L(z)-L(z^4).
$$
Using the duplication formula (1.8) to remove the terms with negative
arguments, we obtain the value zero for the $RHS$ of the last equation,
as it was stated.

Finally, let us prove the identity (1.15). We use the following series
of equalities which are a direct consequence of the five-term relation
(1.4)
$$\eno{
\bullet ~-&L\left( -{(1+z)^3\over z^9}\right) -L\left({-z^4\over
(1-z)(1+z)^2}\right) + L\left({1+z\over z^5(1-z)}\right) +
L\left({1+z\over z^4(1+z^2)}\right)\cr
&+L\left({(1+z+z^2)(1+z+z^3)\over (1+z)^2(1+z^2)}\right) =0;\cr
\bullet ~-&L(-z(1-z)(1-z^2))-L\left({-z^4\over (1-z)(1+z)^2}\right) +
L\left({z^5(1-z)\over 1+z}\right) +L\left({-z^4\over 1-z^4}\right)\cr
&+L\left( -{z(1-z)\over 1+z^2}\right) =0;\cr
\bullet ~-&L\left({1+z\over z^4(1+z^2)}\right) -L\left( -{z^3\over 1+z}\right)
+
L\left({-1\over z(1+z^2)}\right) +L\left({1-z^3\over 1+z}\right) +
L\left({1\over z^3}\right) =0;\cr
\bullet ~-&L\left({z(1-z)\over 1+z}\right) -L\left({-z^4\over 1-z^4}
\right) +L\left({-z^5\over (1+z)^2(1+z^2)}\right)\cr
&+L\left({z\over (1+z+z^2)(1+z+z^3)}\right) +
L\left( 1-{1+z\over (1-z^3)(1+z+z^3)}\right) =0;\cr
\bullet ~L&\left({1+z\over (1-z^3)(1+z+z^3)}\right) +L\left({1+z+z^3
\over 1+z}\right)
-L\left({1\over 1-z^3}\right) -L\left({1\over 1+z+z^3}\right)\cr
&-L\left({z(1+z^2)\over 1+z}\right) =0;\cr
\bullet ~L&\left({z\over 1+z+z^2}\right)  +L\left({1\over 1+z+z^3}
\right) -L\left({z\over (1+z+z^2)(1+z+z^3)}\right)\cr
&-L\left({1\over (1+z)(1+z^2)}\right)
-L\left({z^2\over (1+z)(1+z^2)}\right) =0;\cr
\bullet ~L&\left({z(1+z^2)\over1+z}\right) +L\left({1\over
(1+z)(1+z^2)}\right) -L\left({z\over (1+z)^2}\right) -\left({1-z\over
1+z^2}\right) -L(z^2)=0;\cr
\bullet ~-&L(1-z)-L\left({z\over 1+z}\right) +L\left({z(1-z)\over 1+z}
\right) +
L\left({z^2\over 1+z^2}\right) +L\left({-1\over z(1+z^2)}\right) =0;\cr
\bullet ~L&\left({z^2\over (1+z)(1+z^2)}\right) +L\left({1+z^2\over
z^2}\right)  -L\left({1\over 1+z}\right) -L\left({1+z+z^3\over z^3}
\right)\cr
&-L\left({1-z\over 1+z^2}\right) =0;\cr
\bullet ~L&\left( -{1+z\over z^2}\right) +L\left({z\over (1+z)^2}
\right)  -L\left({-1\over z(1+z)}\right) -L\left({1\over 1+z}\right) -
L\left({-1\over z}\right) =0;\cr
\bullet ~L&\left({1+z+z^2\over 1+z}\right) +L\left({1+z\over 1-z^3}
\right) -L\left({1\over 1-z}\right) -L\left({z\over 1+z+z^2}\right)  -
L\left({1+z^2\over 1+z}\right) =0;\cr
\bullet ~2&L(z^3)+2L\left({-z\over 1-z}\right) +2L\left({-z^2\over
1-z^2}\right) -2L(-z(1+z))-2L\left({-z^2\over 1+z}\right) =0.}
$$
Taking the sum of these, simplifying and using the identities
(1.3), (1.8) and (1.9), we obtain
$$\eno{
&LHS(1.15)-RHS(1.15)=-L\left({1\over 1-z^3}\right) -L(z^2)+L(z)-
L\left({1\over 1+z}\right)-L\left({-1\over z}\right)\cr
&-L\left({1\over 1-z}\right) +L(z^3)+2L\left({-z\over 1-z}\right)
+2L\left({-z^2\over 1-z^2}\right) +3L(z^2)+2L(z)+2L(1).}
$$
Using the duplication formula (1.8), it is easy to see that the $RHS$
of the last equality is equal to zero, as we set out to prove.

Note that the following identities are crucial in the proof of (1.14)
and (1.15)
$$\eno{
&(1+z)^3+z^9=(1+z+z^2)(1+z+z^3)(1+z(1-z)(1-z^2)),\cr
&1+z-z^5+z^6=(1+z^2)(1+z(1-z)(1-z^2)),\cr
&1+z-z^4-z^6=(1-z^3)(1+z+z^3).}
$$
\qed

Now let us return back to the Coxeter and Lewin identities (1.11) and
(1.12) respectively. We start with a proving of accessibility of the
first two identities among (1.11). The last Coxeter and Lewin
identities are a
direct consequence of the functional equations (1.13)-(1.15) when
$z:=\rho$ and duplication formula (1.8). Using the relations
$$1+\rho +\rho^2=2,~~1+\rho^3=2\rho,~~2+2\rho^2-\rho^4=\rho^{-2},
$$
and functional equation (1.4), it can be shown that
$$\eno{
&L(\rho )+L(\rho^2)=L(\rho^3)+L\left({\rho^2\over 2}\right)
+L\left({1\over 2}\right);\cr
&L\left({\rho^3\over 1+\rho^3}\right) =L\left({\rho^2\over 2}\right) ;
{}~~L\left({\rho^6\over 1+\rho^6}\right) =L\left({\rho^4\over
2(1+\rho^2)}\right) ;\cr
&L\left({\rho^2\over 2}\right) +L\left({\rho^2\over 1+\rho^2}\right) =
L\left({\rho^4\over 2(1+\rho^2)}\right) +L(\rho^3)+L(\rho^4).}
$$
Hence,
$$\eno{
i)&~L(\rho^6)-4L(\rho^3)-3L(\rho^2)+6L(\rho )=
-2(L(\rho^3)+L\left({\rho^2\over 2}\right))-3L(\rho^2)+6L(\rho )\cr
&=4L(\rho )-5L(\rho^2)+L(1)={7\over 5}L(1);\cr
ii)&~L(\rho^{12})-2L(\rho^6)-3L(\rho^4)-4L(\rho^3)+6L(\rho^2)\cr
&=-2L\left({\rho^4\over 2(1+\rho^2)}\right) -3L(\rho^4)-4L(\rho^3)+
6L(\rho^2)\cr
&=-2(L\left({\rho^2\over 2}\right)
+L(\rho^3))+4L(\rho^4)-2L(\rho^2)+2L(\rho^4)-3L(\rho^4)+6L(\rho^2)\cr
&=2L(\rho^2)-2L(\rho )+L(1)={3\over 5}L(1).}
$$
Thus the accessibility of Coxeter's and Lewin's dilogarithm identities
are proved.
\medbreak

{\bf 1.2.2.} Accessibility of Watson's, Loxton's and Lewin's identities.
\medbreak

Let $\al,~-\beta$ and $-\displaystyle{1\over\gamma}$ be the three roots
of the cubic $x^3+2x^2-x-1=0$, so that $\al ={\displaystyle{1\over
2}\sec{2\pi\over 7},~\beta ={1\over 2}\sec{\pi\over 7}}$ and $\gamma
=2\cos\displaystyle{3\pi\over 7}$ all lie between 0 and 1. Then
Watson's relations [W2] are
$$L(\al )-L(\al^2)={1\over 7}L(1),~~2L(\beta )+L(\beta^2)={10\over 7}
L(1),~~2L(\gamma )+L(\gamma^2)={8\over 7}L(1).
$$

To prove these relations, first note that
$$\beta ={1\over 2}\sec{\pi\over 7}={1\over 1+\al},~~\gamma
=2\cos{3\pi\over 7}={\al\over 1+\al}.
$$
Now we use the fact that $\al$ satisfies the equation $\al^3+2\al^2-
\al -1=0$. Consequently,
$$\eno{
&{1\over 1+\beta}={1+\al\over 2+\al}=\al^2,~~\beta +\gamma =1,\cr
&{\gamma\over 1+\gamma}={\al\over 1+2\al}={1\over (1+\al
)^2}=\beta^2,~~\al +\gamma^2=1.}
$$
{}From the duplication formula (1.7),
$$\eno{
&L(\al^2)=2L(\al )-2L(\gamma ),~~L(\beta^2)=2L(\beta )-2L(1-\al^2),\cr
&L(1-\al )=L(\gamma^2)=2L(\gamma )-2L(\beta^2);}
$$
and Watson's relations follow by using (1.3) and some easy elimination.
It is interesting to compare the Watson results with the following one
(see [Kir7])
$$\eno{
&\sum_{n=1}^{r-1\over 2}L\left(\left({\sin\displaystyle{(j+1)\pi\over
r+2}\over\sin\displaystyle{(n+1)(j+1)\pi\over r+2}}\right)^2\right)=
{(3j+1)r-3j^2-1\over r+2}~L(1),&(1.16)\cr
& \cr
&r\equiv 1~(\mod 2),~~0\le j\le{r-1\over 2},~~{\rm g.c.d.}~(r+2,j+1)=1.}
$$

If $r=3$,~(1.16) is reduced to
$$L\left(\left({1\over 2}\sec{(j+1)\pi\over 5}\right)^2\right)
={9j-3j^2+2\over 5}~L(1),~~j=0,1.\eqno (1.17)
$$
But $\rho=\displaystyle{{1\over 2}\sec{\pi\over 5},~~\rho^{-1}={1\over
2}\sec{2\pi\over 5}}$ and (1.17) is equivalent to the Landen result
(1.10).

If $r=5$, (1.16) is reduced to ($j=0,1,2$)
$$L\left(\left({1\over 2}\sec{(j+1)\pi\over 7}\right)^2\right) +
L\left(\left({{1\over 4}\sec^2\displaystyle{(j+1)\pi\over 7}
\over 1-{1\over 4}\sec^2\displaystyle{(j+1)\pi\over 7}}\right)^2
\right)={15j-3j^2+4\over 7}~L(1).\eqno (1.18)
$$

Now if $j=0$ then (1.18) is reduced to ($\beta =\displaystyle{1\over 2}
\sec{\pi\over 7}$)
$$L(\beta^2)+L\left(\left(\displaystyle{\beta^2\over 1-\beta^2}
\right)^2\right) ={4\over 7}L(1).\eqno (1.19)
$$
But $\displaystyle{\beta^2\over 1-\beta^2}=1-\beta$ and
$\displaystyle{1-\beta\over 2-\beta}=\beta^2$. Consequently,
$${\rm LHS}~(1.19)=L(\beta^2)+\hbox{$2L(1-\beta )$}-2L\left(\ds
{1-\beta\over 2-\beta}\right) =2L(1)-2L(\beta )-L(\beta^2)
$$
and (1.19) is equivalent to
the second Watson relation. If $j=1$ then (1.18) is equivalent to
$$L(\al^2)+L\left(\left({\al^2\over 1-\al^2}
\right)^2\right) ={4\over 7}L(1). \eqno (1.20)
$$
But $\displaystyle{\al^2\over 1-\al^2}=1+\al$ and $\displaystyle{\al
+1\over\al +2}=\al^2$. Consequently,
$$\eno{
{\rm LHS}~(1.20)&=L(\al^2)+
\hbox{$2L(1+\al )$}-2L\left(\displaystyle{\al +1\over\al +2}\right) =
2L(1+\al )-L(\al^2)\cr
&=2L(\al )-L(\al^2) +2L(1)-L(\al^2)=2L(1)-
2\left( L(\al )+L(\al^2)\right)}
$$
and (1.20) is reduced to the first Watson's identity.

Now we are going to prove the accessibility of the following identities due
to Loxton [Lo1], [Lo2] and Lewin [Le2]
$$\eno{
&L(x^3)-3L(x^2)-3L(x)=-{7\over 3}L(1),&({\rm Loxton})\cr
&L(y^6)-2L(y^3)-9L(y^2)+6L(y)=-{2\over 3}L(1),&({\rm Lewin})\cr
&L(z^6)-2L(z^3)-9L(z^2)+6L(z)={2\over 3}L(1).&({\rm Lewin})}
$$
Here $x=\displaystyle{{1\over 2}\sec{\pi\over 9},~y={1\over
2}\sec{2\pi\over 9}={1\over 1+x}}$ and $z=2\displaystyle{\cos
{4\pi\over 9}={x\over 1+x}}$ lie in ~$(0,1)$~ and $x,-y,-\ds{1\over z}$ are
the three roots of the cubic equation
$$x^3+3x^2-1=0.
$$

The history of the discovery of the dilogarithm identities involving the
three roots
$x,-y,-\ds{1\over z}$ of the cubic $x^3+3x^2=1$ is amusing. The first was
found by Loxton [Lo1], who applied the Richmond-Szekeres method ([RS] or
Section 2.1.2, Lemma 4) to some special partition identity of the
Rogers-Ramanujan type (see identity 92 from [Sl]). It was L.~Lewin [Le2]
who observed the parallel with Watson's three identities and conjectured
the second and third relations of the triple. The second was proved by
Loxton in the same paper [Lo1], but the third one was a numerical fact
and for some time had no analytic derivation. Situation is changed only
in 1990 when H. Gangl (non published) proved the last Loxton-Lewin
identity, using methods based on Bloch groups. However, a direct proof of
accessibility of the Loxton-Lewin three relations was open problem till
now. The main purpose of this Section is to solve this problem and
present an algebraic proof of Loxton-Lewin's dilogarithm identities,
using the five-term relation (1.4) only.
Our proof is based on the following identities which are a direct
consequence of functional equation (1.4)
$$\eno{
&L(x^2)=2L(x)-2L(z);\cr
&L(y^2)=2L(y)-2L(a);\cr
&L(a^2)=2L(a)-2L(x^2);\cr
&L(z^2)=2L(z)-2L\left({x\over 1+2x}\right) ;\cr
&L(x^2)+L(y^2)=L(z^2)+L(z)+L(b);\cr
&L(y)+L(a)=L\left({x\over 1+2x}\right) +L(x)+L(b);\cr
&L(x)+L(x^2)=L(x^3)+L(a^2)+L(1)-L(c^2);\cr
&L(c^2)=2L(c)-2L(y^2);\cr
&L(c)=L(y)+{1\over 2}L(z^2);\cr
&L\left({y\over 3}\right) =L(a)+L(c^2)-L(y^2)-L(d);\cr
&L(d)=2L(z^2)-2L(x)+L(1);\cr
&L\left({z^2\over 3}\right)  =L(a^2)+L\left({x\over 1+2x}\right)
-L(z^2)-L(d);\cr
&L(y^6)=2L(y^3)-2L\left({y\over 3}\right) ;\cr
&L(z^6)=2L(z^3)-2L\left({z^2\over 3}\right) .}
$$
Here the following notations and relations are used
$$\eno{
&a:={1\over x+2},~~b:={x^2\over 1+x},~~c:={1\over x(2+x)},~~d:={x\over
(1+x)^2},\cr
&{z^3\over 1+z^3}={z^2\over 3}={x\over (1+2x)(2+x)^2},~~
{y^3\over 1+y^3}={1\over x^2(2+x)^3}={y\over 3}.}
$$
After elimination of $a,b,c$ and $d$ from the relations above, one can find
$$\eno{
&L(x^2)=2L(x)-2L(z);\cr
&L(a)=L(y)-{1\over 2}L(y^2);\cr
&L(a^2)=2L(y)-L(y^2)-2L(x^2);\cr
&L\left({x\over 1+2x}\right) =L(z)-{1\over 2}L(z^2);\cr
&L(b)=L(x^2)+L(y^2)-L(z^2)-L(z);\cr
&L(z^2)=L(y^2)+{2\over 3}L(x^2)+{2\over 3}L(x)-{4\over 3}L(y);\cr
&L(x^3)=3L(x^2)+L(z^2)-L(y^2)+L(x)-L(1);\cr
&L\left({y\over 3}\right) =-{7\over 2}L(y^2)-L(z^2)+2L(x)+3L(y)
-L(1);\cr
&L\left({z^2\over 3}\right) =-2L(x^2)-L(y^2)-{7\over 2}L(z^2)+2L(x)
+2L(y)+L(z)-L(1).}
$$
Consequently,
$$\eno{
\bullet~&3L(x)+3L(x^2)-L(x^3)=3L(x)+3L(x^2)-3L(x^2)-L(z^2)+L(y^2)
-L(x)+L(1)\cr
&=2L(x)-{2\over 3}L(x^2)-{2\over 3}L(x)+{4\over 3}L(y)+L(1)=
{4\over 3}L(x)-{2\over 3}L(x^2)+{4\over 3}L(y)+L(1)\cr
&={4\over 3}L(1)+L(1)={7\over 3}L(1);\cr
\bullet~&L(y^6)-2L(y^3)-9L(y^2)+6L(y)=-2L\left({y\over 3}\right)
-9L(y^2)+6L(y)\cr
&=7L(y^2)+2L(z^2)-4L(x)-9L(y^2)+2L(1)=2L(z^2)-2L(y^2)-4L(x)+2L(1)\cr
&={4\over 3}L(x^2)+{4\over 3}L(x)-{8\over 3}L(y)-4L(x)+2L(1)={4\over
3}\left(L(x^2)-2L(x)\right) -{8\over 3}L(y)+2L(1)\cr
&=2L(1)-{8\over 3}L(1)=-{2\over 3}L(1);\cr
\bullet~&L(z^6)-2L(z^3)-9L(z^2)+6L(z)=-2L\left({z^2\over 3}\right)
-9L(z^2)+6L(z)\cr
&=4L(x^2)+2L(y^2)+7L(z^2)-4L(x)=4L(y)-2L(z)-9L(z^2)+6L(z)\cr
&=4L(x^2)+2L(y^2)-2L(z^2)-4L(x)-4L(y)+4L(z)+2L(1)\cr
&=4L(x^2)-{4\over 3}L(x^2)-{4\over 3}L(x)+{8\over
3}L(y)-4L(x)-4L(y)+4L(z)+2L(1)\cr
&={8\over 3}\left( L(x^2)-2L(x)\right) -{16\over
3}L(y)+6L(1)=6L(1)-{16\over 3}L(1)={2\over 3}L(1).}
$$

Finally, let us prove the accessibility of (1.16) for the case $r=7$
and $j=0$. Under this assumption, (1.16) is reduced to
$$L(x^2)+L(a^2)+L(z^2)={2\over 3}L(1).\eqno (1.21)
$$
Using the relation $L(a^2)=2L(y)-L(y^2)-2L(x^2)$, one can obtain
$$\eno{
{\rm LHS}~(1.21)&=L(x^2)+2L(y)-L(y^2)-2L(x^2)+L(z^2)\cr
&=2L(y)-L(x^2)-L(y^2)+L(y^2)+{2\over 3}L(x^2)+{2\over 3}L(x)-{4\over
3}L(y)\cr
&=-{1\over 3}\left( L(x^2)-2L(x)\right) +{2\over 3}L(y)={2\over 3}
\left( L(y)+L(z)\right) ={2\over 3}L(1).}
$$
\qed

In fact, it can be shown that the Loxton-Lewin identities follow from
(1.16) with $r=7$, $j=0,1,2$ and vice versa.
\medbreak

{\bf 1.2.3.} Lewin's dilogarithm identity related with $x=\ds{1\over
2}(\sqrt{13}-3)$.
\medbreak

In the paper [Le] based on the numerical calculations the following
dilogarithm identity  was suggested
$$4L(x)-L(x^2)-{2\over 3}L(x^3)+{1\over 6}L(x^6)={7\over 6}L(1).\eqno
(1.22)
$$
We are going to prove the accessibility of this identity. For this
purpose let us set
$$a={x^3\over 1+x^3},~~b={x\over 1+x},~~c={-x\over 1-x},~~d={-1\over
1+x}.
$$
Having in mind the fact that $x$ satisfies the quadratic equation
$x^2+3x-1=0$, one can check
$${x^3\over 1+x^3}={x^2\over 2(1+x)},~~2+4x+x^2=x^{-1},~~x(1+2x)=
(1-x)^2,~~(1+x)^2=2-x.
$$
Now, using the duplication formula (1.7) and functional equation (1.4),
one can check the following relations
$$\eno{
&L(x^2)=2L(x)-2L(a),\cr
&L(x^2)=2L(-x)-2L(c),\cr
&L(x^6)=2L(x^3)-2L(a),\cr
&L\left({1\over 2}\right)+L(b^2)=L(a)+L(x^3)+L((1-x)^2),\cr
&L(b^2)=2L(b)-2L(c^2),\cr
&L(c^2)=2L(c)-2L(d),\cr
&L(d^2)=2L(d)-2L\left({-1\over x}\right),\cr
&L((1-x)^2)=2L(d^2)-2L(x).}
$$
After elimination $a,b,c,d$ from these relations, one can find
$$L(x^6)=4L(x^3)-2L(x^2)-8L(x)+8L(-x)-8L\left({-1\over x}\right) -L(1).
$$
Finally, if we substitute this expression for $L(x^6)$ to the LHS of
(1.22) and use (1.9) to remove the terms with negative arguments, the
result is
$$6{\rm LHS}(1.22)=-16L(-1)-L(1)=7L(1),
$$
as it was stated.
\qed
\medbreak

{\bf 1.2.4.} Browkin's dilogarithm identities.
\medbreak

Let $x=\displaystyle{1\over 6}(\sqrt{13}-1)$ and $z=\displaystyle
{1\over 6}(\sqrt{13}+1)$ be the roots of the quadratic
equations $3x^2+x=1$ and $3z^2-z=1$ correspondently. In the paper [Bro2],
p.261 the following dilogarithm identities were suggested
$$\eno{
&L(x^6)-6L(x^3)+L(x^2)+18L(x)=8L(1),&(1.23)\cr
&L(z^6)-3L(z^3)-6L(z^2)+9L(z)=2L(1).&(1.24)}
$$
The main aim of this section is to prove the accessibility of (1.23)
and (1.24). Let us start with a proof of the Browkin first identity. It can
be easily checked by using the functional equation (1.4) and relations
$${x^3\over 1+x^3}={x\over 4(1+x)},~~1-x+x^2=4x^2,~~3x+4=x^{-2},~~
1-x^3=2x^2(x+2)
$$
that the following identities are valid
$$\eno{
&L(x^6)=2L(x^3)-2L\left({x^3\over 1+x^3}\right) ,\cr
&L(x)+L(x^2)=L(x^3)+L\left({x\over 2(1+2x)}\right) +L\left({1+x\over
2(1+2x)}\right) ,\cr
&L\left({1\over 2}\right)+L\left({x\over 1+2x}\right) =L\left({x\over
2(1+2x)}\right) +L\left({x^2\over 1+x}\right) +L(x),\cr
&L\left({1\over 2}\right) +L\left({1+x\over 1+2x}\right)
=L\left({1+x\over 2(1+2x)}\right) +L(x(1+x))+L(x^2),\cr
&L\left({1\over 4}\right) +L\left({x\over 1+x}\right) =L\left({x\over
4(1+x)}\right) +L(x^2)+L(x(1-x)),\cr
&L(x)+L\left({x\over 1+x}\right) =L\left({x^2\over 1+x}\right)
+L\left({ 1-x\over 2(1+x)}\right) +L\left({1\over 2(1+x)}\right),\cr
&L\left({1\over 2}\right) +L\left({1-x\over 1+x}\right) =L\left({1-x
\over 2(1+x)}\right) +L(x(1-x))-L(x(1+x))+L(1),\cr
&L\left({1\over 2}\right) +L\left({1\over 1+x}\right) =L\left({1\over
2(1+x)}\right) +L\left({1\over 1+2x}\right) +L\left({x\over
1+2x}\right),\cr
&L(x(1-x))+L\left(\left({1-x\over 2x}\right)^2\right) +L\left({1\over
4}\right) =L(1),&(1.25)\cr
&L\left({1\over 1+x}\right) +L\left({1+x\over 1+2x}\right)
=L\left({1\over 1+2x}\right) +L\left({1\over 2(1+x)}\right)
+L\left({1\over 2}\right),\cr
&L\left(\left({1-x\over 2x}\right)^2\right) =2L\left({1-x\over 2x}
\right) -2L\left({1-x\over 1+x}\right) .&(1.26)}
$$
Using these identities one can find
$$\eno{
L(x^6)&=2L(x^3)+2L(x^2)+2L(x(1-x))-2L\left({x\over 1+x}\right)
-2L\left({1\over 4}\right) ,\cr
L(x^3)&=2L(x^2)+3L(x)+L\left({x\over 1+x}\right)
+L(x(1-x))-L\left({1-x\over 1+x}\right) \cr
&-L\left({1\over 2(1+x)}\right)
-{3\over 2}L(1).}
$$
Consequently,
$${\rm LHS}(1.23)=-2L(x(1-x))-2L\left({1\over 4}\right)
+4L\left({1-x\over 1+x}\right) +4L\left({1\over 2(1+x)}\right) +6L(1).
$$
Using the relations (1.25) and (1.26) one can rewrite the last
expression as
$${\rm LHS}(1.23)=4L(1)+4\left\{L\left({1-x\over 2x}\right)
+L\left({1\over 2(1+x)}\right)\right\} .
$$
But it follows from the equation $3x^2+x=1$, that
$${1-x\over 2x}+{1\over 2(1+x)}=1.
$$
The proof of the Browkin first  dilogarithm relation is over.
\qed

Now let us prove the second Browkin's identity. Having in mind an
observation that $z=\displaystyle{x\over 1-x}$, one can rewrite the LHS
of (1.24) in the following form (hint: $1+z+z^2=4x^2$)
$$-6L\left({x\over 1-x}\right)+14L(x)+L\left({1\over
4}\right)+L\left({1\over 4x}\right) -2L\left({x^3\over 1-3x+3x^2}
\right) .\eqno (1.27)
$$
In order to prove that the expression (1.27) is equal to $2L(1)$, we
are going to use the relations
$$\eno{
&{x^3\over 1-3x+3x^2}={x(1+x)\over 2},~~3x-1={x\over
1+x},~~(2-3x)x=3x-1,\cr
&1-2x=(3x-1)x,~~(2+x)x^2=(1-x)(1-x^2),~~{1\over 1+2x}={x\over 1-x^2},}
$$
which are a direct consequence of the quadratic equation for $x$,
namely, $3x^2+x-1=0$, and, additionally, the following identities
which can be obtained immediately from the functional equation (1.4)
$$\eno{
&2L\left({x(1+x)\over 2}\right) +2L\left({1+x\over 2+x}\right)
+2L\left({x\over 2+x}\right) -2L(x)-2L\left({1+x\over 2}\right) =0,\cr
&2L\left({1\over 2}\right) +2L(1-x)-2L\left({1-x\over 1+x}\right)
-2L\left({ x\over 1+x}\right) -2L\left({1-x\over 2}\right) =0,\cr
&4L\left({x\over 1-x}\right) -4L(x)-2L\left({1+x\over 2+x}\right) =0,~~
({\rm hint:}~~{1+x\over 2+x}=\left({x\over 1-x}\right)^2),\cr
&L\left({4x-1\over 4x}\right) -2L\left({1-x\over 2x}\right)
+2L\left({1-x\over 1+x}\right) =0,~~({\rm hint:}~~{4x-1\over
4x}=\left({x-1\over 2x}\right)^2),\cr
&2L\left({x\over 1-x}\right) +2L\left({1-x\over 2x}\right)
-2L\left({1\over 2}\right) -2L\left({x\over 1+x}\right)
-2L\left({3x-1\over 1-x}\right) =0,\cr
&2L(1-x^2)+2L\left({x\over 1-x^2}\right) -2L(x)-2L\left({x\over
2+x}\right) -2L\left({2\over 3}\right) =0.}
$$
Taking the sum of these, simplifying and using the relation
$\displaystyle{x\over 1-x^2}={3x-1\over 1-x}$, one can find
$$4L(x)-4L\left({x\over 1+x}\right) +2L(1-x^2)+L\left({1\over 4}\right)
-2L\left({2\over 3}\right) +L(1).
$$
The last expression can be simplified by using the duplication formula
(1.7). The result is $2L\left({1\over 2}\right) +L(1)=2L(1)$, as it was
claimed.
\qed

\bigbreak
{\bf 1.3. Dilogarithm identities and conformal weights.}
\medbreak

Let us consider the following dilogarithm sum
$$\sum^{n-1}_{k=1}\sum^r_{m=1}L\left({\sin k\varphi\cdot\sin (n-k)
\varphi\over \sin (m+k)\varphi\cdot\sin (m+n-k)\varphi}\right):=
{\pi^2\over 6}s(j,n,r),
$$
where $\varphi ={\displaystyle{(j+1)\pi\over n+r}}$,
$0\le 2j\le n+r-2$ and g.c.d. $(j+1,~n+r)=1$.

\medbreak
{\bf Theorem 2} (A.N. Kirillov [Kir7]).
1. {\it We have
$$s(j,n,r)=6(r+n)\sum_{k=0}^{[{n-1\over 2}]}\{{1\over 6}-{\overline B_2}\left(
(n-1-2k) \theta\right)\}-{1\over 4}\{2n^2+1+3(-1)^n\},\eqno(1.28)
$$
where ~${\displaystyle\theta ={j+1\over n+r}}$, ~and
{}~~${\overline B_2}(x)=-\ds{1\over \pi^2}\sum_{k=1}^{\infty}{\cos 2k\pi
x\over k^2}$ ~~is the second modified Bernoulli polynomial.}

2. (Level - rank duality)
$$s(j,n,r)+s(j,r,n)=nr-1
$$

{\bf Corollary 1.} {\it We have
$$s(j,n,r)=c_r^{(n)}-24h_j^{(r,n)}+6{\bf Z}_+,
$$
where
$$c_r^{(n)}={(n^2-1)r\over n+r}, \ \ h_j^{(n,r)}={n(n^2-1)\over 24}
\cdot{j(j+2)\over r+n}, \ \ 0\le j\le r+n-2
$$
are the central charge and conformal dimensions of the $SU(n)$ level
$r$ $WZNW$ primary fields, respectively.}

Sketch of a proof of Theorem 2.

We are going to show that the dilogarithm
identity (1.28) follows from the functional equation (1.4) and the
well-known results (F. Newman, 1847)
$${\rm Re}{\rm L}(e^{i\theta})=\pi^2{\overline B_2}\left({\theta\over
2\pi}\right) ,~~{\rm Re}{\rm L}\left({e^{i\theta}\over
2\cos\theta}\right)=-\pi^2{\overline B_2}\left({\theta\over\pi}+{1\over
2}\right).\eqno(1.29)
$$
(Hint: take $x:=\ds e^{2i\theta}$ in the Abel duplication formula (1.7)
and use functional equation for the modified Bernoulli polynomial:
${\overline B_2}(nx)=n\ds\sum_{k=0}^{n-1}{\overline B_2}\left( x+
{k\over n}\right)$.)

We start with a proving of the following accessible dilogarithm
identity
$$\eno{
&{\rm L}\left({1-a^{-1}\over 1-b^{-1}}\cdot{1-ac\over 1-bc}\right) =
{\rm L}\left({b-a\over 1-a}\right) +{\rm L}\left({(b-a)c\over 1-ac}
\right) -{\rm L}\left({(b-1)ac\over 1-ac}\right)&(1.30)\cr
&-{\rm L}\left({(bc-1)a\over 1-a}\right) +{\rm L}(abc)+{\rm
L}(ba^{-1})-{\rm L}\left({-b\over 1-b}\right) -{\rm L}\left({-bc\over
1-bc}\right) ,}
$$
where $a,b,c\in{\bf C}$ are the complex numbers such that $a\ne 1,
c^{-1}$, and $b\ne 1,c^{-1}$.

For this purpose let us apply three times the five-term relation (1.4):
$$\eno{
&{\rm L}\left({1-a^{-1}\over 1-b^{-1}}\cdot{1-ac\over 1-bc}\right)
+{\rm L}\left({1-bc\over 1-ac}\right)  -{\rm L}\left({1-a^{-1}\over
1-b^{-1}}\right) -{\rm L}\left({(a-1)bc\over 1-bc}\right) -{\rm
L}\left({1-abc\over 1-ac}\right) =0,\cr
&{\rm L}\left({(a-1)bc\over 1-bc}\right) +{\rm L}\left({(bc-1)a\over
1-a}\right) -{\rm L}(abc)-{\rm L}\left({-a\over 1-a}\right) -{\rm
L}\left({-bc\over 1-bc}\right) =0,\cr
&{\rm L}\left({1-a^{-1}\over 1-b^{-1}}\right) +{\rm L}\left({1-b\over
1-a}\right) -{\rm L}(ba^{-1})-{\rm L}\left({-b\over 1-b}\right)
+{\rm L}\left({-a\over 1-a}\right) -{\rm L}(1)=0.}
$$
Taking the sum of these,
one can obtain the relation
(1.30). The next step is to consider a specialization
$$a\longrightarrow\exp (2i\theta ),~~b\longrightarrow\exp (2i\varphi ),
{}~~c\longrightarrow\exp (2i\psi )
$$
of the dilogarithm identity (1.30). One can
easily check that under this specialization the following rules are
valid
$$\eno{
{1-a^{-1}\over 1-b^{-1}}\cdot{1-ac\over
1-bc}&\longrightarrow{\sin\varphi\over\sin\theta}\cdot{\sin (\varphi
+\psi )\over\sin (\theta +\psi )},\cr
{b-a\over 1-a}&\longrightarrow -{\sin (\varphi -\theta
)\over\sin\theta}\exp (i\varphi ),\cr
{(b-a)c\over 1-ac}&\longrightarrow -{\sin (\varphi -\theta )\over\sin
(\theta +\psi )}\exp (i(\varphi +\psi )),\cr
{(b-1)ac\over 1-ac}&\longrightarrow -{\sin\varphi\over\sin (\theta
+\psi )}\exp (i(\varphi +\theta +\psi )),\cr
{(bc-1)a\over 1-a}&\longrightarrow -{\sin (\varphi +\psi )\over\sin
\theta}\exp (i(\varphi +\theta +\psi )),\cr
{-b\over 1-b}&\longrightarrow{e^{i(\varphi +{\pi\over 2})}\over 2\cos
(\varphi +{\pi\over 2})}.}
$$
Finally, let us introduce a single-valued function (see e.g. [Le1], [KR1])
$${\rm L}(x,\theta ):={\rm Re}{\rm L}(xe^{i\theta}).
$$
Taking the real part of the both sides of specialized dilogarithm
identity (1.30) and using the Newman formulae (1.29), one can obtain
$$\eno{
&L\left({\sin\theta\over \sin\varphi}\cdot{\sin (\theta +\psi
)\over\sin (\varphi +\psi )}\right) ={\rm L}\left( -{\sin (\varphi
-\theta)\over \sin\theta},\varphi\right) +{\rm L}\left( -{\sin (\varphi
-\theta )\over\sin (\theta +\psi )},\varphi +\psi\right)&(1.31)\cr
& -{\rm L}\left( -{\sin\varphi\over\sin (\theta +\psi )},\theta
+\varphi +\psi\right) -{\rm L}\left( -{\sin (\varphi +\psi )
\over\sin\theta},\theta +\varphi +\psi\right)\cr
&+\pi^2\left\{{\overline B_2}\left({\varphi +\theta
+\psi\over\pi}\right) -{\overline B_2}\left({\varphi +\psi\over
\pi}\right) +{\overline B_2}\left({\varphi -\theta\over\pi}\right)
-{\overline B_2}\left({\varphi\over\pi}\right)\right\}.}
$$
As it was shown in [Kir7], the dilogarithm identity (1.28) follows from
(1.31) and properties of the modified Bernoulli polynomial
${\overline B_2}(x)$.
\medbreak

{\bf Remark.} One can show that the five-term relation (1.4) follows from
the nine-term relation (1.30). Namely, let us take $x:=a=b$ and $y:=bc$
in (1.30) ($a,b,c\in{\bf R}$). Then (1.30) is reduced to the relation
$${L}(xy)={L}(x)+{L}(y)+{L}\left(-x{1-y\over 1-x}\right)
+{L}\left(-y{1-x\over 1-y}\right) ,
$$
which is a form of the five-term relation due to C. Hill (see e.g. [Le1]).

Hint: using the Landen result ${\rm Re}\left(L(x)+L\left(\ds{-x\over
1-x}\right)\right)=0$, replace the terms $L(a)$ and $L(b)$ in RHS of
(1.6) on $-L\left(\ds{-a\over 1-a}\right)$ and $-L\left(\ds{-b\over
1-b}\right)$ correspondently. It is easy to check that $\ds{-a\over
1-a}=-x{1-y\over 1-x}$ and $\ds{-b\over 1-b}=-y{1-x\over 1-y}$.
\bigbreak

{\bf 1.4. Kuniba-Nakanishi's dilogarithm conjecture.}
\medbreak

Recall that the dilogarithm function is defined by
$${\rm Li}_2(z)=\sum_{n=1}^{\infty}{z^n\over n^2},~~(|z|<1).
$$
By using the integral representation
$${\rm Li}_2(z)=-\int_0^z{\log (1-t)\over t}dt,\eqno (1.32)
$$
${\rm Li}_2(z)$ can be analytically continued as a multivalued complex function
to the complex plane cut along the real axis from 1 to $+\infty$.
In the sequel the principal branch of dilogarithm which can be defined
by taking the principal branch of logarithm function
$$-\pi <{\rm Im}(\log (z))=\arg (z)\le\pi
$$
in the previous integral is used. The integration contour in (1.32)
does
not pass through the branch cut of $\log z$. Let us consider also the
principal branch of Rogers' dilogarithm using the integral
representation
$${\rm L}(z)=-{1\over 2}\int_0^z\left({\log (1-x)\over x}+{\log x\over
1-x}\right) dx,\eqno(1.33)
$$
where the principal branch of logarithm function is taken. The
integration contour in (1.33) is assumed to be located in the branch of
$\log x$.

Let us set
$$Q_m(\varphi )={\sin (m+1)\varphi\over\sin\varphi},~~f_m(\varphi
)=1-{Q_{m+1}(\varphi )Q_{m-1}(\varphi )\over Q_m^2(\varphi )}={1\over
Q_m^2(\varphi )},
$$
where $\varphi =\displaystyle{\pi (j+1)\over r+2}$, $1\le j\le r$ and g.c,d.
$(j+1,r+2)=1$.

Following [KN] and [KNS], let us define $c_0(\varphi )$, $d_m(\varphi )$, and
$y(\varphi )$ as
$$\eno{
&{\pi^2\over 6}c_0(\varphi )=\sum_{m=1}^{r-1}\left(
{\rm L}(f_m(\varphi ))-{\pi i\over 2}d_m(\varphi )\log (1-f_m(\varphi ))
\right) ,\cr
&\pi id_m(\varphi )=\log f_m(\varphi )-2\sum_{k=1}^{r-1}A_{mk}
\log (1-f_k(\varphi )),~~{\rm where}\cr
&A_{mk}=\min (k,m)-{km\over r},~~1\le m,k\le r-1,\cr
&y(\varphi )={1\over\pi i}\left\{\sum_{k=1}^{r-1}k\log (1-f_k(\varphi
))+r\log Q_{r-1}(\varphi )-(r-1)\log Q_r(\varphi )\right\}.}
$$
\medbreak

{\bf 1.4.1.} Kuniba-Nakanishi's conjecture for $sl(2)$.
\medbreak

$$N(j,r):={1\over 24}\left\{{3r\over r+2}-1-
{6j(j+2)\over (r+2)}+{6y^2\over r}-c_0(\varphi )\right\}\in{\bf Z}.
$$

{\bf Remarks.} 1) One can show that $y:=y(\varphi )$ is an integer for
any $\varphi$, whereas $d_m(\varphi )$ is an integer if $\varphi =
\displaystyle{\pi (j+1)\over r+2}$. Namely, it follows from definition
that ($f_m:=f_m(\varphi ),~~Q_m:=Q_m(\varphi )$)
$$1-f_m={Q_{m+1}Q_{m-1}\over Q_m^2}.
$$
Consequently,
$$\prod_{k=1}^{r-1}(1-f_k)^kQ_{r-1}^rQ_r^{-(r-1)}=1,
$$
that is equivalent to $\exp (2\pi iy(\varphi ))=1$. Now, if $\varphi =
\displaystyle{\pi (j+1)\over r+2}$, then $Q_r=f_r=1$ and
$$(1-f_m)^2={f_m^2\over f_{m-1}f_{m+1}}.
$$
Using the fact that the matrix $A:=(A_{mk})$ appears to be the inverse to
the
Cartan matrix $C_{r-1}=(2\delta_{i,j}-\delta_{i+1,j}-\delta_{i,j+1})$,
$1\le i,j\le r-1$, one can check
$$f_m=\prod_{k=1}^{r-1}(1-f_k)^{2A_{mk}}.
$$
Consequently, $\log (2\pi id_m(\varphi ))=1$, as it was claimed.

2) If $\varphi =\displaystyle{\pi (j+1)\over r+2}$, $0\le j\le r$,
and g.c.d.$(j+1,r+2)=1$, it can be shown that
$$y(\varphi )=r\min (j,r-j+(-1)^j).
$$

3) It follows from the definitions that ($x\in{\bf R}$)
$$\eno{
&{\rm L}(x)={\pi^2\over 3}-{\rm L}(x^{-1})-{\pi i\over 2}\log x,~~
{\rm if}~~x\ge 1,&(1.34)\cr
&{\rm L}(x)=-{\pi^2\over 6}+{\rm L}\left({1\over 1-x}\right) +
{\pi i\over 2}\log (1-x),~~{\rm if}~~ x\le 0.}
$$
Thus (see (1.5)), the single-valued dilogarithm is the real part of
${\rm L}(x)$, if $x\in{\bf R}$.

Before stating our main results of this section, let us consider an
explanatory example.
\medbreak

{\bf Example.} Assume $r=3$ and $j=1$, then we have $\varphi
=\displaystyle{2\pi\over 5}$ and
$$f_1=f_2=\left(2\sec\displaystyle{2\pi\over
5}\right)^2=\rho^{-2}={3+\sqrt5\over 2};~~1-f_1=-\rho^{-1}.
$$
First of all, $\pi id_1=\log f_1-2\log (1-f_1)=-2\pi i$. Thus,
$d_1=d_2=-2$. Further, using the fact $Q_2=1-Q_1^2=\displaystyle{1\over
f_1-1}$, one can find
$$y={1\over \pi i}\left\{\log (1-f_1)+2\log (1-f_2)+3\log Q_2\right\}
={3\over\pi i}(\log (1-f_1)-\log (f_1-1))=3.
$$
Consequently,
$$\eno{
{\pi^2\over 6}c_0(\varphi )&=2({\rm L}(f_1)+\pi i\log (1-f_1))=
2({\pi^2\over
3}-L(\rho^2)-{\pi i\over 2}\log f_1+\pi i\log (f_1-1)-\pi^2)\cr
&=2\pi^2\left({1\over 3}-{1\over 15}-1\right)=-{44\over 5}L(1).}
$$
Here we used (1.34) and Landen's result
$L(\rho^2)=\displaystyle{\pi^2\over 15}$. Finally,
$$N(1,3)={1\over 24}\left\{{9\over 5}-1-{18\over 5}+18+{44\over
5}\right\} =1.
$$
Even more, using the result (1.16) with $j=0,1,r-1,r$ and
$r\equiv 1~({\rm mod}2)$,
it can be shown that $N(1,r)=\displaystyle{r-1\over 2}$, $N(r-1,r)=r$
and $N(0,r)=N(r,r)=0$.

\medbreak
{\bf Theorem 3.} {\it Let $r$ be an odd positive integer and $j$ be an
integer such that\break $0\le 2j+(-1)^{j+1}\le r$. Then}
$$N(r-j,r)=N(j,r)+j(-1)^{j-1}{r+(-1)^{j+1}\over 2}.
$$

To obtain an additional information about the numbers $N(j,r)$ let us
introduce a function
$$b_j(r):={j^2(r-1)\over 2}-{j^3-j\over 6}\left[{r\over j+1}\right]
-N(j,r),
$$
where $[x]$ is the integer part of a real number $x$.

Note that according to Theorem 3, it is sufficient to consider a
behavior of the function $b_j(r)$ only in the "stable region", namely
when $r\ge 2j+(-1)^{j+1}$.

\medbreak
{\bf Theorem 4.} {\it Let $r$ and $j$ be the positive integers such
that
$r\ge 2j+(-1)^{j+1}$ and $r\equiv 1({\rm mod}~2)$. Then $b_j(r)$ is
an integer depending only on the residue of $r$ modulo $j+1$.}

If $r\equiv s({\rm mod}~j+1)$ and $0\le s\le j$, let us denote by
${\wt b_j}(s)$ the "stable value" of the function $b_j(r)$.
Thus for a given
positive integer $j$ we reduced a problem of computing the infinite
family of numbers $N(j,r)$, $r\ge 2j+1$, to that for finite collection
of quantities ${\wt b_j}(s)$, $0\le s\le j$, only.
\medbreak

{\bf Theorem 5.} $i)$ (Duality) If $0\le s\le j-3$, then ${\wt
b_j}(s)+{\wt b_j}(j-3-s)=\pmatrix{j\cr 3}-\pmatrix{j\cr 2}$,

$ii)$ ${\wt b_j}(j)={\wt b_j}(j-2)=\pmatrix{j\cr 3}$.

In principle, using the identity (1.16) it is possible to compute the
functions ${\wt b_j}(s)$ exactly (and consequently, to find the
remainder term $N(j,r)$ in Kuniba-Nakanishi's dilogarithm sum
$c_0(\varphi )$). We give here only partial results concerning the
computation of the numbers ${\wt b_j}(s)$.
\medbreak

{\bf Theorem 6.} $i)$ ${\wt b_j}(0)=-\left[\left(\ds{j-1\over
2}\right)^2\right]$,

$ii)$ ${\wt b_j}(1)=\left[\ds{2j\over 3}\right]\left[\ds{2j+2\over
3}\right]-\ds{j(j-1)\over 2}$,

$iii)$ ${\wt b_j}(2)=2\left[\ds{j\over 4}\right]\left[\ds{3j+3\over
4}\right]-\left[\left(\ds{j-1\over 2}\right)^2\right]$.

{\it On the other hand ($r\equiv 1({\rm mod}~2)$)},

$iv)$ $b_1(r)=0$,

$v)$ $b_2(r)=0$, if $r\ge 3$,

$vi)$ $b_3(r)=1$, if $r\ge 7$.
\medbreak

{\bf Corollary 2} (of Theorem 4). {\it If $r$ is an odd positive integer,
then $N(j,r)$ is an integer for any admissible value of $j$
(i.e. g.c.d. $(j+1,r+2)=1$).}
\medbreak

{\bf Corollary 3} (of Theorem 6). {\it We have ($r\equiv 1({\rm mod}~2)$)}

$i)$ $N(2,r)=2(r-1)-\left[\ds{r\over 3}\right]$, if $r\ge 3$,

$ii)$ $N(3,r)=9\ds{r-1\over 2}-4\left[\ds{r\over 4}\right]-1$, if
$r\ge 7$.
\medbreak

{\bf Conjecture 1.} If g.c.d. $(j+1,r+2)=1$ and $r\not\equiv 0({\rm
mod}~j)$, then $N(j,r)$ is a positive integer.

\bigbreak
{\bf 1.5. General $A_1$-type dilogarithm identity} [Kir7].
\medbreak

Let $p$ be a rational number and $k$ be a natural number. Let us consider a
decomposition of $p/k$ into continued fraction
$${p\over k}=b_r+{1\over\displaystyle b_{r-1}+
{\strut 1\over\displaystyle\cdots +{\strut 1\over\displaystyle b_1+
{\strut 1\over
\displaystyle b_0}}}},\eqno (1.35)
$$
where $b_i\in{\bf N},~~0\le i \le r-1$ and $b_r\in{\bf Z}$. Using the
decomposition (1.35) we define:
$$\eqalignno{
&y_{-1}=0,~~y_0=1,~~y_1=b_0,\ldots ,y_{i+1}=y_{i-1}+b_iy_i,\cr
&m_0=0,~~m_1=b_0,~~m_{i+1}=|b_i|+m_i,\cr
&r(j):=r_k(j)=i,~~~{\rm if}~~km_i\le j<km_{i+1}+\delta_{i,r},\cr
&n_j:=n_k(j)=ky_{i-1}+(j-km_i)y_i,~~{\rm if}~~km_i\le j<km_{i+1}+\delta_{i,r},
}
$$
where $0\le i\le r$. Using the decomposition (1.35) let us consider the
following dilogarithm sum
$$\sum_{j=1}^{km_{r+1}}(-1)^{r(j)}L_k\bigg( y_{r(j)}\theta ,~(n_j+y_{r(j)})
\theta\bigg) =(-1)^r{\pi^2\over 6}s(l,k+1,p),\eqno (1.36)
$$
where ${\displaystyle \theta ={(l+1)\pi\over ky_{r+1}+(k+1)y_r}}$, and
$$L_k(\theta ,\varphi ):=2L\left({\sin\theta\cdot\sin k\theta\over\sin\varphi
\cdot\sin (\varphi +(k-1)\theta )}\right)-\sum_{j=0}^{k-1}L\left(\left(
{\sin\theta\over\sin (\varphi +j\theta )}\right)^2\right).\eqno (1.37)
$$
\medbreak

{\bf Theorem 7.} {\it We have}
$$\eqalignno{
&(i)~~~s(0,k+1,p):=c_k={3(p+1-k)\over p+k+1}, ~~k\ge 1,&(1.38)\cr
&(ii)~~s(l,k+1,p)=c_k-{6k~l(l+2)\over p+k+1}+6{\bf Z}, &(1.39)}
$$

If $k=1$ one can obtain an exact formula for the remainder term in (1.39)
(see [Kir7] or Exercise 11).

\bigbreak
{\bf 1.6. Exercises to Section 1.}
\medbreak

{\bf 1.} Using the integral representation (1.1) for the Euler dilogarithm
${\rm Li}_2(x)$, prove the simple single-variable functional equations

\hskip -0.5cm$i)$ ${\rm Li}_2(z)+{\rm Li}_2(-z)=\ds{1\over 2}{\rm Li}_2(z^2)$,

\hskip -0.5cm$ii)$ ${\rm Li}_2(z)+{\rm Li}_2(-z)={\rm Li}-2(1)-\log z\log
(1-z)$,
\hfill(L. Euler, 1768),

\hskip -0.5cm$iii)$ ${\rm Li}_2(-z)+{\rm Li}_2\left(-\ds{1\over z}\right)=2{\rm
Li}_2(-1)-{1\over 2}\log^2(z)$.

\medbreak
{\bf 2.} Prove the following five-term, two-variable functional equations

\vskip 0.3cm
\hskip -0.5cm$i)$ ${\rm Li}_2\left[\ds{x\over 1-x}\cdot{y\over
1-y}\right]={\rm Li}_2\left[\ds{x\over 1-y}\right]+{\rm
Li}_2\left[\ds{y\over 1-x}\right]-{\rm Li}_2(x)-{\rm Li}_2(y)-\log
(1-x)\log (1-y)$,
\vskip 0.2cm
\line{\hfil (W. Spence, 1809; N. Abel, 1830)}
\vskip 0.3cm

\hskip -0.5cm$ii)$ ${\rm Li}_2(x)-{\rm Li}_2(y)+{\rm Li}_2\left(\ds{y\over x}
\right)+{\rm Li}_2
\left[\ds{1-x\over 1-y}\right]-{\rm Li}_2\left[\ds{y(1-x)\over
x(1-y)}\right]={\rm Li}_2(1)-\log x\log\left[\ds{1-x\over 1-y}\right]$,
\vskip 0.3cm
\line{\hfil (W. Schaeffer, 1846)}
\vskip 0.5cm

\hskip -0.5cm$iii)$ ${\rm Li}_2\left[\ds{x(1-y)^2\over y(1-x)^2}\right]={\rm
Li}_2\left[-x\cdot\ds{1-y\over 1-x}\right]+{\rm Li}_2\left[-\ds{1\over
y}\cdot{1-y\over
1-x}\right]+{\rm Li}_2\left[\ds{x(1-y)\over y(1-x)}\right]+{\rm Li}_2\left[
\ds{1-y\over 1-x}\right]$

$+\ds{1\over 2}\log^2(y)$,\hfill (E. Kummer, 1840)
\vskip 0.3cm

\hskip -0.5cm $iv)$ $L(x)+L(y)-L(xy)=L\left(\ds{x(1-y)\over 1-xy}\right)
+L\left(\ds{y(1-x)\over 1-xy}\right) +\log\left(\ds{1-x\over
1-xy}\right)\log\left(\ds{1-y\over 1-xy}\right)$.
\vskip 0.3cm
\line{\hfil (L.~Rogers, 1906)}

\medbreak
{\bf 3.} Prove the following nine-term, three-variable functional equations

\hskip -0.5cm$i)$ ${\rm Li}_2\left(\ds{vw\over xy}\right)={\rm Li}_2\left(
\ds{v\over x}\right)
+{\rm Li}_2\left(\ds{w\over y}\right)+{\rm Li}_2\left(\ds{v\over y}\right)
+{\rm Li}_2\left(\ds{w\over x}\right)+{\rm Li}_2(x)+{\rm Li}_2(y)-{\rm
Li}_2(v)-$

${\rm Li}_2(w)+\ds{1\over 2}\log^2\left(\ds{-x\over y}\right)$,\ \  subject
to the constraint $(1-v)(1-w)=(1-x)(1-y)$ \ (\hbox{W. Mantel,} 1898),

\vskip 0.3cm
\hskip -0.5cm$ii)$ $L\left(\ds{(1-x)(1-y)\over (1-v)(1-w)}\right)+
L\left(\ds{1-x\over 1-vy^{-1}}\right)+L\left(\ds{1-x\over 1-wy^{-1}}\right)
-L\left(\ds{1-v\over 1-vy^{-1}}\right)-L\left(\ds{1-w\over
1-wy^{-1}}\right)$

\vskip 0.3cm
$+L(x)+L(y)-L(v)-L(w)=L(1)$,
subject to the constraint $xy=vw$, $x,y,v,w\in{\bf R}$

\vskip 0.2cm
(compare with (1.30)).

Prove the accessibility of these identities.

\medbreak
{\bf 4.} Prove (${\rm L}(r,\theta ):={\rm Re}{\rm L}(re^{i\theta})$, $r,\theta
\in{\bf R}$)
$${\rm L}\left({\sin\theta\over\sin (\vp +\theta )},\vp\right)+{\rm L}\left(
{\sin\vp\over\sin (\vp +\theta )},\theta\right)=\pi^2\left\{\ol B_2(\vp
+\theta )-\ol B_2(\theta )-\ol B_2(\vp )\right\}.
$$

\medbreak
{\bf 5.} Prove the accessibility of Rogers' functional equation (Proposition
D) for $n=3$.

\medbreak
{\bf 6.} Prove the accessibility of the following dilogarithm identities

\vskip 0.2cm
\hskip -0.5cm$i)$ $\ds 6L\left({1\over 3}\right)-L\left({1\over 9}\right)
={\pi^2\over 3}$,

\vskip 0.3cm
\hskip -0.5cm$ii)$ $\ds\sum_{k=2}^nL\left({1\over k^2}\right)+2L\left({1\over
n+1}\right)={\pi^2\over 6}$,

\vskip 0.3cm
\hskip -0.5cm$iii)$ If $\al =\sqrt 2-1$, then
$$\eno{
&4L(\al )-L(\al^2)={\pi^2\over 4},&({\rm L.~Lewin})\cr
&4L(\al )+4L(\al^2)-L(\al^4)={5\pi^2\over 12},&({\rm L.~Lewin})\cr
&5L(\al^2)-L(\al^4)={\pi^2\over 6}.&({\rm L.~Lewin})}
$$
$iv)$ Let us take $a=\ds{\sqrt 3-1\over 2}$ and $c=\sqrt 3-1$, then
$$\eno{
&12L(a)+3L(a^2)-2L(a^3)={5\pi^2\over 6},&({\rm J.~Loxton})\cr
&12L(c)-9L(c^2)-2L(c^3)+L(c^6)={\pi^2\over 2}.&({\rm J.~Loxton})}
$$
$v)$ Let $u$ be the solution in $(0,1)$ of the quintic $u^5+u^4-u^3+u^2=1$.
Then
$$L(u^6)+L(u^5)-2L(u^3)+8L(u^2)-5L(u)={\pi^2\over 6}.
$$

{\bf 7.} Prove the accessibility of the so-called $\theta$-family of
single-variable functional equations (see [Le4], Chapter 6, Section (6.4)).

\medbreak
{\bf 8.} Prove the Coxeter and Lewin relations (1.11) and (1.12)
using the Ray multivariable functional equation (Proposition E).

Hint:
take $r(x,t)$ in the following forms
$$\eno{
(1-\rho x)(1+\rho^3x)&-(1-\rho^3x)(1-t),~~t:=\rho^2;\cr
(1+\rho^6x)&-(1+\rho^2x)(1-t),~~t:=\rho^3;\cr
(1-\rho^2x)(1+\rho^{10}x)&-(1-\rho^4x)^2(1-t),~~t:=\rho^4;\cr
(1+\rho^{12}x)(1+\rho^3x)&-(1+\rho^4x)^2(1-t),~~t:=\rho^6,~~{\rm or}\cr
(1+\rho^{12}x)(1+\rho^6x)&-(1-\rho^8x)(1+\rho^4x)(1-t),~~t:=\rho^6.}
$$

{\bf 9.} Prove the following properties of the Wigner-Bloch
dilogarithm $D(z)$.

\hskip -0.5cm$i)$ For $0<\theta <\pi$ the function $D(z)$ is positive and
attains its maximum value on the half line $L_{\theta}:=\{z~|~\arg
z=\theta\}$ at $z=e^{i\theta}$.

\hskip -0.5cm$ii)$ The function $D(z)$ attains its maximum on the complex
plane ${\bf C}$  at $z=\ds e^{2\pi i\over 3}$.

\hskip -0.5cm$iii)$ $\ds\lim_{z\to\infty}D(z)=0$.

\hskip -0.5cm$iv)$ $D(z)=\ds{1\over 2}\{{\rm Cl}_2(2\theta )+{\rm
Cl}_2(2w)-{\rm Cl}_2(2\theta +2w)\}$, where $\theta =\arg z$, $w=\arg
(1-\ol z)$.

Hint: using the five-term relation for $D(z)$, to show
$$D\left(\ds{z\over\ol z}\right)+D\left(\ds{1-\ol z\over
1-z}\right)=2D(z)+D\left(\ds{z(1-\ol z)\over\ol z(1-z)}\right).
$$

{\bf 10.} Prove the accessibility of the following dilogarithm
identities (the Lewin Conjectures, [Le3]).

\hskip -0.5cm$i)$ Let $\al=\ds{5-{\sqrt 21}\over 2}$ be the root of the
quadratic equation $x^2-5x+1=0$. Then
$$42L(\al )-3L(\al^2)-6L(\al^3)+L(\al^6)=\ds{5\pi^2\over 3};
$$

\hskip -0.5cm$ii)$ Let $\beta =4-{\sqrt 15}$ be the root of the quadratic
equation $x^2-8x+1=0$. Then
$$30L(\beta )+2L(\beta^2)-2L(\beta^3)-L(\beta^4)={5\pi^2\over 6};
$$

\hskip -0.5cm$iii)$ Let $\gamma =5-2\sqrt 6$ be the root of the quadratic
equation $x^2-10x +1=0$. Then
$$46L(\gamma )-15L(\gamma^2)-2L(\gamma^3)+L(\gamma^6)=\pi^2.
$$

{\bf 11.} (We use the notation of Section 1.5). Let us define for a given
positive rational number $p$ the set of integers $\{s_k\}$,
$k=,1,2,\ldots$ such that $\left[\ds{j+1\over p+2}\right]=k$ iff $s_k\le
j<s_{k+1}$, $s_0:=0$. For dilogarithm sum (1.37) to show
$$s(j,2,p)={3p\over p+2}-{6j(j+2)\over p+2}+6t(j,p),
$$
where $t(j,p)=(2k+1)j+k-2\ds\sum_{a=0}^ks_a$\ \  iff $s_k\le j<s_{k+1}$.

\medbreak
{\bf 12.} Let us define Rogers' trilogarithm as
$$L_3(x)={\rm Li}_3(x)-\log |x|L(x)-{1\over 6}\log^2|x|\log (1-x).
$$
\eject
Prove

\hskip -0.5cm$i)$ (Landen's third-order functional equations)
$$\eno{
&L_3(x)+L_3(1-x)+L_3\left({-x\over 1-x}\right)=\zeta (3),\cr
&L_3(x)+L_3(-x)={1\over 4}L_3(x^2).}
$$

\hskip -0.5cm$ii)$\ \  $L_3\left(\ds{1\over 2}\right)=\ds{7\over 8}\zeta (3)$,
\
$L_3(\rho^2)=\ds{4\over 5}\zeta (3)$.\hfill (J.~Landen, 1780)

\hskip -0.5cm$iii)$ (Nine-term relation)
$$\eno{
&2L_3(x)+2L_3(y)+2L_3\left({x(1-y)\over x-1}\right)+2L_3\left({y(1-x)\over
y-1}\right)+2L_3\left({1-x\over 1-y}\right)\cr
&+2L_3\left({x(1-y)\over y(1-x)}\right)-L_3(xy)-L_3\left({x\over
y}\right)-L_3\left({x(1-y)^2\over y(1-x)^2}\right)=2\zeta (3).}
$$
\hfill(E.~Kummer,~1840)

Does this functional equation define the trilogarithm uniquely?

\hskip -0.5cm$iv)$ Let $\al =\ds{1\over 2}(5-\sqrt{21})$. Then
$${1\over 36}L_3(\al^6)-{1\over 4}L_3(\al^2)+7L_3(\al )=
{77\over 18}\zeta (3).
$$

\bigbreak

{\bf 2.1. Dilogarithm and partitions.}
\medbreak

{\bf 2.1.1.} Rogers-Ramanujan and Gordon-Andrews identities
\medbreak

{\bf Theorem C.} {\it Let $a$ be an integer, $1\le a\le k+1$. Then}

\hskip -0.5cm $i)$ (Gordon-Andrews)
$$ \eno{\sum_{n_1,\ldots ,n_k}{q^{N_1^2+\ldots +N_k^2+N_a+\ldots +N_k}
\over (q)_{n_1}
\ldots (q)_{n_k}}&=
\prod_{n\not\equiv 0,~\pm a~{\rm mod}~(2k+3)}(1-q^n)^{-1}=&(2.1.a)\cr &\cr
&=(q)_{\infty}^{-1}\sum_{m\in{\bf Z}}
(-1)^mq^{{1\over 2}[m(m+1)(2k+3)-am]},&(2.1.b) }
$$
$ii)$ (G\"ollnitz-Gordon-Andrews)
$$\eno{
&\sum_{n_1,\ldots ,n_k}{(-q;q^2)_{N_1}q^{N_1^2+N_2^2+\cdots
+N_k^2+N_a+\ldots +N_k}
\over (q^2;q^2)_{n_1}\ldots (q^2;q^2)_{n_k}}\cr
&=\prod_{\matrix{n\not\equiv
2({\rm mod}~4)\cr n\not\equiv 0,~\pm (2a-1)~({\rm mod}~4k+4)}}(1-q^n)^{-1},
& (2.2)}
$$
{\it where $N_i=n_i +\ldots +n_k$, $(q)_n:=(q;q)_n$ and $(x;q)_n=
(1-x)(1-qx)\cdots (1-q^{n-1}x)$. }

The Rogers-Ramanujan identities correspond to $k=1$, $a=1$, or 2 in $(2.1a)$.

Note that the second equality (2.1.b) follows from the Jacobi
triple-product formula
$$\sum_{m\in{\bf Z}}x^mq^{{1\over 2}m(m+1)}=\prod_{m\ge
1}(1-q^m)(1+xq^m)(1+x^{-1}q^{m-1}),
$$
which follows from the Cauchy identity (2.6) after passing to the
limit $N\to\infty$.

There exist essentially four
methods of proving the partition identities of Gordon-Andrews' type,
namely, analytical one, using the transformation properties of $q$-series
(L.~Rogers, {G.~Watson,} I. Schur, W.~Bailey, L.~Slater, G. Andrews,  D.
Bressoud,
J.~Stembridge, ...), algebraic one, based on the
recurrence relations technique (L. Rogers, S.~Ramanujan, G.~Andrews, R.~Baxter,
P.~Forrester, ...),
combinatorial one, based on an explicit construction of a bijection
between some sets of partitions (\hbox{F. Franklin,} G.~Andrews,
W. Durfee, A. Garsia,
S. Milne, D. Bressoud, W.~Burge, ...) and group-theoretical one, based on an
investigation of the (integrable) highest-weight modules over the Kac-Moody or
Virasoro algebras and the Weyl-Kac character formula (I. Macdonald,
\hbox{A. Feingold,}
\hbox{J. Lepowsky,} R.~Wilson, \hbox{M. Primc,} V.~Kac, S.~Milne, B.~Feigin,
D.~Fuchs, E. Frenkel, ...). We give here an
analytical proof of (2.1) due to D. Bressoud [Bre2].

\medbreak
{\bf Proof of the Gordon-Andrews identity} (for $a=k+1$).

It is convenient to divide the proof into three steps.

$1^0$. Reduction to finite dimensional (polynomial) case. Note that it is
sufficient to prove the following polynomial identity (finite
dimensional analog of (2.1))
$$\eno{
&(q)_N\sum_{n_1,\ldots ,n_k}q^{n_1^2+\cdots +n_k^2}\left[\matrix{2N\cr
N-n_1,n_1-n_2,\ldots ,n_{k-1}-n_k,n_k}\right]_q\cr &  \cr
&=\sum_m(-1)^m
q^{{1\over 2}((2k+3)m^2+m)}\left[\matrix{2N\cr N+m}\right]_q,&(2.3)_N}
$$
where $\ds\left[\matrix{M\cr m_1,\ldots ,m_k}\right]_q:={(q)_M\over
(q)_{m_1}\ldots ,(q)_{m_k}(q)_{M-m_1-\ldots -m_k}}$ is the $q$-analog
of multinomial coefficient.

Indeed, it is easy to check that
$$\lim_{N\to\infty}(2.3)_N=(2.1.b).
$$

$2^0$. Generalization. It is clear that one can deduce $(2.3)_N$
from the more general identity
$$\eno{
&\sum_{n_1,\ldots ,n_{k+1}}{q^{n_1^2+\cdots +n_k^2}
{\ds\prod_{m=1}^{n_{k+1}}
(1+xq^m)(1+x^{-1}q^{m-1})}\over (q)_{N-n_1}(q)_{n_1-n_2}\cdots
(q)_{n_k-n_{k+1}}(q)_{2n_{k+1}}}\cr &  \cr
&={1\over (q)_{2N}}\sum_mx^m
q^{{1\over 2}((2k+3)m^2+m)}\left[\matrix{2N\cr N+m}\right]_q.&(2.4)}
$$
Indeed, let us take $x=-1$ in (2.4). Then the non zero terms in the
LHS(2.4) happen to appear only if $n_{k+1}=0$.

$3^0$. Induction. The proof of the identity (2.4) is based on the
following two Lemmas.

\medbreak
{\bf Lemma 1} (Classical identities).

$i)$ ($q$-binomial theorem)
$$(qx;q)_n=\sum^n_{k=1}(-1)^kq^{k(k+1)}x^k\left[\matrix{n\cr k}
\right]_q,\eqno (2.5)
$$

$ii)$ (Euler identity)
$${1\over (qx;q)_n}=\sum^n_{k=0}\left[\matrix{n\cr k}\right]_q
{x^kq^k\over (xq;q)_k},\eqno (2.6)
$$

$iii)$ (Cauchy identity)
$$\sum^n_{k=-n}x^kq^{{1\over 2}k(k+1)}\left[\matrix{2n\cr n+k}
\right]_q=\prod^n_{k=1}(1+xq^k)(1+x^{-1}q^{k-1}).\eqno (2.7)
$$

\medbreak
{\bf Lemma 2.} {\it Given $N\in{\bf N}$, $a\in {\bf C}$, then}
$$\sum^N_{m=-N}{x^mq^{am^2}\over (q)_{N-m}(q)_{N+m}}=\sum^N_{l=0}
{q^{l^2}\over (q)_{N-l}}\sum^l_{n=0}{x^nq^{(a-1)n^2}\over (q)_{l-n}
(q)_{l+n}}.\eqno (2.8)
$$
The statements of Lemma 1 are well-known (see e.g. [An1] or [GR]).
Let us
prove the \hbox{Lemma 2.} For this purpose let us take $n:=N-m$,
$x:=q^{2m}$ in (2.6) and
multiply the both sides of Euler's identity (2.6) on $(q)_{2m}^{-1}$.
We get
$$(q)^{-1}_{n+m}=\sum_k\left[\matrix{n-m\cr k}\right]_q{q^{k^2+2km}
\over (q)_{k+2m}}.\eqno (2.9)
$$
Now let us substitute the expression for $(q)_{n+m}^{-1}$ from (2.9) to
the LHS of (2.8):
$${\rm LHS}(2.8)=\sum_m{x^mq^{am^2}\over (q)_{N-m}}\sum_k{q^{k^2+2mk}
(q)_{N-m}\over (q)_{k+2m}(q)_k(q)_{N-m-k}}=\sum_{m,k}{q^{(m+k)^2}\over
(q)_{N-m-k}}\cdot{x^mq^{(a-1)m^2}\over (q)_k(q)_{k+2m}}.
$$
Finally, let us take $l:=m+k$ in the last expression.
\qed

Now let us return back to a proving of (2.4). For this purpose we
apply Lemma 2 to the RHS of (2.4):
$$\eno{
&{\rm RHS}(2.4)=\sum_m{(xq^{{1\over 2}})^mq^{{1\over 2}(2k+3)m^2}
\over (q)_{N-m}(q)_{N+m}}=\sum_{n_1}{q^{n_1^2}\over (q)_{N-n_1}}
\sum_m{(xq^{{1\over 2}})^mq^{{1\over 2}(2k+1)m^2}\over (q)_{n_1-m}
(q)_{n_1+m}}\cr &  \cr
&=\ldots ~~k~{\rm times}\ldots
=\sum_{n_1,\ldots ,n_{k+1}}{q^{n_1^2+\cdots +n^2_{k+1}}\over
(q)_{N-n_1}
(q)_{n_1-n_2}\ldots (q)_{n_k-n_{k+1}}}\sum_m{(xq^{{1\over 2}})^m
q^{{1\over 2}m^2}\over (q)_{n_{k+1}-m}(q)_{n_{k+1}+m}}.}
$$
Now let us apply the Cauchy identity (2.6) to simplify the sum over
$m$. The result is the LHS(2.4), as we set out to prove.
\qed

\bigbreak
{\bf 2.1.2.} Partitions.
\medbreak

Let us consider the following classes of partitions:
$$\eno{
&A_{k,n,a}=\left\{(m_1,\ldots ,m_p)\in{\bf Z}_+^p~\Big|~\matrix{m_1\ge
\ldots\ge m_p\ge 1,~~ m_i-m_{i+k}\ge 2,~~i+k\le p,\cr \sum m_i=n,~~
{\rm and~at~most}~ a-1~{\rm of}~ b_i~{\rm equal}~1}\right\},\cr \cr
&B_{k,n,a}=\left\{(r_1,\ldots ,r_s)\in{\bf Z}_+^s~\Big|~r_i=0,~{\rm if}~
i\equiv 0,~\pm a({\rm mod}~(2k+3)),~~\sum ir_i=n~\right\},\cr \cr
&C_{n,k,a}=\left\{ (f_1,\ldots ,f_l)\in{\bf Z}_+^l~|~\sum jf_j=n,~f_1\le
a-1,~f_j+f_{j+1}\le k,~\forall j\right\},\cr \cr
&D_{k,n,a}=\left\{ (\ld_1,\ldots ,\ld_r)\in{\bf Z}_+^r~|~\ld\vdash
n;~{\rm if}~\ld_j\ge j~{\rm then}~-a+2\le\ld_j-\ld_j'\le 2k-a+1\right\}.}
$$

\medbreak
{\bf Lemma 3.}
$$\sum_n  \# |A_{k,n,a}|q^n={\rm LHS}~~{\rm of~~ Gordon-Andrews'~~ identity},
$$
$$\sum_n\# |B_{k,n,a}|q^n={\rm RHS}~~{\rm of~~Gordon-Andrews'~~ identity}.
$$
Proof. The second statement of Lemma 3 is clear. Let us prove the first
one for $k=1$ and $a=2$. In fact, we are going to prove a slightly more
general result ($k=1,a=2$). Namely, let us consider the set
$$C_{1,N}^d=\left\{(m_1,\ldots ,m_p)\in{\bf Z}_+^p~\Bigg\vert~
\matrix{m_1\ge m_2
\ge\cdots\ge m_p\ge d,\cr m_i-m_{i+1}\ge d,~~{\rm if}~~i+1\le p,
{}~~\sum m_i=N}\right\}.
$$
Then
$$\sum_{n\ge 0}{q^{dn^2}\over (q)_n}=\sum_{(m)\in C_{1,N}^d}q^{\sum m_i}.
$$
Indeed, one can check
$$\sum_{n\ge 0}{q^{dn^2}\over (1-q)\ldots (1-q^n)}=\sum_{\{k_i\ge 0\}}
q^{dn^2+k_1+2k_2+\cdots +nk_n}.\eqno (2.10)
$$
Now let us consider the collection of integer numbers
$\{k_i\ge 0\}^n_{i=1}$
appearing in the RHS of (2.10) and define the partition
$m=(m_1,m_2,\ldots ,m_n)$ by the following rules
$$\eno{
&m_1=d(2n-1)+k_1+\ldots +k_n,~~m_2=d(2n-3)+k_2+\ldots +k_n,
{}~~\ldots ,\cr
&m_i=d(2n-2i+1)+k_i+\ldots ,k_n,~~\ldots,~~m_n=d+k_n,~~N=
m_1+\cdots ,m_n.}
$$
It is easy to check that the partition $m$ constructed by this way
belongs to the set $C_{1,N}^d$ and vice versa.
\qed

\medbreak
{\bf Theorem D.} (A.~Garsia and S.~Milne [GM], W.~Burge [Bur]).
 {\it There exist the natural bijections}
$$A_{k,n,a}\longleftrightarrow B_{k,n,a}\longleftrightarrow C_{k,n,a}
\longleftrightarrow D_{k,n,a}.
$$

{}From Theorem D there follows a combinatorial proof of the Gordon-Andrews
identity.

\medbreak
{\bf Remark.} Bijection $A_{k,n,a}\leftrightarrow C_{k,n,a}$ is obvious;
bijection $C_{k,n,a}\leftrightarrow D_{k,n,a}$ was constructed by
W.~Burge (see e.g. [Bur]). Further information can be found in Exercise~7.

\medbreak
{\bf Example.}
$$\eno{&A_{2,6,3}=\{~(3,2,1),~(4,1,1),~(6),~(5,1),~(4,2),~(3,3)~\},\cr
&B_{2,6,3}=\{~(1^6),~(2,1^4),~(2^2,1^2),~(2^3),~(5,1),~(6)~\},\cr
&C_{2,6,3}=\{~(4,1^2),~(3^2),~(3,2,1),~(4,2),~(5,1),~(6)~\},\cr
&D_{2,6,3}=\{~(3,1,1,1),~(4,1,1),~(3,2,1),~(2,2,2),~(3,3),~(4,2)~\}.}
$$
The next step (S. Ramanujan, G. Hardy, G. Meinardus, G. Andrews, B. Richmond,
\hbox{G. Szekeres,} TBA (Thermodynamic Bethe's Ansatz, Al. Zamolodchikov,
...) ...) is
to examine the asymptotic behavior of the numbers $\# |C_{k,n,a}|$ and
$\# |B_{k,n,a}|$ as $n\to\infty$. It is well known from the theory of
partitions (see e.g. [An1]) that if we introduce the partition function
$p(n)$ using the expansion of the generating function
$$\sum_{n=0}^{\infty}p(n)q^n=\prod_{n=1}^{\infty}(1-q^n)^{-1},$$
then $\log p(n)=\pi\ds\sqrt{2n\over 3} +o({\sqrt n})$ (more exactly,
$p(n)\sim\ds{1\over 4n\sqrt 3}\exp\left(\pi\sqrt{2n\over 3}\right)$).

Consequently,
$$\log (\# |B_{k,n,a}|)=\pi\sqrt{2k\over 2k+3}\cdot\sqrt{2n\over 3}
+o({\sqrt n}).
$$
In order to find the asymptotic behavior of $\log (\# |C_{k,n,a}|)$
as $n\to +\infty$ one can use the saddle point method.

\medbreak
{\bf Lemma 4.} {\it Assume that
$$\sum_{n=(n_1,\ldots ,n_k)\in{\bf Z}_+^k}{q^{nBn^t}\over (q)_{n_1}
\ldots (q)_{n_k}}=\sum_{N=0}^{\infty}a_Nq^N,\eqno (2.11)
$$
where $B$ is a symmetric and positively definite rational matrix. Then
$$\log^2a_N=4N\sum^k_{i=1}L(z_i)+o(N),
$$
where $z_i,~~1\le i\le k$,  satisfy the following system of algebraic
equations}
$$z_i=\prod^k_{j=1}(1-z_j)^{2B_{ij}}.
$$

Proof. Using Cauchy's theorem, $a_{N-1}$ can be expressed as the integral
$$a_{N-1}=\sum_{n=(n_1,\ldots ,n_k)}\oint{q^{nBn^t-N}\over (q)_{n_1}\cdots
(q)_{n_k}}.\eqno (2.12)
$$
In the sequel we are follow to the paper [RS] (see also [NRT], [Tr2],
[DKMM]). In order to analyze the behavior of the integral (1.12), we are
going to use the saddle point approximation, i.e. a crude estimate of the
integral can be obtained from the integrand evaluated at its saddle
point. First of all, we rewrite each summand of (2.12) in the exponential
form
$$\exp ((nBn^t-N)\log q-\sum^k_{j=1}\sum^{n_k}_{i=1}\log (1-q^i)).\eqno
(2.13)
$$
Using the Euler-Maclaurin formula, we first approximate
$$\log (q)_{n_k}=\sum^{n_k}_{i=1}\log (1-q^i)\simeq\int_0^{n_k}\log
(1-q^t)dt.\eqno (2.14)
$$
Hence, we replace (2.13) on
$$\exp ((nBn^t-N)\log q-\sum^k_{j=1}\int^{n_k}_0\log (1-q^t)dt).\eqno
(2.15)
$$
Furthermore, we replace the summation in (2.12) by the integration over
$dn$, treating the $n_i$'s as continuous variables, so that the RHS of
(2.12) is approximately given by
$$\oint{dq\over 2\pi i}\int dn\exp (F(q,n)),\eqno (2.16)
$$
where $F(q,n)=(nBn^t-N)\log q-\ds\sum^k_{j=1}\int_0^{n_j}\log
(1-q^t)dt$.
Now the saddle point conditions with respect to $n$, namely,
$\partial_{n_i}F(q,n)=0$, $1\le i\le k$, give the set of constraints
$$2(nB)_i\log q-\log (1-q^{n_i})=0.\eqno (2.17)
$$
If we put $x_i:=q^{n_i}$, then we obtain a system of algebraic equations
on $x_i$:
$$1-x_i=\prod^k_{j=1}x_j^{2B_{ij}}.\eqno (2.18)
$$
Further, the value of the function $F(q,n)$ at the critical point
$x=(x_1,\ldots ,x_k)$ can be found using the formulae:
$$\eno{
&\int_0^{n_i}dt\log (1-q^t)~~{\buildrel z=q^t\over =}~~{1\over\log
q}\int_1^{x_i}\log (1-z){dz\over z}
={1\over\log q}\left\{ Li_2(1-x_i)+\log x_i\cdot\log (1-x_i)\right\}\cr \cr
&({\rm hint}:~~Li_2(x)+Li_2(1-x)=L(1)-\log x\log (1-x),~~{\rm L. Euler});}
$$
$$nBn^t\log q={1\over 2\log q}\sum_i\log x_i\cdot\log (1-x_i).
$$
The result is
$$F(q,n_{\rm crit})=-N\log q-{1\over\log q}\sum^k_{j=1}L(1-x_j).
$$
Now the equation $\partial_qF(q,n_{\rm crit})=0$ fixes $q$ at the saddle
point
$$(\log q)^2={1\over N}\sum^k_{j=1}L(1-x_j),
$$
so that finally, the asymptotic behavior of $a_N$ is given by
$$a_N\sim\exp \{2(N\sum^k_{j=1}L(1-x_j))^{1\over 2}\}.
$$
Now if we put $z_j:=1-x_j$, then $z_j$ satisfy the following system of
algebraic equations
$$z_i=\prod^n_{j=1}(1-z_j)^{2B_{ij}}.
$$
\qed

\medbreak
{\bf Corollary 4} ( L. Lewin, B. Richmond, G. Szekeres, A.N. Kirillov, N.
Reshetikhin ).
$$\sum^k_{n=1}L\left(\left({\sin\ds{\pi\over k+2}\over \sin\ds{(n+1)
\pi\over k+2}}\right)^2\right) ={3k\over k+2}\cdot{\pi^2\over 6}.
$$

The interesting applications of the Corollary 4 to the study of
thermodynamic properties of the $XXX$ model one can find in [KR1] and [BR].

\vfill\eject
\bigbreak

{\bf Exercises to Section 2.1.}
\medbreak

{\bf 1.} Using the Jacobi triple product identity, prove

\hskip -0.5cm $i)$ (Euler's pentagonal number theorem)
$$\prod_{n=1}^{\infty}(1-q^n)=\sum_{m\in{\bf Z}}(-1)^mq^{{1\over
2}m(3m-1)}.
$$

\hskip -0.5cm $ii)$ (Gauss' identities)
$$\eno{
&\sum_{n\in{\bf Z}}(-1)^nq^{n^2}=\prod_{m=1}^{\infty}{1-q^m\over 1+q^m},\cr
&\sum_{n=0}^{\infty}q^{n(n+1)\over 2}=\prod_{m=1}^{\infty}{1-q^{2m}\over
1-q^{2m-1}}.}
$$

\hskip -0.5cm $iii)$  (Gauss' identity)
$$\prod_{n=1}^{\infty}(1-q^n)^3=\sum_{m\ge 0}(-1)^m(2m+1)q^{m(m+1)\over 2}.
$$
More generally, [FeSt], (we use the notation of Section 2.2)
$$\sum_{n=(n_a)\in{\bf Z}_+^{2(k-1)}}{q^{{1\over 2}n(A_2\otimes
T_{k-1}^{-1})n^t}\over\ds\prod_a(q)_{n_a}}={1\over (q)^3_{\infty}}
\sum_{n\in{\bf Z}}((2k+2)n+1)q^{(k+1)n^2+n}.\eqno (1)
$$
The Gauss identity $iii)$ corresponds to $k=1$. If $k=2$, then (1) takes
the form
$$\sum_{m\in{\bf Z}}q^{m^2}={1\over (q)^2_{\infty}}\left(\sum_{m\in{\bf Z}}
(6m+1)q^{3m^2+m}\right).
$$
Proofs and further details see in [FeSt].

\hskip -0.5cm $iv)$ (Problem). Find a polynomial analog for the identity
(1) (compare with Exercise 5).

\medbreak
{\bf 2.} Prove a $q$-analog of the binomial series
$$\sum_{n=0}^{\infty}{(a;q)_nz^n\over (q;q)_n}={(az;q)_{\infty}\over
(z;q)_{\infty}},~~|z|<1,~|q|<1. \eqno({\rm E.~Heine,~1847})
$$
Hint: both sides satisfy the functional equation $(1-z)f(z)=(1-az)f(qz)$.

\medbreak
{\bf 3.} Using the result of Exercise 2, prove the Ramanujan $~_1\psi_1$
summation formula
$$\sum_{n\in{\bf Z}}{(a;q)_nz^n\over
(b;q)_n}={(b/a;q)_{\infty}(az;q)_{\infty}(q/az;q)_{\infty}(q;q)_{\infty}
\over (q/a;q)_{\infty}(b/az;q)_{\infty}(b;q)_{\infty}(z;q)_{\infty}}.
\eqno({\rm S.~Ramanujan,~1915})
$$

\medbreak
{\bf 4.} Deduce from Ramanujan's $~_1\psi_1$-summation formula the following
identities

\vskip 0.3cm
\hskip -0.5cm $i)$\hskip 0.5cm $\ds\sum_{n\in{\bf Z}}{(-1)^nq^{n(n-1)\over
2}z^n\over
(b;q)_n}={(z;q)_{\infty}(q/z;q)_{\infty}(q;q)_{\infty}\over
(b/z;q)_{\infty}(b;q)_{\infty}}$.

\vskip 0.3cm
\hskip -0.5cm $ii)$ (Gauss-Jacobi's identity)
$$\sum_{k\in{\bf Z}}(-1)^kz_1^{k(k+1)\over 2}z_2^{k(k-1)\over 2}=
\prod_{n\ge 1}(1-z_1^nz_2^n)(1-z_1^nz_2^{n-1})(1-z_1^{n-1}z_2^n).
$$
$iii)$ (Watson's identity)
$$\eno{
&\sum_{k\in{\bf Z}}z_1^{3k^2+k\over
2}\left(z_1^{3k^2-2k}-z_2^{3k^2-4k+1}\right)=\cr \cr
&=\prod_{n\ge 1}(1-z_1^nz_2^{2n})
(1-z_1^nz_2^{2n-1})(1-z_1^{n-1}z_2^{2n-1})(1-z_1^{2n-1}z_2^{4n-4})
(1-z_1^{2n-1}z_2^{4n}).}
$$
\hfill({\rm G.~Watson,~1928})

\hskip -0.5cm $iv)$ (Kac-Van der Leur-Wakimoto's identity, [KW3])
$${1\over (q)^2_{\infty}}\left(\sum_{m,n=0}^{\infty}-\sum_{m,n=-1}^{-\infty}
\right)x^ny^mq^{mn}={(xy)_{\infty}(qx^{-1}y^{-1})_{\infty}\over
(x)_{\infty}(y)_{\infty}(qx^{-1})_{\infty}(qy^{-1})_{\infty}}.
$$
Hint: consider the Laurent series $F_1(x,y)=(q)^2_{\infty}\cdot$RHS and
$F_2(x,y)=(q)^2_{\infty}\cdot$LHS. Show that these series satisfy
the same functional equations ($i=1,2$)
$$F_i(qx,y)=y^{-1}F_i(x,y),~~F_i(x,qy)=x^{-1}F_i(x,y).
$$
$v)$ (Problem) Find the polynomial analog for Watson's and
Kac-Van~der~Leur-Wakimoto's identities.

\medbreak
{\bf 5.} Let $[x]$ denote the largest integer $\le x$. Prove 

\hskip -0.5cm $i)$ (Polynomial analog of Euler's pentagonal number theorem)
$$ \sum_{k\in{\bf Z}}(-1)^kq^{k(3k-1)\over 2}\left[\matrix{N\cr \cr
\left[\ds{N+1-3k\over 2}\right]}\right]_q=1,~~N\in{\bf Z}_+.\eqno({\rm
I.~Schur,~1917})
$$
$ii)$ (Polynomial analog of Rogers-Ramanujan's identities, $a=0$ or 1).
$$\sum_{j\ge 0}q^{j^2+aj}\left[\matrix{N-a-j\cr
j}\right]_q=\sum_{k\in{\bf Z}}(-1)^kq^{{1\over 2}k(5k+1)-2ak}\left[
\matrix{N\cr \cr\ds{1\over 2}(N-5k)+a}\right]_q.
$$
\line{\hfill (I.~Schur,~1917;~~G.~Andrews,~1970)}

Hint: check that each side satisfies the recurrence relation
$f_N=f_{N-1}+q^{N-1}f_{N-2}$.

\medbreak
{\bf 6.} Deduce from Exercise 5 the Rogers-Ramanujan identities ($a=0$ or 1)
$$\sum_{j\ge 0}{q^{j^2+aj}\over (q)_j}=\prod_{j\ge 0}{1\over (1-q^{5j+1+a})
(1-q^{5j+4-a})}.\eqno (2)
$$

$\bullet$~Prove,  \ \ RHS(2)=$\ds\lim_{N\to\infty}\sum_{\matrix{\sigma_i\in\{
0,1\},~\sigma_i\sigma_{i+1}=0\cr \sigma_1=a,~\sigma_{N+1}=0}}q^{
\sum_{j=1}^{N-1}j\sigma_{j+1}}$
$$={1\over (q)_{\infty}}\sum_{k\in{\bf Z}}
\left(q^{k(10k+1+2a)}-q^{(2k+1)(5k+2-a)}\right).\eqno (3)
$$

{\bf 7. } (Fusion rules and Rogers-Ramanujan's type identities).

{\bf A.} Let $r\ge 3$ be an odd integer and $l:=\ds{r-3\over 2}$. Let us
introduce

\hskip -0.5cm $i)$ matrix $M:=M(x)\in{\rm Mat}_{l+1\times l+1}({\bf
Z}[x])$, where ($1\le i,k\le l+1$)
$$M_{ik}:=\cases{0,& if $i+k<l+2$;\cr x^{l-k+1},& if $i+k\ge l+2$;\cr}
$$
$ii)$ the highest weight vector $|j>=e_{l-j+1}+\ldots +e_{l+1}$, where
$0\le j\le l$ and $e_i$ is the basis vector in $({\bf R}^l)^*$, namely,
$e_i^t=(\delta_{ik})$, $1\le k\le l+1$;

\hskip -0.5cm $iii)$ vector $a_{j,N}:=a_{j,N}(x;q)=(a_{j,N}^{(1)},\ldots
,a_{j,N}^{(l+1)})$ by the following rule
$$M(q^{N-1}x)M(q^{N-2}x)\cdots M(qx)M(x)|j>=a_{j,N}^t.
$$

$\bullet$~Prove, \hskip 0.5cm $a_{j,N}^{(p)}(x,q)=$
$$\sum_{m=(m_1,\ldots ,m_l)\in{\bf
Z}_+^l}x^{N_1+\ldots +N_l}q^{N_1^2+N_2^2+\ldots +N_l^2-N_1-\ldots -N_j}
\prod^l_{k=1}\left[\matrix{P_k(m;j,p)+m_k\cr m_k}\right]_q,
$$
where $N_i:=m_i+m_{i+1}+\ldots +m_l$, $1\le p\le l+1$, \ \ and
$$P_k(m;j,p):=kN+\min (k,l-j)-\max (k+1-p,0)-2\sum_{i=1}^l\min (i,k)m_i.
$$

It is well-known (see e.g. [FrSz1]) that
$$\lim_{N\to\infty}a_{j,N}^{(p)}(q,q)=\chi_{1,1+j}^{(2,r)}(q).
$$
Using the Feigin--Fuchs--Rocha-Caridi character formula for
$\chi_{1,1+j}^{(2,r)}(q)$ (see e.g. [FF] or formula (5)), and the exact
expression for
$a_{j,N}^{(p)}(x,q)$, we obtain the group-theoretic proof/ex\-pla\-nation of
the Gordon-Andrews identity. Even more, one can consider the polynomial
$(x;q)_N\cdot a_{j,N}^{(l+1)}(x,q)$ as a "natural finitization" of the
Watson-Andrews identity (9). However, the exact computation of
polynomials $(x;q)_N\cdot a_{j,N}^{(l+1)}(x,q)$ seems to be very difficult
(even for $x=1$). In this direction one can obtain the following partial
result.
\medbreak

{\bf Theorem 8} (A.N. Kirillov). {\it We have} \hskip 0.3cm $(x;q)_N
a_{j,N}^{(l+1)}(x,q)\equiv$
$$\eno{
&\sum_{m\ge
0}(-1)^mx^{(l+1)m}q^{{1\over 2}(2l+3)m(m+1)-(l+2+j)m}\left[
\matrix{(l+1)(N-2m)+l+2+j\cr m}\right]_q\cr
&\cdot \left(1-(q^{2m}x)^{1+j}\right){(x;q)_{m}\over  (q;q)_{m}}~({\rm
mod~deg}_qN).}
$$
Taking the limit $N\to\infty$ one can obtain the Watson--Andrews identity
(9).

{\bf B.} (Fusion algebra). Let us define the level $r$ fusion algebra
${\cal F}_r$ as a finite dimensional algebra over rational numbers ${\bf
Q}$ with generators $\{v_j~|~j=0,\ds{1\over 2},1,{3\over 2},\ldots ,
{r-2\over 2}\}$ and the
following multiplication rule (the level $r$ Clebsch-Gordan series):
$$v_{j_1}\wh{\otimes}v_{j_2}=\sum_{j=|j_1-j_2|,~j-j_1-j_2\in{\bf Z}}^{\min
(j_1+j_2,~r-2-j_1-j_2)}v_j.\eqno (4)
$$
It is well known (see e.g. [Kac]), that the fusion algebra ${\cal F}_r$
is a commutative and associative one. Note also, that the fusion rules (4)
correspond to a decomposing the tensor product
$V_{j_1}\wh{\otimes}V_{j_2}$ of the restricted representations [Ros]
$V_{j_1}$ and $V_{j_2}$ of the Hopf algebra $U_q(sl(2))$ when $q$ is the
root of unity $q=\exp \left(\ds{2\pi i\over
r}\right)$ into the irreducible parts  (see e.g. [Lu3]).

Further, let us denote by
Mult$_{V_k}(V_{j_1}\wh{\otimes}\cdots\wh{\otimes}V_{j_N})$ the coefficients
which appear in the decomposition of the product $v_{j_1}\wh{\otimes}\cdots
\wh{\otimes}v_{j_N}$ in the fusion algebra ${\cal F}_r$:
$$v_{j_1}\wh{\otimes}\cdots\wh{\otimes}v_{j_N}=\sum_k{\rm Mult}_{V_k}(v_{j_1}
\wh{\otimes}\cdots\wh{\otimes}v_{j_N})\cdot v_k.
$$
$i)$ Prove the following relation (see Exercise 7, part A, $0\le j\le l$)
$$a_{j,N}^{(l+1)}(1,1)={\rm
Mult}_{V_{r(j)}}(V_s^{\wh{\otimes}(N+1)}),~~s:=\left[{l+1\over 2}\right],
$$
where for given $j$, $r(j)$ is the unique integer such that
$$0\le r(j)\le{r-3\over 2}~~{\rm and}~~j{r-3\over 2}\equiv\pm r(j)~({\rm
mod}~(r-2)).
$$
$ii)$ Prove the following multiplicity formula (cf. [Kir6]):
$${\rm Mult}_{V_k}(V_{j_1}\wh{\otimes}\cdots\wh{\otimes}V_{j_N})=
\sum_{\{\nu\}}\prod_{n\ge 1}\pmatrix{P_{n,r}(\nu_ij)+m_n(\nu )\cr m_n(\nu
)},
$$
where the summation is taken over all partitions $\nu
=(\nu_1\ge\nu_2\ge\cdots\ge 0)$ such that

$a)$ $|\nu |:=\nu_1+\nu_2+\ldots =\ds\sum_{s=1}^Nj_s-k$,

$b)$ (inequalities for vacancy numbers)
$$P_{n,r}(\nu ;j):=\sum_{s=1}^N\min (n,2j_s)-\max (n+2k+2-r,0)-2Q_n(\nu
)\ge 0.
$$
Here $Q_n(\nu ):=\ds\sum_{j\ge 1}\min (n,\nu_j)$ and $m_n(\nu )$ is the
number of parts of partition $\nu$ which are equal to $n$.

{\bf C.} (Restricted Kostka-Foulkes polynomials).

For the given natural number $l$, partition $\ld =(\ld_1\ge\ld_2\ge 0)$
and composition $\mu$, we define the level $l$ (or restricted)
Kostka-Foulkes polynomial $K_{\ld ,\mu}^{(l)}(q)$ by the following way
$$K_{\ld ,\mu}^{(l)}(q):=\sum_{\nu,~l(\nu')\le l}q^{2n(\nu )}\prod_{n\ge 1}
\left[\matrix{P_n(\nu ;\mu)+m_n(\nu )\cr m_n(\nu )}\right]_q,
$$
where summation is taken over all partitions $\nu$ such that

$i)$ $|\nu |=\nu_1+\nu_2+\cdots  =\ld_2$, $\nu_1\le l$,

$ii)$ $P_n(\nu ;\mu ):=\ds\sum_j\min (n,\mu_j)-2Q_n(\nu )\ge 0$, $\forall
n$.

It follows from the Exercise 7, A, that
$$a_{0,N}^{(l+1)}(x;q)=\sum_{0\le m\le{Nl\over 2}}(qx)^m
K_{(Nl-m,m),(l^N)}^{(l)}(q).
$$

$\bullet$~Prove, that if $m\le l$, then
$$K_{(Nl-m,m),(l^N)}^{(l)}(q)=\ol{K}_{(Nl-m,m),(l^N)}(q)=
\left[\matrix{m+l-1\cr l-1}\right]_q.
$$

$\bullet$~Using the definition of restricted Kostka polynomials $K_{\ld
,\mu}^{(l)}(q)$, prove that the $q$-binomial coefficients
$\ds\left[\matrix{m\cr n}\right]_q$ are the unimodal and symmetric
polynomials (cf. [Kir5]).
\medbreak

{\bf Conjecture 2.} {\it Polynomials $K_{(Nl-m,m),(l^N)}^{(l)}(q)$
are unimodal.}

Let us consider a simple example with $l=2$, $N=m=4$. One can compute the
Kostka polynomial $\ol{K}_{(4,4),(2^4)}(q)$ by using the Theorem 10.
Namely, there exist three configurations
\vskip 0.2cm
$$\matrix{\Hthreebox(,,)~0,&~~{\overline c}=1+1+1+1-4=0,\cr \cr \cr
	\Threeone(,,,),&~~{\overline c}=3+1+1+1-4=2,\cr \cr \cr
	\Twotwo(,,,)~0,&~~{\overline c}=3+3+1+1-4=4. }
$$
\vskip 0.2cm
Consequently, $\ol{K}_{(4,4),(2^4)}(q)=1+q^2+q^4$ and this is a
non-unimodal polynomial, whereas the level 2 Kostka polynomial
${K}_{(4,4),(2^4)}^{(2)}(q)=q^4$  is the unimodal one.

{\bf D.} (Polynomial analog of Gordon-Andrews' identities).

Let us remind the basic facts about the minimal series irreducible
representations of the Virasoro algebra. Given two (coprime) positive
integers $p'>p\ge 2$, then central charge and highest weights of the
corresponding irreducible representations ${\cal V}_{r,s}^{(p,p')}$ of
the Virasoro algebra are (see e.g. [FF])
$$\eno{
&c^{(p,p')}=1-{6(p'-p)^2\over pp'}~~{\rm and}\cr
&\Delta_{r,s}^{(p,p')}={(rp'-sp)^2-(p'-p)^2\over 4pp'},~~1\le r\le
p-1,~1\le s\le p'-1.}
$$
The characters of these representations are ([R-C], [FF])
$$\chi_{r,s}^{(p,p')}(q):=q^{-\Delta_{r,s}^{(p,p')}}{\rm Tr}_{{\cal
V}_{r,s}^{(p,p')}}q^{L_0}={1\over (q)_{\infty}}\sum_{k\in{\bf Z}}\left(
q^{k(kpp'+rp'-sp)}-q^{(kp+r)(kp'+s)}\right).\eqno (5)
$$
Note the symmetry of the "conformal grid"
$$\eno{
&\{(r,s)~|~1\le r\le p-1,~1\le s\le p'-1\},&(6)\cr
&(r,s)\leftrightarrow (p-r,p'-s):~\Delta_{r,s}^{(p,p')}=
\Delta_{p-r,p'-s}^{(p,p')}\Rightarrow\chi_{r,s}^{(p,p')}(q)=
\chi_{p-r,p'-s}^{(p,p')}(q).}
$$
Note also that the RHS(3) is precisely $\chi_{1,2-a}^{(2,5)}(q)$ as given
by the RHS(5).

We are interested in finding a natural polynomial analog for the
character $\chi_{r,s}^{(p,p')}(q)$. For this purpose let us introduce the
polynomials (cf. [Me2])
$$\eno{B_{p,p';r,s}^{(N)}(q)&:=\sum_{k\in{\bf Z}}\left\{q^{k(kpp'+rp'-sp)}
\left[\matrix{N\cr \cr \left[\ds{N+s-r-d(p,p')\over 2}\right]-p'k}\right]_q
\right.\cr
&\left. -q^{(pk+r)(p'k+s)}\left[\matrix{N\cr \cr \left[\ds{N-s-r-d(p,p')
\over 2}\right]-p'k}\right]_q\right\},& (7)}
$$
where $d(p,p'):=\left[\ds{p'-p\over 2}\right]$.

It is clear that in the "physical region" (6)
$$\lim_{N\to\infty}B_{p,p';r,s}^{(N)}(q)=\chi_{r,s}^{(p,p')}(q).
$$

\medbreak
$\bullet$~~Prove the following properties of the "bosonic" polynomials (7)

\hskip -0.5cm $i)$~ (Symmetry) If $p\equiv p'~({\rm mod}~2)$,
$1\le r\le p-1$ and $1\le s\le p'-1$, then
$$B_{p,p';r,s}^{(N)}(q)=B_{p,p';p-r,p'-s}^{(N)}(q).
$$
$ii)$~ (Positivity) All coefficients of the polynomials
$B_{p,p':r,s}^{(N)}(q)$ in the "physical region" (6) are the non-negative
integers.

\hskip -0.5cm $iii)$ (Recurrence relations and initial conditions)
$$\eno{
&B_{p,p';r,s}^{(N)}(q)=B_{p,p';r.s}^{(N-1)}(q)+\cases{q^{N+s-r-d(p,p')\over
2}B_{p,p';r-1,s}^{(N-1)}(q),&if $N\equiv s-r-d(p,p')({\rm mod}~2)$;\cr \cr
q^{N-s+r+d(p,p')+1\over 2}B_{p,p';r+1,s}^{(N-1)}(q),&if $N\not\equiv
s-r-d(p,p')({\rm mod}~2)$;\cr}\cr \cr
&B_{p,p';r,s}^{(0)}(q)=\delta_{s,r+d(p,p')}+\delta_{s,r+1+d(p,p')}.}
$$

\hskip -0.5cm $iv)$~ (Fusion multiplicities) Assume that $p'\equiv 1(2)$,
$l=(p-3)/2$ and $\{V_0,V_1,\ldots ,V_l\}$ are the odd dimensional
restricted representations of the quantum group $U_q(sl(2))$ at
the root
of unity $q=\exp\left(\ds{2\pi i\over p'}\right)$ (see [Lu3] or Exercise~7,
A and B). If $-{d}\le r\le p+{\wt d}$,
$1\le s\le p'-1$ then $B_{p,p';r,s}^{(N)}(1)$ is a
linear combination with non-negative integer coefficients of the "fusion
multiplicities" ~~Mult$_{V_j}(V_l^{{\wh\otimes}N})$, ~$0\le j\le l$.

\medbreak
$\bullet$~ Prove, for example, that ($1\le s\le l+1$)
$$B_{2,2l+3;1,s}^{(N)}(1)=\sum_{j=1}^s{\rm
Mult}_{V_{r(j;l)}}\left(V_l^{{\wh\otimes}N}\right),
$$
where $r(j;l):=\left[\ds{l+1\over 2}\right]+(-1)^{l+j}\left[\ds{j\over
2}\right]$;
$$B_{p,p';r,1}^{(N)}(1)=\cases{{\rm Mult}_{V_{\left[{r+d\over
2}\right]}}\left(V_l^{{\wh\otimes}N}\right),& $N\equiv 0(2)$,\cr\cr
{\rm Mult}_{V_{\left[{p-r+{\wt d}\over
2}\right]}}\left(V_l^{{\wh\otimes}N}\right),& $N\equiv 1(2)$,\cr}
$$
where ${\wt d}:={\wt d}(p,q)=\ds\left[{p'-p-1\over 2}\right]$,
{}~$-d\le r\le p+{\wt d}$.
\vskip 0.2cm

$\bullet$~Prove that
$$B_{p,p';r,s}^{(N)}(1)=B_{p,p';r+1,s}^{(N)}(1),~~{\rm if}~~s+r\equiv
{1+(-1)^N\over 2}({\rm mod}~2).
$$

\medbreak
$\bullet$~Let us define a matrix ${\cal B}_{p,p'}^{(N)}$ of the size
$p'\times (p'-1)$ by the following manner
$$\left({\cal B}_{p,p'}^{(N)}\right)_{i,j}=B_{p,p';i-1-d,j}^{(N)}(1),~~
1\le i\le p',~~1\le j\le p'-1.
$$

Prove that ($p'\equiv 1(2)$)
$${\cal B}_{2,p'}^{(N)}={\cal B}_{3,p'}^{(N)}=\cdots =
{\cal B}_{p'-1,p'}^{(N)}.
$$

$\bullet$~Let us put ${\cal B}_{p'}^{(N)}:={\cal B}_{p,p'}^{(N)}$, where
$p'\equiv 1~({\rm mod}~2)$ and $1\le p\le p'-1$.

Prove ($1\le i\le p'$, $1\le j\le p'-1$)

\hskip -0.5cm $j)$ If $i+j\equiv N+1~({\rm mod}~2)$, then
$$({\cal B}_{p'}^{(N)})_{ij}=\cases{\ds\sum_{k=1}^{\min (i-1,p'-i+1,j,p'-j)}
{\rm Mult}_{V_{k-1+{1\over 2}|p'-i-j+1|}}(V_l^{\wh\otimes^N}),& if
$N\equiv 1~({\rm mod}~2)$;\cr \cr
\ds\sum_{k=1}^{\min (i-1,p'-i+1,j,p'-j)}
{\rm Mult}_{V_{k-1+{1\over 2}|i-j-1|}}(V_l^{\wh\otimes^N}),& if
$N\equiv 0~({\rm mod}~2)$.}
$$
$jj)$ If $i+j\equiv N~({\rm mod}~2)$, then
$$({\cal B}_{p'}^{(N)})_{ij}=\cases{\ds\sum_{k=1}^{\min (i,p'-i,j,p'-j)}
{\rm Mult}_{V_{k-1+{1\over 2}|p'-i-j|}}(V_l^{\wh\otimes^N}),& if
$N\equiv 1~({\rm mod}~2)$;\cr \cr
\ds\sum_{k=1}^{\min (i,p'-i,j,p'-j)}
{\rm Mult}_{V_{k-1+{1\over 2}|i-j|}}(V_l^{\wh\otimes^N}),& if
$N\equiv 0~({\rm mod}~2)$.}
$$

Let us remark that
$$B_{p,p';r,s}^{(N)}(1)=({\cal B}_{p'}^{(N)})_{r+d+1,s}.
$$

It is an interesting problem to find a pure combinatorial interpretation
in terms of some kind of partitions for the numbers
$B_{p,p';r,s}^{(N)}(1)$ in the region $\{ -d\le r\le p+{\wt d},~1\le s\le
p'-1\}$, as well as their natural $q$-analogs.
\medbreak

{\bf A proof of part $ii)$.} Following the paper [ABBBFV], we give a
combinatorial interpretation of the "bosonic" polynomials
$B_{p,p';r,s}^{(N)}(q)$.
\medbreak

{\bf Definition 4.} {\it Let $\ld$ be a partition/Young diagram. For any box
$x\in\ld$ lying in the $i$-th row and $j$-th column of the Young diagram
$\ld$, let us define the hook difference at the $x$ (notation hd$(x)$) as
follows}
$${\rm hd}(x)=\ld_i-\ld_j'.
$$
\medbreak

{\bf Definition 5.} {\it We say that a box $x:=(i,j)\in\ld$ lies on
diagonal}
$c$, if $i-j=c$.
\medbreak

{\bf Definition 6.} {\it For given  positive integers ~$A,B,~p,p',~r$~
and $s$ let
us define\break ${\cal R}_{p',s}(A,B;p,r;n)$ to be the set of partitions of $n$
into at most $B$ parts each $\le A$ such that the hook differences on
diagonal $1-r$ are $\ge 1+r-s$ and on diagonal $p-r-1$ are $\le
p'-s-p+r-1$.}

The related generating function is, of course, a polynomial
$$D_{p',s}(A,B;p,r;q)=\sum_{n\ge 0}\# |{\cal R}_{p',s}(A,B;p,r;n)|q^n.
$$
\medbreak

{\bf Theorem} ([ABBBFV]). {\it Assume that $1\le r<p$ and $1\le 2s\le
p'$. Then
$$B_{p,p';r,s}^{(N)}(q)=D_{p',s}(N-B,B;p,r;q),
$$
where \ \ $B=\left[\ds{N+s-r-d(p,p')\over 2}\right]$.}
\medbreak

{\bf Remark.} It seems plausible that for given integers $p'>p\ge 2$,
$N\ge 1$ and $r$, the polynomials $B_{p,p';r,s}^{(N)}(q)$ have the
nonnegative coefficients for all $s$, $1\le s\le p'-1$, if and only if
$r$ satisfies the inequalities $0\le r\le p$. For example,
$$B_{4,9;0,1}^{(9)}(q)=q^4(1+q^4)(1+q^3+q^6){1-q^7\over 1-q}.
$$
However, one can prove that
$\ds\lim_{N\to\infty}B_{p,p';r,s}^{(N)}(q)=0$, if $r=0$ or $p$ and $1\le
s\le p'-1$.

$\bullet$~It is an interesting task to find a natural representation of
the "restricted Hecke algebra" ${\wt H}_N(q)$, $q=\exp\ds\left({2\pi
i\over p'}\right)$, $p'\equiv 1({\rm mod}~2)$ (see e.g. [GW]), in the linear
space
generated by the finite set $\ds\coprod_n{\cal R}_{p',s}(N-B,B;p,r;n)$.
\medbreak

$\bullet$~Another interesting problem is to find the "fermionic"
representation (see e.g. [Me2]) for $B_{p,p';r,s}^{(N)}(q)$. Such
representation is known in the following three cases:

$i)$ $(p,p'):=(p,p+1)$, $d(p,p')=0$, E. Melzer [Me2] (conjecture),
A.~Bercovich [Ber] (proof the Melzer conjecture for all pairs $(r,s)$
with $s=1$).

$ii)$ $(p,p'):=(2,2l+3)$, $d(p,p')=l$, Y.-H. Quano, A.N. Kirillov.

$iii)$ $(p,p'):=(2l+2)$, $d(p,p')=l$, Y.H. Quano.

\medbreak
{{\bf Theorem 9}\footnote*{After finishing this work, I was informed by
A.~Kato about the recent preprint of Y.-H.~Quano [Q1] which contains the
proof of Theorem 9. Also in [Q1] a polynomial analog of Bressoud's
identity (10) is given.}} (A.N. Kirillov). {\it We have  the following
equivalent expressions for the RHS(7) (we use the notation from Exercise
7, A,B,C,D. Note also, that in our case $r=1$ and we assume that
$0\le s\le l+1$):\ \  $B_{2,2l+3;1,s}^{(N)}(q)=$
$$\eno{
&=\sum_{(m_1,\ldots ,m_l)\in{\bf Z}_+^l}q^{N_1^2+\cdots +N_l^2+N_s+\cdots
+N_l}\prod_{k=1}^l\left[\matrix{P_k(m;s,N)+m_k\cr m_k}\right]_q &(8)\cr
&=\sum_mq^{2m}\sum_{\nu\vdash m}q^{2n(\nu )-Q_{s-1}(\nu )}\prod_{k=1}^l
\left[\matrix{N-\max (k+1-s,0)-2Q_k(\nu )+m_k(\nu )\cr m_k(\nu
)}\right]_q,}
$$
where ( the so-called vacancy numbers)}
$$P_k(m;s,N):=N-\max (k+1-s,0)-2\ds\sum_{j=1}^l\min (k,j)m_j\ge 0.
$$

Our proof is based on the Theorem [ABBBFV], the combinatorial description
of the RHS(8) in terms of the special type partitions (see e.g. [AB],
[Bre4], [Bre5]) and that in terms of the rigged configurations (see e.g.
[Kir3]).

\medbreak
{\bf Problems.} $i)$ To find the corner-transfer-matrix type
representation for the  "bosonic" sum (7) (see e.g. [ABF], [FB], [Me2]).

$ii)$ To find a natural polynomial analog for the G\"ollnitz-Gordon-Andrews
identity (2.2). It is well-known, that the RHS(2.2) is the character of a
representation of the Neveu-Schwarz algebra.

\medbreak
{\bf 8.} Prove the following generalizations of the Gordon-Andrews
identity (2.1):

\vskip 0.3cm
$i)$ $\ds\sum_{n_1,\ldots ,n_k}{z^{N_1+\cdots +N_k}q^{N_1^2+\cdots
+N_k^2+N_j+\cdots +N_k}\over (q)_{n_1}\ldots (q)_{n_k}}$\hfill (9)

\vskip 0.3cm

\hskip 0.5cm $=\ds{1\over (qz)_{\infty}}\left\{\sum_{m\ge
0}(-1)^mz^{(k+1)m}q^{(2k+3){m(m+1)\over
2}-jm}(1-(q^{2m+1}z)^j){(qz)_m\over (q)_m}\right\}$,

\vskip 0.3cm
\hfill(L. Rogers and S. Ramanujan, 1920; G.~Watson, 1928; G.~Andrews, 1974)
\vskip 0.3cm

$ii)$ $\ds\sum_{n_1,\ldots ,n_k}{(zq^{-1})^{N_1+\cdots
+N_k}q^{N_1^2+\cdots +N_k^2}(a;q^{-1})_{N_1}(b;q^{-1})_{N_1}\over
(q)_{n_1}\ldots (q)_{n_k}}$
\vskip 0.3cm

$=\ds\sum_{m\ge 0}(-1)^mz^{(k+1)m}q^{(2k+3){m(m+1)\over 2}}
{(a;q^{-1})_m(b;q^{-1})_m(azq^m)_{\infty}(bzq^m)_{\infty}(1-zq^{2m-1})\over
(q)_m(zq^{m-1})_{\infty}(abz)_{\infty}}$.

\vskip 0.3cm
\hfill (G.~Andrews, J.~Stembridge)
\vskip 0.3cm

{\bf 9.} Prove the following identity
$$\sum_{n_1,\ldots ,n_l}{q^{N_1^2+\ldots +N_l^2+N_s+\ldots +N_l}\over
(q)_{n_1}\cdots (q)_{n_{l-1}}(q^2;q^2)_{n_l}}=\prod_{n\not\equiv 0,\pm
s~({\rm mod}~2l+2)}(1-q^n)^{-1}.\eqno(10)
$$
{\hfill(D.~Bressoud, 1980)}
\medbreak

$\bullet$~ Prove the polynomial analog of Bressoud's identity (10),
{}~ $1\le s\le l+1$ (see [Q1]),

\vskip 0.3cm
$B_{2,2l+2;1,s}^{(N)}(q)=$
$$\sum_{m=(m_1,\ldots ,m_l)\in{\bf Z}_+^l}q^{N_1^2+\cdot +N_l^2+N_s+\cdot
+N_l}\prod_{k=1}^{l-1}\left[\matrix{P_k(m;s,N)+m_k\cr
m_k}\right]_q\left[\matrix{\left[\ds{N+s-l-1\over
2}\right]-N_{l-1}\cr \cr m_l}\right]_{q^2},
$$
where ~$P_k(m;s,N):=N+1-\min (l-s+1,k)-\delta_{s,1}-\delta_{s,l+1}-
2\ds\sum_{j=1}^l\min (j,k)m_j\ge 0$.

\medbreak
{\bf 10.} Prove the following identity
$$\sum_{n_1,\ldots ,n_l}{q^{2(N_1^2+\cdots +N_l^2+N_s+\cdots +N_l)}\over
(q^2;q^2)_{n_1}\cdots (q^2;q^2)_{n_l}(-q)_{2n_l}}={1\over (-q)_{\infty}}
\prod_{n\not\equiv 0,\pm 2s~({\rm mod}~4k+3)}(1-q^n)^{-1}
$$
\hfill (L.~Rogers, 1894; A.~Selberg, 1936; P.~Paule, 1985)

\medbreak
{\bf 11.} Prove the following polynomial identities ($a=0,1$)
$$\sum_{k\in{\bf Z}}(-1)^kq^{4k^2-(2a+1)k}\left[\matrix{2N+a\cr
N+k}\right]_{q^2}=(q^{2N+2};q^2)_{N+a}\sum_{k=0}^N{q^{2k^2+2ak}\over
(-q;q^2)_{k+a}}\left[\matrix{N\cr k}\right]_{q^2}.\eqno (11)
$$
\hfill (P.~Paule, 1985)

It is easy to see that in the limit $N\to\infty$ these identities become
$$\sum_{k\ge 0}{q^{2k^2+2ak}\over (-q;q^2)_{k+a}(q^2;q^2)_k}={1\over
(q^2;q^2)_{\infty}}\prod_{n\equiv 0,\pm (3-2a)~({\rm mod}~8)}(1-q^n).
$$
\hfill (L.~Slater, 1950; H.~G\"ollnitz, 1967; B.~Gordon, 1961)

\medbreak
{\bf 12.} (Bailey's transform and Rogers-Ramanujan's type identities).

\hskip -0.5cm 1) (Bailey's transform). Let $a$ be indeterminate and $i,j\ge 0$
be
integers. Let us consider the matrices $M$ and $M^*$, where
$$\eno{
&M_{ij}:=(q)_{i-j}^{-1}(aq)_{i+j}^{-1};&(12a)\cr
&M_{ij}^*:=(-1)^{i-j}q^{(i-j)(i-j-1)\over 2}(1-aq^{2i})(aq)_{i+j-1}
(q)_{i-j}^{-1}.&(12b)}
$$

Prove that $M$ and $M^*$ are inverse, infinite, lower-triangle matrices.
That is
$$\sum_{j\le k\le i}M_{ik}M_{kj}^*=\delta_{ij}.
$$

Hint: use the terminating very well-poised $_4\vp_3$ summation theorem
(see e.g. [GR]).

\hskip -0.5cm 2) Let $\al =\{\al_n\}$ and $\beta =\{\beta_n\}$, $n\ge 0$ be
sequences
of functions in $q$. Let $M$ and $M^*$ be as in (12). We say (cf. [An3])
that $\al$ and $\beta$ form a Bailey pair relative to $a$ if
$$\beta_n=\sum_{k=0}^nM_{nk}\al_k,~~{\rm for~all}~~n\ge 0.
$$
It is clear that (Bailey's pair inversion rule)
$$\al_n=\sum_{k=0}^nM_{nk}^*\beta_k.
$$

\medbreak
{\bf Theorem} (G.~Andrews [An2]) (Bailey's lemma). {\it Let the sequences
$\al =\{\al_n\}$ and $\beta =\{\beta_n\}$ form a Bailey pair. If
$\al'=\{\al'_n\}$ and $\beta'=\{\beta'_n\}$ are defined by
$$\eno{
&\al'_n:={(\rho_1)_n(\rho_2)_n\over
(aq/\rho_1)_n(aq/\rho_2)_n}(aq/\rho_1\rho_2)^n\al_n,\cr
&\beta'_n:=\sum_{k=0}^n{(\rho_1)_k(\rho_2)_k(aq/\rho_1\rho_2)_{n-k}\over
(q)_{n-k}(aq/\rho_1)_k(aq/\rho_2)_k}(aq/\rho_1\rho_2)^k\beta_k,}
$$
then $\al'$ and $\beta'$ also form a Bailey pair.}

A proof can be found in [An3].

$\bullet$ ~Prove the following corollary of Bailey's lemma:
$$\sum_{k\ge 0}a^kq^{k^2}\beta_k={1\over (aq)_{\infty}}\sum_{k=0}^{\infty}
a^kq^{k^2}\al_k,\eqno (13)
$$
for any Bailey pair $\al =\{\al_n\}$ and $\beta =\{\beta_n\}$.

Hint: take the limit $n,\rho_1,\rho_2\to\infty$ in the Bailey pair
defining relation
$$\beta'_n=\sum_{k=0}^n{\al'_k\over (q)_{n-k}(aq)_{n+k}}.
$$
More generally, prove that if $\al =\{\al_n\}$ and $\beta =\{\beta_n\}$
is a Bailey pair, then
$$\sum_{m_1,\ldots ,m_k}{a^{N_1+\cdots +N_k}q^{N_1^2+\cdots +N_k^2}\over
(q)_{m_1}\cdots (q)_{m_k}}\beta_{m_k}={1\over (aq)_{\infty}}
\sum_{n\ge 0}q^{kn^2}a^{kn}\al_n.
$$
\hfill (G.~Andrews, 1984)

Hint: use the $k$-fold iteration of Bailey's lemma.

\hskip -0.5cm 3) Prove that the following sequences form the Bailey pair

\vskip 0.3cm
\hskip -0.5cm $i)$  $\beta_n=\ds\cases{1,& $n=0$,\cr 0,& $n>0$,\cr}$\hfill (15)

\vskip 0.3cm
$\al_n=\ds{(-1)^nq^{n(n-1)\over 2}(1-aq^{2n})(a)_n\over (1-a)(q)_n}$.

\vskip 0.3cm
Hint: use the Agarwal identity (see e.g. [An2]):
$$\sum_{k=0}^N{(1-aq^{2k})(q^{-n})_k(a)_kq^{nk}\over
(1-a)(q)_k(aq^{n+1})_k}={(aq)_Nq^{nN}(q^{1-n})_N\over (q)_N(aq^{n+1})_N}.
$$

It is easy to see that the Watson-Andrews identity (9) (with $j=1$ or
$k+1$) follows from (14) and (15).

\hskip -0.5cm $ii)$  Prove that if $\al =\{\al_n\}$ and $\beta
=\{\beta_n\}$ be a Bailey pair relative to $a$, then
$$\eno{
&\al_n':=\cases{\al_0, & $n=0$,\cr &\cr(1-a)a^nq^{n^2-n}\left\{\ds{\al_n\over
1-aq^{2n}}-{aq^{2n-2}\al_{n-1}\over 1-aq^{2n-2}}\right\} , & $n>0$,\cr}\cr
\cr
&\beta'_n:=\sum_{k=0}^n{a^kq^{k^2-k}\over (q)_{n-k}}\beta_k}
$$
is also a Bailey pair relative to $aq^{-1}$.

Proofs and further details see in [AAB], [Bre6], [P2], [GS], [FQ], [Q2], [Sl],
[ML]. It is well-known (see e.g. [W1]) that the classical
Rogers-Ramanujan identities (see e.g. Exercise 6 to Section 2.1) can be
deduced from Watson's [W1] $q$-analog of Whipple's transformation formula
(see e.g. [GR]). It seems very interesting to understand what kind of
partition identitis correspond to the Milne (see e.g. [Ml3], [ML])
multidimensional generalization of Bailey's lemma and the Watson-Whipple
transformation formula.

\bigbreak

{\bf 2.2. Dilogarithm and characters of the affine Kac-Moody algebras.}
\medbreak

{\bf Theorem E} (Kac-Wakimoto [KW2]). {\it Let ${\rm ch } V^k_r$ be the
character of level $k$ representation $V(k\Lambda_0)$ of the affine
Kac-Moody Lie algebra $\widehat{sl_r}$. Then}
$$\lim_{q\to 1}(1-q)\log{\rm ch }V_r^k={\pi^2\over 6}{(r^2-1)k\over
r+k}.$$

\medbreak
{\bf Theorem 10.} {\it We have
$${\rm ch}(V(k\Lambda_0))=\sum_{\lambda\in{\bf Z}_+^{r-1}}
\Theta_{\lambda}^k(z)c_{\lambda}^k(q),$$
where
$$\eno{&\Theta_{\lambda}^k(z):=\Theta_{\lambda}^k(z_1,\ldots ,z_{r-1})=
\sum_{m\in{\bf Z}^{r-1}}z^{km+\ld}q^{{1\over 2}kmA_{r-1}m^t+mA_{r-1}
\lambda^t},& (2.19)\cr
&c_{\ld}^k(q)={1\over (q)_{\infty}^{r-1}}\sum_{n=(n_i^a)}{q^{{1\over 2}
n(A_{r-1}\otimes T_{k-1}^{-1})n^t}\over \prod_{a,i}(q)_{n_i^a}},~~
k\ge 2;~~c_{\ld}^{(1)}(q)={\delta_{\ld ,0}\over (q)_{\infty}^{r-1}}
& (2.20)}
$$
and summation in (2.20) is taken over the sequences of nonnegative
integers $n=(n_i^a)$,\break $1\le a\le r-1$, $1\le i\le k-1$ under
the following constraints}
$$\sum_{i=1}^{k-1}in_i^a=\ld_a,~~1\le a\le r-1.$$

In (2.19) and (2.20) we used the Cartan matrices
$$\eno{&A_l=\pmatrix{2&-1&\cdots &0\cr -1&&\cdots\cr
&\cdots &\cdots \cr
&\cdots &&-1\cr 0&\cdots &-1&2}_{l\times l}\cr
&T_n=\pmatrix{2&-1&\cdots &0\cr -1&&\cdots &\cr
&\cdots &\cdots \cr &\cdots
&2&-1\cr 0&\cdots &-1&1}_{n\times n}=(\min (i,j))^{-1}_{1\le i,~j\le n}}
$$

\medbreak
{\bf Corollary 5.} {\it The constant term of ${\rm ch}(V(k\Lambda_0))$
with respect to $z$ is equal to
$$CT[{\rm ch}(k\Lambda_0)]={1\over (q)_{\infty}^{r-1}}\sum_{n\in(n_i^a)}
{q^{{1\over 2}n(A_{r-1}\otimes A_{k-1}^{-1})n^t}\over
{\displaystyle\prod_{i,a}}(q)_{n_j^a}},\eqno (2.21)
$$
where summation is taken over the sequences of nonnegative integers \
$n~=~(n_i^a)$,\break\hbox{$1\le a\le r-1$,} $1\le i\le k-1$, such that}
$$\sum_{i=1}^{k-1}in_i^a\equiv 0~({\rm mod}~k),~~1\le a\le r-1.
$$

For $r=2$ the formula (2.21) is exactly the result of Lepowsky and Primc
[LP] (see also [FeSt], [DKKMM], [KMM]). Analogously, for any weight $\ld
=(\ld_1\ge\ld_2\ge\cdots\ge\ld_{r-1}\ge 0$), one can find the coefficient
before $z^{\ld}$ in the Laurent series $({\rm ch}V_r^k)(z)$. The result is
given by the RHS(2.21) under the following constraints
$$\sum_{i=1}^{k-1}in_i^a\equiv\ld_a({\rm mod}~k),~~1\le a\le r-1.
$$
The last result about the constant term CT$[z^{-\ld}({\rm ch}V_r^k)(z)]$
coincides with $A_r$-case of the Terhoeven and Kuniba-Nakanishi-Suzuki
conjecture (see [Tr1] and [KNS], Section 2, (9)).

\vfil\eject
\bigbreak

{\bf 2.3. Dilogarithm identities and algebraic $K$-theory}
(A. Suslin, S. Bloch,\break
\hskip 0.8cm D. Zagier, E. Frenkel, A. Szenes).
\medbreak

{\bf 2.3.1.} Bloch group.
\medbreak

For any field $F$ we consider the following exact sequence
$$0\to C(F)\to D(F)\buildrel\lambda\over\to F^*\wedge
F^*\buildrel\chi\over\to K_2(F)\to 0,
$$
where

$i)$ $K_2$ is the $K$-functor of Milnor. By the Theorem of Matsumoto (see
e.g. [Mi1]) we have
$$K_2(F)=(F^*\otimes F^*)/I,
$$
where $I$ is the subgroup of $F^*\otimes F^*$ generated by elements
$x\otimes (1-x)$, $x\in F^*\setminus\{ 1\}$. In other words, $K_2F$ is
generated by symbols $\{ x,y\}$, $x,y\in F^*$, subject to the following
relations:
$$\eno{
&1)~~\{ xy,z\}= \{ x,z\}\{ y,z\},~~\{ x,yz\} =\{ x,y\}\{ x,z\} ,\cr
&2)~~\{ x,1-x\} =1,~~x\in F^*\setminus\{ 1\}.}
$$

$ii)$ $D(F)$ is the group, generated  over ${\bf Z}$ by formal symbols
$[x]$, $x\in F^*\setminus\{ 1\}$, where $F^*$ is the multiplicative group
of $F$.

$iii)$ $F^*\wedge F^*$ is the quotient of group $F^*\otimes_{\bf Z}F^*$
by the subgroup generated by the elements $x\otimes y+y\otimes x$. In other
words, an abelian group $F^*\wedge F^*$ is generated by the elements $x\wedge
y$ subject to the following relations
$$\eno{
&1)~~x\wedge y=-y\wedge x,\cr
&2)~~(xy)\wedge z=x\wedge z+y\wedge z.}
$$
Consequently, we have $(\pm 1)\wedge x=0$ in $F^*\wedge F^*$ for any
$x\in F^*$.

$iv)$ The homomorphism $\ld$ is defined by
$$\ld [x]=x\wedge (1-x),~~x\in F^*\setminus\{ 1\}.
$$

$v)$ The homomorphism $\chi$ is defined by
$$\chi (x\wedge y)=\{ x,y\} ,~~x,y\in F^*.
$$

$vi)$ $C(F):={\rm Ker}~\ld$.

One can check that the elements of the form
$$[x]-[y]+[y/x]-[(1-x^{-1})/(1-y^{-1})]+[(1-x)/(1-y)],~~x\not=y\in
F^*\setminus\{ 1\}\eqno (2.22)$$
are contained in $C(F)$. The quotient $B(F)$ of
$C(F)$ by the subgroup generated by the elements of this form is called
the Bloch group [Bl], [Su].

Let us assume now that $F$ is a totally real field of algebraic numbers.
The element $[x]+[1-x]$, $x\in F$, belongs to the Bloch group $B(F)$,
does not depend on $x$, and has the order 6, [Su]. It is known that for
the rational number field ${\bf Q}$ the group
$B({\bf Q})$ is generated by the element $[x]+[1-x]$ and is isomorphic to
cyclic group ${\bf Z}/6$.

One can use the Rogers dilogarithm function to define a map ${\cal L}
{}~:~B({\bf R})\to{\bf R}/({\bf Z}\pi^2)$. Namely, let ${\ol{\cal L}}$ be a
map $D({\bf R})\to{\bf R}$, which sends $[x]$ to
$L(x)-\ds{\pi^2\over 6}$, i.e. $\ol{\cal L}([x])=L(x)-\ds{\pi^2\over 6}$.
We can restrict it to $C({\bf R})$. Further,
one can show that if $\al$ is an element of $C({\bf R})$ of the form
(2.22), then ${\ol{\cal L}}(\al )=0~({\rm mod}~\pi^2)$. More exactly,
${\ol{\cal L}}(\al )=0$, except the case $x<0$ and $y>1$, when we have
${\ol{\cal L}}(\al )=-\pi^2$. Hence this map gives rise to a well-defined
homomorphism ${\ol{\cal L}}~:~B({\bf R})\to{\bf R}/({\bf Z}\pi^2)$.
Following [FrSz2], one can use the homomorphism $\ol{\cal L}$ to study a
torsion in the Bloch group $B(F)$ for totally real number fields, using
the dilogarithm identities (1.16), or (1.28). So, let $\zeta_{k+2}$ be a
primitive $k+2$-th root of unity and ${\bf Q}(\zeta_{k+2})^+$ be the
maximal real subfield of the cyclotomic field ${\bf Q}(\zeta_{k+2})$.
Consider the elements (see Section 1.4)
$$f_n=f_n^{(k)}:={\sin^2\ds{\pi\over k+2}\over\sin^2\ds{\pi (n+1)\over
(k+2)}},~~n=1,\ldots k.
$$
It is clear that $f_n^{(k)}\in{\bf Q}(\zeta_{k+2})^+$. Let us define
$$\Delta_{k+2}=2\sum_{n=1}^{k-1}\left[ f_n^{(k)}\right] .
$$
Using the relations (see Section 1.4) $(1-f_m)^2=\ds{f_n^2\over
f_{n-1}f_{n+1}}$, $f_0=f_k=1$ (here $f_n:=f_n^{(k)}$), one can show that
$\Delta_{k+2}\in C({\bf Q}(\zeta_{k+2})^+)$. Indeed, we have to check
$\ld (\Delta_{k+2})=0$. Using the properties of the elements $f_n$, one
can find
$$\eno{
\ld (\Delta_{k+2})&=\sum_{n=1}^{k-1}f_n\otimes
(1-f_n)^2=\sum_{n=1}^{l-3}f_n\otimes (f_n^2/f_{n-1}f_{n+1})\cr
&=\sum_{n=1}^{k-1}2f_n\otimes f_n -\sum_{n=1}^{k-2}(f_n\otimes
f_{n+1}+f_{n+1}\otimes f_n)=0,}
$$
as we set out to prove.

Let $B'(F)$ be the quotient of $B(F)$ by the subgroup generated by
$[x]+[1-x]$. A map sending $[x]$ to $L(x)$ gives rise to a well defined
homomorphism ${\cal L}'~:~B'({\bf R})\to{\bf R}/\left({\bf
Z}\ds{\pi^2\over 6}\right)$. Now we are able to formulate Frenkel-Szenes'
result.

\medbreak
{\bf Theorem F} ([FrSz2]). {\it Let  $F$ be a totally real number field and
$m_p$ be the maximal number $m\ge 0$ such that $F$ contains ${\bf
Q}(\zeta_{p^m})^+$. Then

$i)$ The symbols $\Delta_{p^{m_p}}$ generate the Bloch group $B(F)$.}

$ii)$ {\it The symbol $\Delta_{k+2}$ generates the group $B'({\bf
Q}(\zeta_{k+2})^+)$.}

$iii)$ {\it For a totally real number field $F$ the homomorphisms
${\cal L}~:~B(F)\to{\bf R}/({\bf Z}\pi^2)$ and  ${\cal L}'~:~B'(F)\to{\bf
R}/({\bf Z}\ds{\pi^2\over 6})$ are injective.}

A proof of Theorem F uses the identity (1.16), with $j=0$, and the
description of the Bloch group $B(F)$ of a totally real number field,
which is due to Merkuriev and Suslin [MS], and Levine [Lv].
According to this description, the group $B(F)$ is cyclic of order
$b(F)=\ds{1\over 2}\prod_pp^{m_p}$, where product is taken over all
primes. Now, using the dilogarithm identity (1.16) (with $j=0$), one can
construct an element of $B(F)$ of order exactly $b(F)$. Namely, using
the identity (1.16) one can find
$${\cal L}(\Delta_{k+2})-k{\cal L}(\Delta_6)=-{2\over k+2}\pi^2~({\rm
mod}~\pi^2).
$$

The symbol $\Delta_6=4\left[\ds{1\over 3}\right]
+2\left[\ds{1\over 4}\right]\in B({\bf Q})$ belongs to the Bloch group
$B(F)$. The element
$$\Delta_{p^{m_p}}-(p^{m_p}-2)\Delta_6\in B(F)
$$
under the homomorphism
${\cal L}~:~B(F)\to{\bf R}/{\bf Z}\pi^2$ gives an element of ${\bf
R}/{\bf Z}\pi^2$ of the order exactly $p^{m_p}$, if $p\ne 2$, and
$2^{m_2-1}$, if $p=2$. So, these elements of $B(F)$ generate a cyclic
group of order at least $b(F)$, hence they generate the whole group $B(F)$.

Now the group $B'({\bf Q}(\zeta_{k+2})^+)$ is cyclic of order $(k+2)/{\rm
g.c.d.}(12,k+2)$, [MS]. On the other hand, $\Delta_{k+2}$ is an element
$B'({\bf Q}(\zeta_{k+2})^+)$ and according to (1.16),
$${\cal L}'(\Delta_{k+2})=-{12\over k+2}~({\rm mod}~{\pi^2\over 6}).
$$
Thus, $\Delta_{k+2}$ generates a subgroup of $B'({\bf Q}(\zeta_{k+2})^+)$ of
order at least
$(k+2)~/{\rm g.c.d.}(12,k+2)$, and so it generates the
whole group $B'({\bf Q}/\zeta_{k+2})^+)$.

\qed

\medbreak
{\bf 2.3.2.} Goncharov's conjecture.
\medbreak

It is known that torsion subgroup of $B({\bf R})$ is generated by the
images of the groups $B({\bf Q}(\zeta_l)^+)$ of real parts of cyclotomic
fields, and consequently is isomorphic to ${\bf Q}/{\bf Z}$. It follows
from Theorem F, that $B({\bf R})_{\rm tor}$ is generated by the symbols
$\Delta_l$, and that the map ${\cal L}$ is injective on the torsion
subgroup of $B({\bf R})$.

{\bf Conjecture 3} (Goncharov).

$i)$ (Week form). {\it If $\al\in D({\bf
R})$ and ${\cal L}(\al )\equiv 0~({\rm mod}~\pi^2)$, then $\ld(\al )=0$.}

$ii)$ (Strong form). {\it If $\al\in D({\bf R})$ and ${\cal L}(\al
)\equiv 0~({\rm mod}~\pi^2)$, then $\al$ is a linear combination of
five-term elements of the form (2.22) (i.e. the map ${\cal L}$ is injective
on the whole Bloch group $B({\bf R}))$.}

In other words, if we have a dilogarithm identity
$$\sum_iL(x_i)=c{\pi^2\over 6},~~{\rm with}~~c\in{\bf Q}~~{\rm
and}~~x_i\in{\ol{\bf Q}}\cap{\bf R},
$$
then (hypothetically) we must have

$i)$ (week form)\ \  $\ld (\ds\sum_i[x_i])=0$;

$ii)$ (strong form)\ \  the relation $\ds\sum_iL(x_i)=c\ds{\pi^2\over 6}$ is a
linear combination over ${\bf Q}$ of the five-term relations (1.4) with
real-algebraic arguments.

Let us say now a few words about connection between dilogarithm
identities (1.16) and (1.28) and torsion part of $B({\bf R})$. The
Galois group $G:={\rm Gal}({\bf Q}(\zeta_{k+2})^+/{\bf Q})$ acts on the
Bloch group $B({\bf Q}(\zeta_{k+2})^+)$. Namely, let us consider an
automorphism $\sigma_j$ of the field
${\bf Q}(\zeta_{k+2})^+/{\bf Q}$, which is given by
$$\zeta_{k+2}+\zeta^{-1}_{k+2}\to\zeta_{k+2}^{j+1}+\zeta_{k+2}^{-(j+1)},~~
0\le j\le\left[{k+2\over 2}\right]-1,~~{\rm g.c.d.}(j+1,k+2)=1.
$$
The elements $\sigma_j$ generate the Galois group $G$ and naturally act on
the group $B({\bf Q}(\zeta_{k+2})^+)$. It is clear that
$$\eno{
&\sigma_j(f_n^{(k)})={\sin^2\ds{(j+1)\pi\over
k+2}\over\sin^2\ds{(n+1)(j+1)\pi\over k+2}}~~{\rm and}\cr
&\Delta^{(j)}_{k+2}:=\sigma_j(\Delta_{k+2})=2\sum_{n=1}^{k-1}\sigma_j
(f_n^{(k)})\in B({\bf Q}(\zeta_{k+2})^+).}
$$
It follows from (1.16) that
$${\cal L}(\Delta^{(j)}_{k+2})={1\over 3}(c_k-24h_k^{(j)}-k)\pi^2~({\rm
mod}~\pi^2),\eqno (2.23)
$$
where $c_k=\ds{3k\over k+2}$ is the central charge and
$h_k^{(j)}=\ds{j(j+2)\over 4(k+2)}$ is the conformal dimension of the
primary field of spin $j/2$. From (2.23) we deduce
$${\cal L}(\Delta_{k+2}^{(j_1)}-\Delta_{k+2}^{(j_2)})=8(h_k^{(j_2)}-
h_k^{(j_1)})\pi^2~({\rm mod}~\pi^2).
$$

Finally we are going to construct the elements $\Delta_{n,r}$ in $B({\bf Q}
(\zeta_{n+r})^+)$ using the dilogarithm identity (1.28). For this purpose
let us denote
$$g_m^{(k)}:=g_m^{(k)}(j)={\sin k\vp \sin (n-k)\vp\over\sin (m+k)\vp\sin
(m+n-k)\vp},~~1\le k\le n-1,
$$
where $\vp =\ds{(j+1)\pi\over n+r}$, $0\le j\le n+r-2$ and
g.c.d.$(j+1,n+r)=1$. Clearly, $g_m^{(k)}\in{\bf Q}(\zeta_{n+r})^+$. Let
us introduce elements
${\wt\Delta_{n,r}}=2\ds\sum_{k=1}^{n-1}\sum_{m=1}^{r-1}[g_m^{(k)}(0)]$.
\medbreak

{\bf Lemma 5.} ${\wt\Delta_{n,r}}\in C({\bf Q}(\zeta_{n+r})^+)$.

Proof. We have to check $\ld ({\wt\Delta_{n,r}})=0$. First of all, one
can easily prove that for any $\vp$ ($1\le k\le n-1$)
$${(1-g_m^{(k)})^2\over (1-g_m^{(k-1)})(1-g_m^{(k+1)})}={(g_m^{(k)})^2
\over g_{m-1}^{(k)}g_{m+1}^{(k)}},~~g_0^{(k)}:=1,~g_m^{(n)}:=0.\eqno (2.24).
$$
If now $\vp=\ds{(j+1)\pi\over r+n}$, then $g_r^{(k)}=1$, $1\le k\le n-1$,
and it follows from (2.24) that the elements $g_a:=g_m^{(k)}$, $a:=(k,m)$,
satisfy the Bethe-ansatz-like equations
$$g_a=\prod_b (1-g_b)^{2B_{a,b}},\eqno (2.25)
$$
where $B=(B_{a,b})=A_n\otimes A_r^{-1}$.

Consequently, $\ld({\wt\Delta_{n,r}})=\ds\sum_aB_{a,a}g_a\otimes
g_a+\sum_{a<b}B_{a,b}(g_a\otimes g_b+g_b\otimes g_a)=0$.

\qed

Let $\Delta_{n,r}$ be an element of $B({\bf Q}(\zeta_{n+r})^+)$
corresponding to ${\wt\Delta_{n,r}}\in C({\bf Q}(\zeta_{n+r})^+)$. It
follows from (1.28) that
$$\eno{
&{\cal L}(\Delta_{n,r})=-{(n-1)r(r-1)\over 3(n+r)}\pi^2~~({\rm
mod}~\pi^2),~~{\rm and}&(2.26)\cr
&{\cal L}(\Delta_{n,r}+\Delta_{r,n})=-{(r-1)(n-1)\over 3}\pi^2~~({\rm
mod}~\pi^2)~~{\rm (level-rank~~duality)}.&(2.27)}
$$
Furthermore, (2.27) admits a generalization. Namely, let us consider an
automorphism $\sigma_j\in{\rm Gal}({\bf
Q}(\zeta_{n+r})^+/{\bf Q})$, which is given by
$$\eno{
&\sigma_j(\zeta +\zeta^{-1})=
\zeta^{j+1}+\zeta^{-(j+1)},~~{\rm where}\cr
&\zeta :=\zeta_{n+r},~~ 0\le j\le\ds\left[{n+r\over 2}\right]~~
{\rm and~~g.c.d.}(j+1,n+r)=1.}
$$
Let us introduce the elements
$$\Delta_{n,r}^{(j)}:=\sigma_j(\Delta_{n,r})=2\sum_{k=1}^{n-1}
\sum_{m=1}^{r-1}g_m^{(k)}(j),
$$
which also belong to the Bloch group $B({\bf Q}(\zeta_{n+r})^+)$. Then it
follows from (1.28) that

$i)$~~${\cal L}(\Delta_{n,r}^{(j)})=\ds{1\over
3}(c_r^{(n)}-24h_j^{(r,n)}-(n-1)r)\pi^2~({\rm mod}~\pi^2)$,
where $c_r^{(n)}:=\ds{(n^2-1)r\over n+r}$ is the central charge and
$h_j^{(r,n)}:=\ds{n(n^2-1)\over 24}\cdot{j(j+2)\over r+n}$,
$0\le j\le r+n-2$
is the conformal dimension of the primary field of "spin" $j/2$
for $sl_n$ level $r$ WZNW model.

$ii)$ (Level-rank duality) \ ${\cal
L}(\Delta^{(j)}_{n,r}+\Delta^{(j)}_{r,n})=-\ds{(n-1)(r-1)\over
3}\pi^2~({\rm mod}~\pi^2)$.

(Hint: if g.c.d,$(j+1,n+r)=1$, then $j(j+2)(n^2-nr+r^2-1)\equiv 0~({\rm
mod}~6)$).

There exists a puzzling connection
between the special linear combinations of symbols $\Delta_{n,r}^{(j)}
\in B({\bf R})$ and the central charges and conformal dimensions of
primary fields of the coset Conformal Fields Theories
obtained by the Goddard-Kent-Olive construction [GKO] (see e.g. [Kir7]).

\medbreak
{\bf Remark.} In the paper [FzSz2], Section 5.2, for a totally real field
of algebraic numbers $F$, a very interesting construction of elements in
$K_3^{\rm ind}(F)$ (the indecomposable part of $K_3(F)$, i.e. the
cokernel of the product map $K_1(F)^{\otimes 3}\to K_3(F)$) using the
relative group $K_2$ of projective line over $F$ modulo two points and
the solutions of the Bether-ansatz-like system of algebraic equations is
given.

\bigbreak
{\bf 2.4. Connection with crystal basis.}
\medbreak

{\bf 2.4.1.} Level 1 vacuum representation $\Lambda_0$ of the affine Lie
algebra $\wh{sl_n}$.

\medbreak
{\bf Theorem 11.} {\it Let $\mu =(\mu_1,\ldots ,\mu_n)$ be a composition.
Then
$$\eno{&\sum_{\ld,~l(\ld )\le n
}K_{\ld ,\mu}\cdot
K_{\ld,~(1^{|\ld |})}(q)=q^{n(\mu')}\left[\matrix{|\mu |\cr \mu_1,
\ldots ,\mu_n}\right]_q,~~&(2.28)\cr
& \cr
&{\rm where}~~\left[\matrix{N\cr m_1,\ldots ,m_n}\right]_q:=
{(q)_N\over (q)_{m_1}\ldots (q)_{m_n}}}
$$
is the $q$-analog of multinomial coefficient $(N=m_1+\cdots +m_n)$.}

Let us explain notation. Here

$i)~~K_{\ld ,\mu}$ is the so-called Kostka number, which is equal to
the number of (semi)stan\-dard Young tableaux of shape $\ld$ and weight
$\mu$, or, equivalently, to the dimension of weight $\mu$ subspace
$V_{\ld}(\mu )$ of the irreducible highest weight $\ld$ representation
$V_{\ld}$ of the Lie algebra $\g l_n$:
$$K_{\ld ,\mu}={\rm dim}~V_{\ld}(\mu ).
$$

$ii)~~K_{\ld ,\mu}(q)$ is the so-called Kostka-Foulkes polynomial (see
e.g. [Ma]), which can be defined from a decomposition of the Schur
functions in terms of the Hall-Littlewood ones:
$$s_{\ld}(x)=\sum_{\mu}K_{\ld ,\mu}(q)P_{\mu}(x;q).
$$
It is well-known (see e.g. [Lu2] or [Ma]), that polynomial $\ol K_{\ld
,\mu}(q):=q^{n(\mu )-n(\ld )}K_{\ld ,\mu}(q^{-1})$ coincides with
Lusztig's $q$-analog of weight multiplicity dim~$V_{\ld}(\mu )$.

$iii)$ $n(\ld ):=\ds\sum^n_{i=1}(i-1)\ld_i=\sum^n_{i=1}\pmatrix{\ld'_j\cr
2}=\sum_{1\le i<j\le n}\min (\ld_i,\ld_j)$, if
$\ld =(\ld_1,\ldots ,\ld_n)\in{\bf Z}^n_+$.

{\bf The proof of Theorem 11} is based on the well-known results from the
theory
of symmetric functions. We will use the notation and terminology involving
symmetric functions from Macdonald [Ma].
\medbreak

{\bf Lemma 6.}
$$\sum_{\ld}s_{\ld}(x)K_{\ld ,\mu}(q)=Q_{\mu}\left({x\over 1-q}\right).
$$

Here we used the standard $\ld$-ring theory notation, namely, for any
symmetric function $f(x):=f(x_1,x_2,\ldots )$ and a new set of variables
$y=(y_1,y_2,\ldots )$ one can define $f(xy)=f(\ldots ,x_iy_j\ldots )$.

Proof. Let us remind that the Hall-Littlewood functions $P_{\ld}$ and
$Q_{\ld}$ satisfy the following orthogonality condition (see [Ma],
Chapter II, (4.4))
$$\sum_{\ld}Q_{\ld}(x;q)P_{\ld}(y;q)=\prod_{i,j}{1-qx_iy_j\over 1-x_iy_j}.
$$
Consequently,
$$\sum_{\ld}Q_{\ld}\left({x\over 1-q}\right) P_{\ld}(y,q)=\prod_{
{\matrix{i,j\cr k\ge 0}}}{1-q^{k+1}x_iy_j\over 1-q^kx_iy_j}=\prod_{i,j}
(1-x_iy_j)^{-1}=\sum_{\ld}s_{\ld}(x)s_{\ld}(y).
$$
Here we used the orthogonality of Schur's functions ([Ma], Chapter I,
(4.3)). It remains to remind the definition of Kostka-Foulkes polynomials:
$$s_{\ld}(y)=\sum_{\mu}K_{\ld ,\mu}(q)P_{\mu}(y;q).
$$
\qed

To continue, let us remark that if $m\ge n$, then
$$Q_{(k^n)}(x_1,\ldots ,x_m;q)=(q;q)_n(e_n(x))^k,~~{\rm
see~~[Ma],~Chapter~III,~~(2.8)}.
$$
Therefore, in order to prove Theorem 9 we have to decompose the symmetric
function
$$\sum_{\ld}s_{\ld}(x)K_{\ld ,(1^n)}(q)=Q_{(1^n)}\left(
{x\over 1-q}\right) =(q;q)_ne_n\left({x\over 1-q}\right)
$$
in terms of the monomial symmetric ones $m_{\ld}(x)$. For this purpose,
let us remind that by definition ($CT$:= constant term)
$$e_n\left({x\over 1-q}\right)= CT~\left[t^{-n}\prod_j(-tx_j;q)_{\infty}
\right].
$$
Thus, using the Euler result
$$(-z;q)_{\infty}=\sum_{n=0}^{\infty}{z^nq^{n(n-1)\over 2}\over (q;q)_n},
$$
one can obtain
$$e_n\left({x\over 1-q}\right)=\sum_{\mu\vdash n}{q^{n(\mu')}\over (q;q)_{\mu}}
m_{\mu}(x),
$$
where for a partition $\mu =(\mu_1,\mu_2,\ldots ,\mu_n)$, we set
$(q;q)_{\mu}:=\ds\prod^n_{j=1}(q;q)_{\mu_j}$. Consequently,
$$\sum_{\ld}s_{\ld}(x)K_{\ld ,(1^n)}(q)=\sum_{\mu\vdash n}q^{n(\mu')}\left[
\matrix{n\cr \mu_1\ldots\mu_n}\right]_qm_{\mu}(x).
$$
\qed

{\bf Remark.} The identity (2.28) is implicitly contained in the Terada
preprint [Te]. The proof given in [Te] is based on a study of the
cohomology groups of the variety of $N$-stable flags. Our proof is pure
algebraic and based on the theory of symmetric functions. Later
B.~Leclerc and J.-Y.~Thibon gave almost the same proof (unpublished). It
is possible to give a pure combinatorial proof of (2.28) using the
properties of the Robinson-Schensted correspondence. Finally, the
identity (2.28) can be extracted from [DJKMO1].

\medbreak
{\bf Corollary 6} (of Theorem 11).

$$\sum_{\ld ,~l(\ld )\le n}s_{\ld}(x)K_{\ld ,(1^{|\ld |})}(q)=
\sum_{m\in{\bf Z}_+^n}q^{\Sigma\pmatrix{m_i\cr 2}}x^m\left[\matrix{
|m|\cr m_1,\ldots ,m_n}\right]_q .\eqno (2.29)
$$

Now we are going to consider an appropriate limit $|m|\to\infty$ in
(2.29). More exactly, let us assume that $|m|=nN$, and consider the
following variant of (2.29):
$$\eno{
&q^{-{(N^2-N)n\over 2}}\sum_{\ld ,l(\ld )\le n}{s_{\ld}(x_1,\ldots ,
x_n)\over (x_1\ldots x_n)^N}K_{\ld ,(1^{|\ld |})}(q)\cr
&=\sum_{\matrix{k\in
{\bf Z}^n,\cr |k|=0,k_i\ge -N,~\forall i}}x_1^{k_1}\ldots
x_n^{k_n}q^{{1\over 2}\sum k_i^2}\left[\matrix{nN\cr k_1+N,\ldots
,k_n+N}\right]_q .&(2.30)_N}
$$
First of all, \ \
$\ds\lim_{N\to\infty}{\rm RHS}(2.30)_N=$
$${1\over (q)_{\infty}^{n-1}}\sum_{\matrix{m\in{\bf Z}^n\cr
|m|=0}}x^mq^{{1\over 2}\sum m_i^2}={1\over
(q)_{\infty}^{n-1}}\sum_{k\in{\bf Z}^{n-1}}z_1^{k_1}\ldots
z_{n-1}^{k_{n-1}}q^{{1\over 2}kA_{n-1}k^t},
$$
where $z_i=\ds{x_i\over x_{i-1}}$, $1\le i\le n-1$, $x_0:=x_n$.
On the other hand,
$$\lim_{N\to\infty}{\rm LHS}(2.30)_N=
\sum_{\matrix{\ld =(\ld_1\ge\ld_2\ge\cdots\ge\ld_n)\in{\bf Z}^n\cr
|\ld |=0}}s_{\ld}(x_1,\ldots ,x_n)b_{\ld}(q),
$$
where for given weight $\ld$  we set ($\ld_N:=\ld +(N^n)$)
$$b_{\ld}(q):=\lim_{N\to\infty}q^{-{n(N^2-N)\over 2}}K_{\ld_N,
(1^{|\ld_N|})}(q)={q^{n(\ld')}\over (q)_{\infty}^{n-1}}
\prod_{1\le i\le j\le n}(1-q^{\ld_i-\ld_j-i+j}).\eqno(2.31)
$$
Finally, it is follow from $(2.30)_N$ that
$$\eno{
&\sum_{\ld ,~l(\ld )\le n}s_{\ld }(x_1,\ldots ,x_n)b_{\ld}(q)={\Theta(x)\over
(q)_{\infty}^{n-1}},~~{\rm where}\cr & \cr
&\Theta (x)=\sum_{\matrix{m=(m_1,\ldots ,m_n)\in{\bf Z^n},\cr |m|=0}}
x^mq^{{1\over 2}(m_1^2+\ldots +m_n^2)},}
$$
is the theta-function corresponding to the basic representation
$V(\Lambda_0)$ of $\widehat{sl_n}$.

Hence, $b_{\ld}(q)$ is the branching function for representation
$\Lambda_0$ of the affine algebra $\wh{sl_n}$.

\medbreak
{\bf Remark.} It is interesting to compare the formula (2.31) with the
following result of V.~Kac ([Kac], (14.12.10)): for the infinite rank
affine algebra of type  $A_{\infty}$
$${\rm dim}_qL(\Lambda_{s_1}+\Lambda_{s_2}+\cdots +\Lambda_{s_n})=
{1\over (q)_{\infty}^n}
\prod_{1\le i<j\le n}(1-q^{s_i-s_j+j-i}),
$$
where $s_1\ge s_2\ge\cdots\ge s_n$ are arbitrary integers.

\medbreak

{\bf 2.4.2.} Kostka-Foulkes polynomials.
\medbreak

We start with reminding some basic properties of the Kostka polynomials
(see e.g. [Ma], [Kir3]).
\medbreak

{\bf Proposition F} (Hook-formula). {\it Let $\ld\vdash n$ be a partition of
natural number $n$. Then we have
$$K_{\ld ,(1^n)}(q)=q^{n(\ld')}{(q)_n\over\ds\prod_{x\in\ld}\left(1-q^{h(x)}
\right)},\eqno(2.32)
$$
where $h(x):=\ld_i+\ld'_j-i-j+1$ is the hook-length corresponding to the
box $x=(i,j)\in\ld$.}

Before stating the next result about Kostka polynomials, let us explain
some notation. Namely, let STY$(\ld ,\mu )$ be a set of all
(semi)standard Young tableaux of shape $\ld$ and weight (or content, or
evolution) $\mu$. It is well-known (see e.g. [Ma]) that $\# |{\rm STY}(\ld
,\mu )|={\rm dim}~V_{\ld}(\mu )$, i.e. the number of (semi)standard Young
tableaux of shape $\ld$ and weight $\mu$ is equal to the dimension of
weight $\mu$ subspace of the highest weight $\ld$ irreducible
representation $V_{\ld}$ of the Lie algebra $\g l_n$. Following A. Lascoux
and M.-P. Sch\"utzenberger, for given partitions $\ld$ and $\mu$ we denote
by $c(T)$ (correspondently $\ol c(T)$) the charge (cocharge) of a
tableau $T\in{\rm STY}(\ld ,\mu )$.
\medbreak

{\bf Proposition G} (Lascoux-Sch\"utzenberger [LS]). {\it If $\ld$ and
$\mu$ are partitions, then}
$$K_{\ld ,\mu}(q)=\sum_{T\in{\rm STY}(\ld ,\mu )}q^{c(T)}.
$$

In the remaining part of this section we are going to explain a new
combinatorial formula for Kostka polynomial $K_{\ld ,\mu}(q)$ coming from
the Bethe ansatz technique (see e.g. [Kir2], [Kir3]).

{}From the representation theoretic point of view,  Bethe's ansatz
method for the generalized Heisenberg magnet gives a very powerful and
convenient algorithm for decomposing the tensor product of some special
representations of semi-simple Lie algebra $\g$ into irreducible parts
(see e.g. [KR2], [Ku]). In the case of the Lie algebra $\g l_n$, the Bethe
ansatz method gives for a tensor product multiplicity ${\rm Mult}_{V_{\ld}}
(V_{\mu_1}\otimes\cdots\otimes V_{\mu_N})$
an equivalent description as the number of solutions to some special system
of algebraic equations (the so-called Bethe equations, see Exercise 2).
Using the
so-called "string conjecture" one can find the number of "string
solutions" to the system of Bethe's equations. In spite of the well-known
fact (see e.g. [EKK]) that solution to the Bethe equations does not have a
"string nature" in general, the total number of solutions (with
multiplicities) to the system of Bethe's equations given by using the "string
conjecture", appears to be correct. The last statement (the so-called
combinatorial completeness of Bethe's states) was proven in [Kir2] (see
also [KL]) for the case when all weights $\mu_i$ have a rectangular shape
(i.e. each weight $\mu_i$ is a proportional to some fundamental weight
$w_{a}$).

A possibility to apply the Bethe ansatz technique to combinatorics of
Young tableaux is based on the well-known fact (see e.g. [Lit] or [Stn])
that for the Lie algebra $\g L_n$ (more generally, for the Lie superalgebra
$\g l(N/M)$) a weight multiplicity can be expressed in terms of tensor
product multiplicity. Namely, if $l(\ld )\le n$ and $\mu=(\mu_1,\ldots
,\mu_N)$, then
$$\#|{\rm STY}(\ld ,\mu )|=K_{\ld ,\mu}={\rm dim}~V_{\ld}^{(\g l(N))}(\mu )
={\rm Mult}_{V_{\ld}^{(\g l(n))}}\left(V_{\mu_1w_1}^{(\g
l(n))}\otimes\cdots\otimes V_{\mu_Nw_1}^{(\g l(n))}\right).
$$
More generally (see e.g. [Se], [KR2]), if $\ld ,\mu$ as above and $\eta
=(\eta_1,\ldots ,\eta_M)$, then
$${\rm dim}~V_{\ld}^{(\g l(N/M))}(\mu |\eta)
={\rm Mult}_{V_{\ld}^{(\g l(n))}}\left(V_{\mu_1w_1}^{(\g
l(n))}\otimes\cdots\otimes V_{\mu_Nw_1}^{(\g l(n))}\otimes V_{w_{\eta_1}}
^{(\g l(n))}\otimes\cdots\otimes V_{w_{\eta_M}}^{(\g l(n))}\right).
$$
Now we are going to state the main result about the tensor product
multiplicities which follows from Bethe's ansatz technique. Thus, let
$\mu_1,\ldots ,\mu_N$ be the rectangular shape partitions (i.e,
$\mu_j=m_{i,j}w_i$ for some $i$, $1\le i\le n$, $1\le j\le N$) and $\ld$
be a partition such that $|\ld |=\sum_j|\mu_j|$ and $l(\ld )\le n$.
Let us denote by $\mu^{(i)}$, $1\le i\le n$, a composition composed by
those $m\ne 0$ for which there exists $\mu_j$ ($1\le j\le N$) such that
$\mu_j=mw_i$.
\medbreak

{\bf Proposition 1} ([KR2]).
$${\rm Mult}_{V_{\ld}^{(\g l(n))}}\left(V_{\mu_1}^{(\g
l(n))}\otimes\cdots\otimes V_{\mu_N}^{(\g l(n))}\right)
=\sum_{\{\nu\}}K_{\{\nu\}},\eqno(2.33)
$$
{\it where summation is taken over all configurations $\{\nu \}$ of type}
$(\ld ;\{\mu\})$.

Let us explain the notation in (2.33). By definition, a configuration
$\{\nu\}$ of type $(\ld ;\{\mu\})$ is a collection of partitions (or
Young diagrams) $\{\nu\}=\{\nu^{(1)},\nu^{(2)},\ldots\}$ such that

$i)$ $|\nu^{(k)}|:=\nu_1^{(k)}+\nu_2^{(k)}+\ldots =\ds\sum_{j\ge k+1}
(\ld_j-(j-k)|\mu^{(j)}|)$,

$ii)$  $P_r^{(k)}(\nu ;\{\mu\}):=Q_r(\mu^{(k)})+Q_r(\nu^{(k-1)})-
2Q_r(\nu^{(k)})+Q_r(\nu^{(k+1)})\ge 0$, for all $k,r\ge 1$, where
$\nu^{(0)}:=0$ and $Q_r(\nu ):=\ds\sum_{j\le r}\nu'_j$. Let us remind
that for a partition $\ld$ we denote by $\ld'=(\ld'_1,\ld'_2,\ldots
)$ the conjugate partition, i.e. $\ld'_i=\{j~|~\ld_j\ge i\}$.

We define the Kostka number $K_{\{\nu\}}$ corresponding to a
configuration $\{\nu\}$ as follows
$$K_{\{\nu\}}:=\prod_{k,r}\pmatrix{P_r^{(k)}(\nu;\{\mu\})+m_r(\nu^{(k)})\cr
\cr m_r(\nu^{(k)})},\eqno (2.34)
$$
where $m_r(\nu^{(k)})$ is a number of length $r$ parts in partition
$\nu^{(k)}$ and $\pmatrix{N\cr n}$ is the binomial coefficient:
$$\pmatrix{N\cr n}=\cases{\ds{N!\over n!(N-n)!},&if $0\le n\le N$;\cr \cr
0&otherwise.\cr}
$$

An analytical proof of Proposition 1 is given in [Kir1] and [Kir2].
Another proof of
Proposition 1 (see [Kir3] and [KK]) is based on a combinatorial
interpretation of the both sides of equality (2.34). The left hand side
of (2.34) admits an interpretation as the number of special kind of
tableaux (see [KK]), whereas the right hand side admits that in terms of
rigged configurations (see [Kir3]). The main step of a combinatorial proof
of Proposition 1 is to construct a bijection between the set of rigged
configurations and that of special kind of tableaux. The bijection
constructed in [Kir3] has many intriguing and mysterious properties such as
agreements with cocharge construction, the Sch\"utzenberger involution
and so on. For example, let us define a $q$-analog of the RHS(2.34) by
the following way
$$\ol{K}_{\ld ,\{\mu\}}(q):=\sum_{\{\nu\}}\ol{K}_{\{\nu\}}(q),~~
\ol K_{\{\nu\}}(q):=q^{\ol c(\nu )}\prod_{k,r}\left[\matrix{
P_r^{(k)}(\nu ;\{\mu\})+m_r(\nu^{(k)})\cr  \cr
m_r(\nu^{(k)})}\right]_q,\eqno(2.35)
$$
where summation is taken over all configurations $\{\nu\}$ of type $(\ld
;\{\mu\})$ and the cocharge $\ol c(\nu )$ of a configuration $\{\nu\}$ is
defined by
$$\ol c(\nu ):=\sum_{k,r\ge 1}\pmatrix{(\nu^{(k-1)})'_r-(\nu^{(k)})'_r\cr
2}-n(\ld ),~~\nu^{(0)}:=0.
$$
\medbreak

{\bf Theorem 12} ([Kir3]). $i)$ {\it Let us assume that $\mu :=\mu^{(1)}$
is a partition, $l(\mu )\le N$ and $\mu^{(i)}=0$, if $2\le i\le n$. Then
$$\ol K_{\ld ,\{\mu\}}(q)=\ol K_{\ld ,\mu}(q),
$$
where $\ol K_{\ld ,\mu}(q)$ is the Lusztig $q$-analog of weight
multiplicity for the Lie algebra $\g l_N$.}

$ii)$ {\it Let us assume that $\mu :=\mu^{(1)}$ is a partition, $l(\mu )
\le N$ and $\mu_k^{(i)}=0$ for all $k\ge 2$ and $2\le i\le n$. Let us define
the partition $\eta =(\mu_1^{(2)},\mu_1^{(3)},\ldots ,\mu_1^{(n)})^+$
corresponding to a composition $(\mu_1^{(2)},\mu_1^{(3)},\ldots
,\mu_1^{(n)})$. If $l(\eta')\le M$ then
$$\ol K_{\ld ,\{\mu\}}(q)=\ol K_{\ld ,\mu |\eta}(q),
$$
where $\ol K_{\ld ,\mu |\eta}(q)$ is the $q$-analog of weight
multiplicity ${\rm dim}~V_{\ld}^{(\g l(N|M))}(\mu |\eta )$ for the Lie
superalgebra $\g l(N|M)$ (see e.g. [Ser]).}

The formula (2.35) is very convenient in many combinatorial applications
(see e.g. [Kir3], [Kir5], [Kir8], [F]). However, there exist other forms for
(2.35), using the different parameterizations of the same type configurations
set. For example, it is possible (and useful~!) to rewrite the formula (2.35)
for $\ol K_{\ld ,\{\mu\}}(q)$ using the vacancy numbers $\{P_r^{(k)}(\nu
;\{\mu\})\}$ instead of parameters $\{\nu_r^{(k)}\}$. But having in mind
the applications of our formula (2.35) for (dual) Kostka polynomials to a
problem of computing the branching functions for integrable highest weight
representations of the affine Lie algebras, it is
more convenient to rewrite the RHS(2.35) in terms of parameters
$m_r(\nu^{(k)})$. Namely, let us fix a natural number $l$ such that
$m_p(\nu^{(k)})=0$, if $p\ge l$ and $1\le k\le n$. We define a vector
$m:=m(\nu )=$
$$(m_1(\nu^{(1)}),\ldots ,m_l(\nu^{(1)}),m_1(\nu^{(2)}),
\ldots ,m_l(\nu^{(2)}),\ldots ,m_1(\nu^{(n)}),\ldots ,m_l(\nu^{(n)}))
\in{\bf Z}_+^{ln}.
$$
 Then it is almost trivial exercise to show that

1) $P_r^{(k)}(\nu ;\{\mu\})+m_r(\nu^{(k)})=(m(I-B)+m_0)^{(k)}_r$, where

\vskip 0.3cm
\hskip -0.5cm$i)$ $B=C_n\otimes T_l^{-1}\in{\rm Mat}_{ln\times ln}
({\bf Z})$, and
$$\eno{
&C_n=\pmatrix{2&-1&\cdots &&0\cr -1& &\cdots\cr
&\cdots&\cdots&\cdots\cr&&\cdots&&-1\cr 0&&\cdots&-1&2}_{n\times n},\cr
\cr &T_l^{-1}=\pmatrix{2&-1&\cdots &&0\cr -1&2 &\cdots\cr
&\cdots&\cdots&\cdots\cr&\cdots&-1&2&-1\cr 0&\cdots &&-1&1}_{l\times l}^{-1}
=\left(\min (a,b)\right)_{1\le a,b\le l};}
$$
\vskip 0.3cm
\hskip -0.5cm $ii)$ vector $m_0:=(m_{0,r}^{(k)})$, where $m_{0,r}^{(k)}
=\ds\sum_{j\le
r}(\mu^{(k)})'_j$, and $I:=I_{ln}=(\delta_{a,b})_{1\le a,b\le ln}$.

2) $\ol c(\nu )={1\over 2}mBm^t-n(\ld )$;
\vskip 0.3cm

3) vector $m\in{\bf Z}_+^{ln}$ satisfies the following constraint $m\cdot
Q=\gamma$, where

\hskip -0.5cm $i)$ $\gamma :=(\gamma_r^{(k)})$, $\gamma_r^{(k)}=\ds\sum_{j\ge
k+1}\left\{\ld_j-(j-k)~|~\mu^{(j)}|\right\}$;

\vskip 0.3cm
\hskip -0.5cm $ii)$ $Q:=\ds I_n\otimes\pmatrix{1&\ldots &1\cr&\ldots\cr
l&\ldots &l}_{l\times l}$.

Consequently, we have
$$q^{n(\ld )}\ol K_{\ld ,\{\mu\}}(q)=\sum_{\matrix{m\in{\bf Z}_+^{ln}\cr
m\cdot Q=\gamma}}q^{{1\over 2}mBm^t}\prod_a\left[\matrix{(m(I-B)+m_0)_a\cr
m_a}\right]_q. \eqno(2.36)
$$

\medbreak
{\bf Example} (Melzer's conjecture [Me1]). Let us consider a particular
case of our formula (2.36) for (dual) Kostka polynomials when $\ld =({1\over
2}L+S,~{1\over 2}L-S)$,  $\mu^{(1)}=(1^L)$ and $\mu^{(k)}=0$, if $k\ge 2$.
Here a "spin" $S$ is a half-integer such that $M:={1\over 2}L-S\in{\bf Z}_+$.

First of all, in our case we have $B=2T_M^{-1}$, a vector $m_0=(L,\ldots
,L)\in{\bf Z}^M_+$ and the vacancy numbers for given configuration
{}~$m\in{\bf Z}_+^M$~ are the following  ones ~$P_a(m;\{\mu\})=$\break$L-
2(mT_M^{-1})_a$. Consequently (compare [Me1], (3.10)),
$${\cal F}_S^{(L)}(q):=q^{-n(\ld )}\sum_{\matrix{m_a\in
{\bf Z}_+\cr
\sum_{a=1}^Mam_a=M}}q^{{1\over 2}mBm^t}\prod^M_{a=1}\left[
\matrix{L+(m(I-B))_a\cr
m_a}\right]_q=\ol K_{\ld ,(1^L)}(q).\eqno (2.37)
$$
On the other hand, using the hook-formula (see Proposition F) for Kostka
polynomials, one can easily find
$$\eno{
\ol K_{\ld ,(1^L)}(q)&={1-q^{2S+1}\over 1-q^{{1\over 2}L+S+1}}\left[
\matrix{L\cr {1\over 2}L-S}\right]_q &(2.38)\cr \cr
&=\left[\matrix{L\cr {1\over 2}L-S}\right]_q-q^{2S+1}\left[\matrix{L\cr
{1\over 2}L-S-1}\right]_q\equiv{\cal B}_S^{(L)}(q).}
$$
Thus, ${\cal F}_S^{(L)}(q)={\cal B}_S^{(L)}(q)$, as we set out to prove.

Now if $L\to\infty$, $S\to\infty$ and $M={1\over 2}L-S$ is fixed
(Thermodynamic limit), then it follows from (2.37) and (2.38) that
$$\sum_{\matrix{m=(m_a)\in{\bf Z}_+^M\cr\sum_{a=1}^Mam_a=M}}\ \
{q^{{1\over 2}mBm^t}\over
\ds\prod^M_{a=1}(q)_{m_a}}={q^M\over (q)_M}.
$$

Another interesting limit is $L\to\infty$, $M\to\infty$ and $S$ being fixed
(Thermodynamic Bethe ansatz (TBA) limit). In the last case one can find
$${1-q^{2S+1}\over 1-q}\sum_{\{m_a\in{\bf Z}_+\}}{q^{{1\over 2}mBm^t+|m|}\over
\ds\prod_{a\ge 1}(q)_{m_a}}=
{1-q^{2S+1}\over (q)_{\infty}},\eqno(2.39)
$$
where summation in (2.39) is taken over all sequences $\{m_a\in{\bf
Z}_+\}_{a=1}^{\infty}$ such that $m_a=0$ for almost all indices $a$, and
$B:=B_{\infty}=2(\min (a,b))$, $1\le a,b<\infty$.

{\bf A proof of the "limiting" identity (2.39).} First of all let us remark
that if a partition $\ld$ consists of only two parts, then for any weight
$\mu$ and any configuration $\nu$ of type $(\ld ;\mu )$ one can compute
the cocharge of $\nu$ by the following rules
$$\ol c(\nu )=mBm^t-n(\ld )=2n(\nu ):=2\sum_{k\ge 1}\pmatrix{\nu'_k\cr 2}.
\eqno (2.40)
$$
Now let us consider a sequence of configurations $\{\nu :=\nu (M)\}$ of
types ($\ld :={(M+2S,M)}$; $(1^{|\ld |})$), $M\ge 1$. It is clear from the
formulae for cocharge (see (2.40)), that under the limit $M\to\infty$ and
$S$ is fixed, the sequence of configurations $\{\nu (M)\}$ gives a
non-zero contribution to the limit
$$\lim_{\matrix{M\to\infty\cr S~~{\rm is~~fixed}}}\ol K_{\ld
=(M+2S,M),(1^{|\ld |})}(q)
$$
if and only if $\ds\lim_{M\to\infty}\ol c(\nu (M))<\infty$.

The last condition is equivalent to the following one:
$\nu (M)=(\nu_1,\wt\nu )$, where $\nu_1:=\nu_1(M)\to\infty$ if $M\to\infty$
and the partition $\wt\nu$ does not depend on $M$. It is clear that under
these assumptions the cocharge $\ol c(\nu)=\sum_{k\ge
1}(\wt\nu_k')^2+|\wt\nu |$ does not depend on $M$ at all. It remains to
consider the limit ($\nu :=\nu (M)$)
$$\lim_{M\to\infty}\ol K_{\{\nu (M)\}}(q)=\lim_{L\to\infty}q^{\ol c(\nu )}
\prod_n\left[\matrix{L-2Q_n(\nu )+m_n(\nu )\cr m_n(\nu )}\right]_q.
$$
It is clear that if $n\le\wt\nu_1$, then $L-2Q_n(\nu (M))\to\infty$, if
$M\to\infty$, and
$$\lim_{M\to\infty}\left[\matrix{L-2Q_n(\nu (M))+m_n(\nu (M))\cr
m_n(\nu (M))}\right] ={1\over (q)_{m_n(\wt\nu )}}.
$$
Now, if $\wt\nu_1<n<\nu_1$, then $\nu_n(\nu )=0$. Finally, if $n=\nu_1$,
then $L-2Q_n(\nu (M))=2S$ and $m_n(\nu (M))=1$. It remains to put
$m_a:=m_n(\wt\nu )$.

\qed

It is well-known (see e.g. [BS]), that the RHS of (2.39) is the character
$\chi_S^{\rm Vir}={\rm Tr}~q^{L_0-S^2}$ of the irreducible highest weight
representation of the Virasoro algebra at central charge $c=1$ and
highest weight $h=S^2$, $S\in{1\over 2}{\bf Z}_{\ge 0}$.

More generally, let us consider the case when $\ld$ is a partition
$l(\ld )\le n$ and $\mu^{(1)}=(1^{|\ld |})$, $\mu^{(k)}=0$,
if $k\ge 2$. Then we have ($L:=|\ld |,\wt L:=|\ld |-\ld_1$)
$$\ol K_{\ld ,(1^L)}(q)=\sum_mq^{{1\over
2}mBm^t}\prod_a\left[\matrix{(m(I-B)+m_0)_a\cr m_a}\right]_q,\eqno(2.41)
$$
where $B=C_{n-1}\otimes T_{\wt L}^{-1}$, $m_0=(m_{0,r}^{(k)})$ and
$m_{0,r}^{(k)}=L\cdot\delta_{k,1}$, $1\le k\le n-1$, $1\le r\le\wt L$. The
summation in (2.41) is taken over all sequences $m\in{\bf
Z}_+^{(n-1)\wt L}$ under the following constraint
$$\sum_{r=1}^{\wt L}rm_r^{(k)}=\sum_{j\ge k+1}\ld_j,~~1\le k\le n-1.
$$
It follows from the hook-formula (see Proposition F) that
$$\ol K_{\ld ,(1^L)}(q)={(q)_L\over\ds\prod_{x\in\ld}(1-q^{h(x)})}.
$$

Now let us consider the limit $L\to\infty$, $\ld_1\to\infty$, but the
partition $\wt\ld =(\ld_2,\ld_3,\ldots ,\ld_n)$ is fixed (Thermodynamic
limit). Then one can obtain the following identity
$$\sum_{m=(m_r^{(k)})}{q^{{1\over 2}mBm^t}\over\ds\prod_{r=1}^{\wt L}
(q;q)_{m_r^{(1)}}}
\prod_{k\ge 2,~r\ge 1}\left[\matrix{(m(I-B))^{(k)}_r\cr m_r^{(k)}}\right]_q
={q^{n(\wt\ld )+|\wt\ld |}\over\ds\prod_{x\in\wt\ld}(1-q^{h(x)})},\eqno(2.42)
$$
where summation on $m$ in (2.42) is the same as in (2.41). Another
interesting limit is $L\to\infty$ all $\ld_i\to\infty$ ($1\le i\le n$),
but the differences ($\ld_{n+1}:=0$) $\ld_i-\ld_{i+1}$ ($1\le i\le n$)
are fixed (TBA limit). In this case one can obtain the following result
$$\lim_{L\to\infty}\ol K_{\ld_L,(1^{|\ld_L|})}(q)
={\ds\prod_{1\le i<j\le n}(1-q^{\ld_i-\ld_j-i+j})\over
(q)_{\infty}^{n-1}},\eqno(2.43)
$$
where $\ld_L=\ld +(L^n)$.

It is well-known (see e.g. [KP] or Section 2.4.1, (2.31)),  that the RHS
of (2.43) is the branching
function for level 1 vacuum representation $\Lambda_0$ of $\wh{sl_n}$.
\medbreak

{\bf 2.4.3.} Level $k$ vacuum representation $k\Lambda_0$ of the affine Lie
algebra $\wh{sl_2}$
\medbreak

{\bf Theorem 13.} {\it Let $\mu =(\mu_1,\mu_2)$ be the length two
composition, $|\mu |=2N$. Then}
$$\sum_{\ld ,~l(\ld )\le 2}K_{\ld ,\mu}K_{\ld ,(2^N)}(q)=\sum_m\left[
\matrix{N\cr
\cr\ds{\mu_1-m\over 2},\ds{\mu_2-m\over 2},m}\right]_q\cdot
q^{{1\over 2}m^2}q^{{1\over 2}(n(\mu')-N)}.\eqno(2.44)_{\mu}
$$

Let us deduce from Theorem 13 some combinatorial applications. First of
all, it is clear that $K_{\ld ,\mu}=1$ if $\ld\succeq\mu$ with respect to
the dominant
order (see e.g. [Ma]) and $K_{\ld ,\mu}=0$ otherwise. Hence ($\ld
=(\ld_1\ge\ld_2)$),
$$K_{\ld ,(2^N)}(q)={\rm RHS}(2.44)_{(\ld_1,\ld_2)}-{\rm
RHS}(2.44)_{(\ld_1+1,\ld_2-1)}.\eqno(2.45)
$$
Using (2.45), one can find
$$\eno{
\sum_{l=0}^Nq^{-l}{1-q^{2l+1}\over 1-q}&K_{(N+l,N-l),(2^N)}(q^2)&(2.46)\cr
&=q^{N^2-2N}\sum_kq^{k^2}\left\{\sum_{l\ge 0}\!\!\!~^{'}~(q^l+q^{-l})q^{l^2}
\left[\matrix{N-k\cr \ds{N-k-l\over 2}}
\right]_{q^2}\right\}\left[\matrix{N\cr k}\right]_{q^2}.}
$$
The next Lemma makes it possible to simplify the RHS(2.46).
\medbreak

{\bf Lemma 7.}
$${\rm RHS}(2.46)=q^{N^2-2N}\sum_{k=0}^Nq^{k^2+k}(1+q)\ldots
(1+q^{N-k})\left[\matrix{N\cr k}\right]_{q^2}.\eqno (2.47)
$$

Analytical proof of Lemma 7 will be given elsewhere.
\medbreak

{\bf Corollary 7} (Milne's conjecture [Ml2]).
$$\eno{
\sum^m_{l=0}q^{-l}{1-q^{2l+1}\over 1-q}&K_{(m+l,m-l),(2^m)}(q^2)&(2.48)\cr
&=\sum^m_{k=0}q^{m^2-2m}q^{k^2+k}(1+q)\ldots (1+q^{m-k})\left[
\matrix{m\cr k}\right]_{q^2}.}
$$
\medbreak

{\bf Corollary 8} (of Theorem 13). {\it Let $\ld =(\ld_1,\ld_2)$ be
a partition, $|\ld |\equiv 0~({\rm mod}~2)$, and $b_{\ld}^{2\Lambda_0}(q)$
be the branching function for the level 2 vacuum representation
$2\Lambda_0$ of the affine Lie algebra $\wh{sl_2}$. Then
$$b_{\ld}^{2\Lambda_0}(q)=\lim_{N\to\infty}q^{-N}\ol K_{\ld_N,\mu_N}(q),
$$
where $\ld_N=(\ld_1+2N,\ld_2+2N)$ and $\mu_N=(2^M)$,
$M=\ds{1\over 2}|\ld |+2N$.}

Finally, let us generalize the identity $(2.44)_{\mu}$ to the higher levels.
\medbreak

{\bf Theorem 14.}
$$\eno{
&\sum_{l(\ld )\le 2}K_{\ld ,\mu}K_{\ld ,(k^N)}(q) &(2.49)\cr
&=\sum_{m=(m_1,m_2,\ldots ,m_{k-1})\in{\bf Z}_+^{k-1}}q^{{1\over k}n(\mu')
-{k-1\over 2}N}q^{mA_{k-1}m^t\over k^2}\left[\matrix{N\cr \cr
\ds{\mu_1-m_1\over k},
\ds{\mu_2-m_{k-1}\over k},{1\over k}(mA_{k-1})_a}\right]_q.}
$$
\medbreak

{\bf Corollary 9} (of Theorem 14). {\it Let $\ld =(\ld_1,\ld_2)$ be a
partition, $|\ld |\equiv 0~({\rm mod}~2k)$, and $b_{\ld}^{k\Lambda_0}(q)$
be the branching function for the level $k$ vacuum representation
$k\Lambda_0$ of the affine Lie algebra $\wh{sl_2}$. Then
$$\lim_{N\to\infty}q^{-k(N^2-N)}K_{\ld_N,\mu_N}(q)=b_{\ld}^{k\Lambda_0}(q),
$$
where $\ld_N:=(\ld_1+kN,\ld_2+kN)$ and $\mu_N=(k^M)$, $M:=M(N)={1\over k}
|\ld |+2N$.}

Let us remind that, by definition, the partition $(k^M)$ is the one
consisting of $M$ parts all equal to $k$.

Proof. Consider the following sum
$$\Sigma_N:=\sum_{\matrix{\ld ~,l(\ld )\le 2\cr |\ld
|=2kN}}{s_{\ld}(x_1,x_2)\over
(x_1x_2)^{kN}}q^{-k(N^2-N)}K_{\ld ,(k^{2N})}(q).\eqno (2.50)
$$
It follows from Theorem 14 that
$$\eno{\Sigma_N=\sum_{\mu_1=0}^{2kN}(x_1/x_2)^{\mu_1-kN}&
\sum_{m=(m_1,\ldots ,m_{k-1})\in{\bf Z}_+^{k-1}}q^{{1\over
k}n(\mu')-(k-1)N-k(N^2-N)}q^{{mA_{k-1}m^t\over k^2}}\cr
&\cdot\left[\matrix{2N\cr
\cr \ds{\mu_1-m_1\over k},\ds{\mu_2-m_{k-1}\over k},\ds{1\over k}(mA_{k-1})_a}
\right]_q.}
$$
Let us put $\mu_1=m_1+(l+N)k$ (thus, $\mu_2=(N-l)k-m_1$) and $z=x_1/x_2$.
It is clear that
$${1\over k}n(\mu')-(k-1)N-k(N^2-N)={m_1^2\over k}+kl^2+2m_1l.
$$
Hence, we have
$$\eno{\Sigma_N=\sum_{\ds m=(m_1,\ldots ,m_{k-1})\in{\bf Z}_+^{k-1}}
&\sum_{\matrix{
l\in{\bf Z}\cr l\ge -N-{m_1\over k}}}z^{kl+m_1}q^{kl^2+2m_1l}
q^{{km_1^2+mA_{k-1}m^t\over k^2}}\cr
&\cdot\left[\matrix{2N\cr \cr l+N,N-l-\ds{m_1+m_{k-1}\over k},
\ds{1\over k}(mA)_a}\right]_q.}
$$
Consequently, \ \ $\ds\lim_{N\to\infty}\Sigma_N=$
$${1\over (q)^2_{\infty}}\sum_{m=(m_1,\ldots ,m_{k-1})\in{\bf Z}_+^{k-1}}
\sum_{l\in{\bf Z}}z^{kl+m_1}q^{kl^2+2m_1l}q^{{km_1^2+mA_{k-1}m^t\over
k^2}}\prod_{a=1}^{k-1}(q)_{s_a}^{-1},\eqno(2.51)
$$
where $s_a:=\ds{1\over k}(mA_{k-1})_{k-a}$.

Now let us rewrite the summation in the RHS(2.51) in terms of parameters
{}~$s_a$,\break$a=1,\ldots ,k-1$. One can easily check that
$$\sum_{a=1}^{k-1}as_a=m_1,~~{km_1^2+mA_{k-1}m^t\over
k^2}=sT_{k-1}^{-1}s^t.
$$
Consequently, \ \ $\ds\lim_{N\to\infty}\Sigma_N=$
$$\eno{
{1\over (q)^2_{\infty}}\sum_{m_1=0}^{\infty}&\sum_{\matrix{s=(s_1,\ldots ,
s_{k-1})\in{\bf Z}_+^{k-1}\cr \sum_aas_a=m_1}}\sum_{l\in{\bf Z}}z^{kl+m_1}
q^{kl^2+2m_1l}{q^{sT_{k-1}^{-1}s^t}\over\ds\prod_a(q)_{s_a}}\cr
&={1\over (q)_{\infty}^2}\sum_{m_1=0}^{\infty}\Theta_{m_1}^k(z)
c_{m_1}^k(q)={\rm ch}(V(k\Lambda_0)).}
$$
On the other hand,
$$\eno{
&\lim_{N\to\infty}\sum_{\matrix{\ld ,l(\ld )\le 2\cr |\ld |=2kN}}{s_{\ld}
(x_1,x_2)\over (x_1x_2)^{kN}}q^{-k(N^2-N)}K_{\ld_N,\mu_N}(q)=
\sum_{\matrix{\ld =(\ld_1\ge\ld_2)\cr |\ld |=0}}s_{\ld}(x)b_{\ld}(q),\cr \cr
&{\rm where}~~b_{\ld}(q)=\lim_{N\to\infty}q^{-k(N^2-N)}K_{\ld_N,(k^{2N})}(q).}
$$
Consequently, $b_{\ld}(q)$ is the branching function of level $k$
vacuum-$\wh{sl_2}$ representation $k\Lambda_0$.

\qed

We assume that any branching function $b_{\ld}^{\Lambda}(q)$ of
integrable highest weight representation $V(\Lambda )$ of the affine Lie
algebra $\wh{sl_n}$ can be constructed as an appropriate limit of the
Kostka-Foulkes polynomials. More exactly:
\medbreak

{\bf Conjecture 4.} {\it Let $\Lambda =k\Lambda_0$ be the level $k$ vacuum
representation of the affine Lie algebra $\wh{sl_n}$. Then
$$b_{\ld}^{\Lambda}(q)=\lim_{N\to\infty}\ol K_{\ld_N,(k^{Nn})}(q),
$$
where $\ld_N:=\ld +((kN)^n)$, $\ld =(\ld_1\ge\ld_2\ge\cdots\ge\ld_n)$ and
$|\ld |=0$.}

\medbreak
{\bf Remark.} There exists an interesting connection between the Milne
polynomials ([Ml1], [Ml2], [Ga])
$$M_{\mu }(x;q):=\sum_{\ld}s_{\ld}(x)K_{\ld
,\mu}(q)=Q_{\mu}\left({x\over 1-q}\right),
$$
where $x=(x_1,\ldots ,x_n)$, and the characters ch$V(\Lambda )$ of
integrable highest weight representations $V(\Lambda )$ of the affine Lie
algebra $\wh sl_n$. Roughly speaking, the character ch$V(\Lambda)$  is an
appropriate limit of Milne's polynomials (see the proofs of Theorems
11,13 and 14). It would be interesting to know the solutions to what kind
of hierarchy (= a system of nonlinear partial differential equations) the
characters of integrable highest weight representations of $\wh sl_n$ are.

\medbreak
{\bf 2.4.4.} Crystal basis and Robinson-Schensted correspondence.
\medbreak

In this section we are going to discuss briefly a connection between a
combinatorial description of the so-called crystal basis [Ka] of an
integrable $\wh{sl_n}$-module $V_n^k$ of level $k$, found by Jimbo,
Misra, Miwa and Okado in [JMMO] (see also [MM], [KKNMMNN] and [DJKMO1],
[DJKMO2]) and the Kostka-Foulkes polynomials via the
Robinson-Schemsted-Knuth (RSK) correspondence (see e.g. [Sa], where a
construction and basic properties of RSK are given).

The starting point for us is the following variant of the
Jimbo-Misra-Miwa-Okado formula for the character ${\rm ch}(V_n^k)$ in the
paths realization of the crystal basis, due to E. Frenkel and A. Szenes
[FrSz2]. Thus (see [FrSz2], Section 6), let us introduce the set
$$S_n^k=\{{\bf a}=(a_1,\ldots ,a_n)~|~a_i\ge 0,~a_1+\cdots +a_n=k\}.
$$
Denote by $\tau :(a_1,a_2,\ldots ,a_n)\to (a_n,a_1,\ldots ,a_{n-1})$ the
cyclic permutation acting on $S_n^k$ and define the element ${\bf a}^k\in
S_n^k$ by ${\bf a}^k=(k,0,\ldots ,0)$. Now the ground state path $\mu$ is
the sequence $({\bf a}^k,\tau{\bf a}^k,\tau^2{\bf a}^k,\ldots )$, and the
set of restricted paths ${\cal R}_n^k$ is the set of sequences $\eta
=(\eta_0,\eta_1,\ldots )$ of elements $\eta_j\in S_n^k$, which coincide
with $\mu$ except for a finite set of indices. Define the weight function
$w$ on ${\cal R}_n^k$ by the formula
$$w(\eta )=\sum_{m=1}^{\infty}m\left[
H(\eta_{m-1},\eta_m)-H(\mu_{m-1},\mu_m)\right] ,
$$
where $H_n({\bf a},{\bf b}):=\ds\max_{1\le i\le n}(h_i({\bf a,b}))$ and
$$h_i({\bf a,b})=\sum_{j=1}^ia_j-\sum_{j=1}^{i-1}b_j.
$$
\medbreak

{\bf Theorem G} ([JMMO]). {\it As formal power series in $q$}
$${\rm ch}~V_n^k=\sum_{\eta\in{\cal R}_n^k}q^{w(\eta )}.
$$

Now let us introduce the sets
$$\eno{
&W_N=\{\eta\in{\cal R}_n^k~|~\eta_j=\mu_j~~{\rm for}~~j\ge nN\},\cr
&W_N^{\bf a}=\{\eta\in{\cal R}_n^k~|~\eta_{nN-1}={\bf a},~\eta_j=\mu_j~~
{\rm for}~~j\ge nN\},~~{\bf a}\in S_n^k,}
$$
and the polynomials (the characters of $W_N$ and $W_N^{\bf a}$)
$$w_N(q)=\sum_{\eta\in W_N}q^{w(\eta )},~~w_N^{\bf a}(q)=\sum_{\eta\in
W_N^{\bf a}}q^{w(\eta )}.
$$

Further, let us consider the column vector ${\bf w}_N$ of characters
${\bf w}_N^{\bf a}$ in a certain order, which is fixed once and for all.
In the paper [FrSz2] the recurrence relation for vectors ${\bf w}_N$ was
obtained. Namely,
$${\bf w}_{N+1}=q^{-knN}M_k(q^{nN+n-1})\ldots M_k(q^{nN+1})M_k(q^{nN})
{\bf w}_N,
$$
where $M_k(x)_{\bf a,b}=x^{H_n({\bf a,b})}$, for ${\bf a,b}\in S_n^k$.
This important recurrence relation gives a way to compute the polynomials
$w_N(q)$.

Two particular cases when either $k=1$ and $n\ge 2$ is arbitrary or $n=2$
and $k\ge 1$ is arbitrary seem to be the most accessible from
combinatorial point of view. Namely, let us consider at first the level
$k=1$ case. Then we can imagine each finite path $\eta\in W_N$ as a word
$w=a_1\ldots a_{nN}$ of length $m=nN$, composed from the numbers
$1,2,\ldots n$. Let $\mu$ be the weight of $w$, i.e. $\mu_i=\{j~|~a_j\in
w,~a_j=i\}$. For given weight $\mu ,l(\mu )\le n$, let us denote by
$M(\mu )$ the set of all words of weight $\mu$. It is well-known (see
e.g. [An1]) that
$${\rm Card}~M(\mu )={|\mu |!\over \mu_1!\ldots\mu_n!}=\pmatrix{|\mu |\cr
\mu_1,\ldots ,\mu_n}.
$$
\medbreak

{\bf Definition 7} ([Fo]). {\it A function $\vp$ on the set $M(\mu )$ is called
to
be mahonian statistics, if
$$\sum_{w\in M(\mu )}q^{\vp (w)}=q^{d(\vp )}\left[\matrix{|\mu |\cr
\mu_1,\ldots ,\mu_n}\right]_q,
$$
where a positive integer $d(\vp )$ depends only on weight $\mu$.}

The classical examples of mahonian statistics (with $d(\mu )=0$) are
the number of inversions (inv$(w)$) and major (or greater) index
(Maj$(w)$) of a word
$w\in M(\mu )$, see e.g. [An1], Section 3.4, or Exercise 1 to Section 2.4.
The next result was proven in [DJKMO1] using the recurrence
relations technique.
\medbreak

{\bf Theorem H} (Date-Jimbo-Kuniba-Miwa-Okado). {\it The weight function
$w(\eta )$ on the set restricted paths ${\cal R}_n^1$ defines on the
weight $\mu$ subsets $M(\mu )\subset W_N$ the mahonian statistics with
$d(\vp )=0$.
In other words}
$$\sum_{\eta\in M(\mu )}q^{w(\eta )}=
\left[\matrix{|\mu |\cr\mu_1,\ldots ,\mu_n}\right]_q.
$$

Proof. First of all, we are going to replace a restricted path
$\eta\in{\cal R}_n^1$ by some word. More exactly,
let $\eta\in{\cal R}_n^1\cap W_N$. The path $\eta$ has the form $\eta
=(\eta_1,\eta_2,\ldots ,\eta_{nN})$, where each $\eta_j\in S_n^1$. But
$S_n^1=\{e_1,e_2,\ldots ,e_n\}$, where $e_i=(0,\ldots ,0,1,0,\ldots ,0)$.
Thus one can replace the path $\eta$ by the word $w$ just changing $e_j$
on $j$. Let us assume that the word $w$ has a weight $\mu$. Our next task
is to rewrite the weight function $w(\eta )$ given on the set of
restricted paths, as a function on the set of words.
The crucial observation is following: if $e_i$ and $e_j$ belong to
$S_n^1$, then $H(e_i,e_j)=\chi (i\ge j)$, where for any statement $P$ the
symbol $\chi (P)$ is equal to 1, if $P$ is true and $\chi (P)=0$
otherwise. Using this observation, one can find
$$w(\eta )=\sum_{m=1}^{nN-1}m\chi (a_m\ge a_{m+1})-{nN(N-1)\over 2},
$$
where $w=a_1a_2\ldots a_{nN}$ is the word corresponding to the restricted
path $\eta\in{\cal R}_n^1\cap W_N$. To go further, let us introduce on
the set of words $w=a_1\ldots a_{nN}$ the statistics $\wt{\rm Maj}(w)$:
$$\wt{\rm Maj}(w)=\sum_{m=1}^{nN-1}m\chi (a_m\ge a_{m+1}).
$$
Note that the classical major index of the word is defined as
$${\rm Maj}(w)=\sum_{m=1}^{nN-1}m\chi (a_m>a_{m+1}).
$$
It is the MacMahon Theorem (see e.g. [An1], Theorem 3.7) that
$$\sum_{w\in M(\mu )}q^{{\rm Maj}(w)}=\left[\matrix{|\mu |\cr
\mu_1,\ldots ,\mu_n}\right]_q.
$$

The proof given in [An1], Theorem 3.7, can be modified to show the
statistics $\wt{\rm Maj}$ is also machonian with $d(\vp )=n(\mu')$.

\qed

It is an interesting task to find a purely combinatorial proof of
equidistribution for the statistics Maj and $\wt{\rm Maj}-n(\mu')$. We are
going to consider this question in more details in a separate publication.

Another interesting question is a connection between the weight function
$w(\eta )$ (or $\wt{\rm Maj}(w)$) and the Robinson-Schensted
correspondence. Namely, applying  the
Robinson-Schensted correspondence to a given word $w\in M(\mu )$, we
obtain the pair $(P,Q)$ of (semi)standard Young tableaux of the same
shape, say $\ld$, such that
$$P\in {\rm STY}(\ld ,\mu ),~~Q\in{\rm STY}(\ld ,(1^{nN})).
$$

It is well-known (see e.g. [Sa]) that if $w{\buildrel{\rm
RS}\over\longleftrightarrow}(P,Q)$, then Maj$(w)={\rm ind}Q$, where the
index of a standard tableau $T\in{\rm STY}(\ld ,1^{|\ld |})$ is defined
as ${\rm ind}(T)=\sum_{j\in{\rm Des}(T)}j$ and Des$(T)$ is the so-called
descent set of tableau $T$ ($j\in{\rm Des}(T)$ iff $(j+1)$ lies in
tableau $T$ strictly below than $j$). This result allows to give a pure
combinatorial proof of Theorem~9. Indeed, it is also well-known result
(see e.g. [Ma]) that
$$K_{\ld ,(1^{|\ld |})}(q)=\sum_{T\in{\rm STY}(\ld ,1^{|\ld |})}q^{{\rm
ind}(T)}.
$$
A relation  between $\wt{\rm Maj}$-statistics and RS-correspondence is
more complicated and will be considered elsewhere.


\bigbreak
{\bf Exercises to Section 2.4.}
\medbreak

{\bf 1.} Let $\mu$ be a composition, $l(\mu )=n$. Consider the set
$$M(\mu )=\left\{w=a_1a_2\cdots a_{|\mu |}~|~a_k\in[1,n],~\#\{
i~|~a_i=j\}=\mu_j\right\}.
$$
Let us define the number of inversions for a word $w$ as
$${\rm inv}(w)=\sum_{1\le i<j\le |\mu |}\chi (a_i>a_j).
$$

$i)$ Prove
$$\sum_{w\in M(\mu )}q^{{\rm inv}(w)}=\left[\matrix{|\mu |\cr
\mu_1,\ldots \mu_n}\right]_q.\eqno({\rm P.~MacMahon,~1914})
$$

$ii)$ Given $w\in M(\mu )$, let $w_{ij}$ be the subword of $w$ formed by
deleting all letters $a_m$ such that $a_m\ne i$ or $j$. For example, if
$w=2411213144321\in M(5,3,2,3)$ then $w_{12}=21121121$, $w_{13}=1113131$,
$w_{14}=41111441$, $w_{23}=22332$, $w_{24}=242442$, $w_{34}=43443$.

The $Z$-index (Zeilberger's index) of $w$ is defined to be the sum of the
major indices of all 2-letter subwords, $w_{ij}$, of $w$. That is
$$Z(w)=\sum_{1\le i<j\le n}{\rm Maj}(w_{ij}).
$$
Analogously, let us introduce the $\wt{Z}$-index of $w$ as follows
$$\wt{Z}(w)=\sum_{1\le i<j\le n}\wt{\rm Maj}(w_{ij})-n(\mu ).
$$

For our example, inv$(w)=29$, Maj$(w)=47$, $\wt{\rm Maj}(w)-n(\mu
)=59-15=44$,\break$Z(w)=12+10+8+4+7+5=46$, $\wt{Z}(w)=19+13+23+11+8+8-15=67$.

Prove
$$\sum_{w\in M(\mu )}q^{Z(w)}=\left[\matrix{|\mu |\cr \mu_1,\ldots
,\mu_n}\right]_q.\eqno({\rm D.~Zeilberger,~D.~Bressoud,~1985})
$$

$iii)$ Give a combinatorial proof of equidistribution for the statistics
$${\rm inv},~~{\rm Maj}~~~{\rm and}~~~Z.
$$
Further details and proofs one can find in [Bre3], [ZB], [Gre] and [Han].

\medbreak
{\bf 2.} (Tensor product multiplicities and Bethe's ansatz equations).

Let $\mu^{(j)}=(\mu_1^{(j)}\ge\cdots\ge\mu_n^{(j)}\ge 0)$, $1\le j\le N$,
and $\ld$ be the highest weights and $V_{\mu^{(j)}}$, $V_{\ld}$ be the
corresponding irreducible finite dimensional representations of the Lie
algebra $\g l_n$.

{\bf Conjecture 5.} {\it The tensor product multiplicity
$${\rm Mult}_{V_{\ld}}(V_{\mu^{(1)}}\otimes\cdots\otimes V_{\ld^{(N)}})
$$
is equal to the number of solutions to the following system of algebraic
equations on variables $v_1=\{v_{1,1},\ldots ,v_{1,m_1}\}$,
$v_2=\{v_{2,1},\ldots ,v_{2,m_2}\}$, ... ,$v_n=\{v_{n,1},\ldots
,v_{n,m_n}\}$ (we assume that $v_0=v_{n+1}=\phi$)
$$\prod_{j=1}^N{v_k-\ld_k^{(j)}i\over v_k-\ld_{k+1}^{(j)}i}=\prod_{v_{k-1}}
{v_k-v_{k-1}+i/2\over v_k-v_{k-1}-i/2}\prod_{v_k'\ne v_k}{v_k-v_k'-i\over
v_k-v_k'+i}\prod_{v_{k+1}}{v_v-v_{k+1}+i/2\over v_k-v_{k+1}-i/2}.
$$
Here $i=\sqrt -1$. The number of equations and that of variables are the
same and are equal
to $m_1+\cdots +m_n$. The relation between the weights $\ld$ and
$\mu^{(1)},\ldots ,\mu^{(N)}$ and the composition $m=(m_1,m_2,\ldots
,m_n)$ is the following}
$$2m_k-m_{k-1}-m_{k+1}=\sum_{l\ge
k+1}\left\{\sum_{j=1}^N\mu_l^{(j)}-\ld_l\right\}.
$$

{\bf 3.} Given two partitions $\ld$ and $\mu$, $|\ld |=|\mu |$, and a
natural number $k$. Let us consider the new partitions
$$\ld_N:=(\ld ,(k^N))^+~~{\rm and}~~\mu_N:=(\mu ,(k^N))^+,
$$
where for any composition $\nu$ we denote by $\nu^+$ the corresponding
partition (for example, if $\nu =(4419583162)$, then $\nu^+=(9865443211)$).

Prove that there exists the limit
$$\lim_{N\to\infty}K_{\ld_N,\mu_N}(q):=G_{\ld ,\mu ;k}(q),
$$
and this limit is a rational function.
\medbreak

$\bullet$~Problem. Find an algebraic/representation theoretic
interpretation of the rational functions $G_{\ld ,\mu ;k}(q)$.

See [Stn2], [Kir3] and [Kir9] for proofs and further results about the
rational functions $G_{\ld ,\mu ;k}(q)$.

\medbreak
{\bf 4.} Prove the following relations

$i)$ $\ds\sum_{\ld}{\rm dim}V_{\ld}^{(n)}\cdot K_{\ld
,\mu}(1)=\prod_j\pmatrix{n+\mu_j-1\cr \mu_j}$,

$ii)$ $\ds\sum_{\ld}\left[\matrix{n\cr \ld'}\right]_q\cdot K_{\ld
,\mu}(1)=\prod_j\left[\matrix{n+\mu_j-1\cr\mu_j}\right]_q$.

Here $\ds\left[\matrix{n\cr\ld}\right]:=\prod_{x=(i,j)\in\ld}{1-q^{n+i-j}
\over 1-q^{h(x)}}$ is the generalized $q$-binomial coefficient;
$V_{\ld}^{(n)}$ is the highest weight $\ld$ irreducible representation of
$\g l(n)$ and $K_{\ld ,\mu}(1)$ is the Kostka number.

Hint: use the Littelwood formula
$$\sum_{\ld} s_{\ld}(x_1,\ldots ,x_n)s_{\ld}(y_1,\ldots
,y_m)=\prod_{i=1}^n\prod_{j=1}^m{1\over 1-x_iy_j}.
$$

{\bf 5.} (Construction of elements in $C(F)$). We use the notation from
the Section 2.3.1.
\medbreak

{\bf Definition 8.} {\it Every equation of the following form
$$X^A\prod_{r\in I}(1-\zeta_m^{k_r}X^r)^{A_r}=1,\eqno (1)
$$
where $A,k_r,A_r$ are integers, $\zeta_m=\exp\left(\ds{2\pi i\over
m}\right)\in F$, and $I$ is a finite set of natural numbers, is called a
(generalized) cyclotomic equation.}

Now, suppose that $a\in F^*$, $a\ne 1$ satisfies the cyclotomic equation
(1), and let a natural number $N$ satisfy

$i)$~ $2~|~NA$,

$ii)$ $r~|~NA$, $m~|~(NA_r/r)k_r$, for $r\in I$.Show that
$$b:=\sum_{r\in I}(NA_r/r)[\zeta_m^{k_r}a^r]\in C(F).
$$

\medbreak
{\bf Remark.} Let $\ol b$ be the image of $b$ in the Bloch group $B(F)$.
It is an interesting problem to find an effective criteria when
${\ol b}\in B(F)_{\rm tor}$.

\vfil\eject
\bigbreak
{\bf 2.5. Quantum dilogarithm.}
\medbreak

Let us remind that the Euler dilogarithm $Li_2(x)$ is defined for $0\le
x\le 1$ by
$$Li_2(x)=\sum_{n\ge 1}{x^n\over n^2}.
$$

{\bf Definition 9} ($q$-analog of Euler's dilogarithm or quantum
dilogarithm). {\it Quantum dilogarithm $Li_2(x;q)$ is the following
formal series}
$$Li_2(x;q)=\sum_{n\ge 1}{x^n\over n(1-q^n)}.\eqno (2.52)
$$

Before stating the basic properties of the quantum dilogarithm, let us
remind the definition of the two $q$-exponential functions (see e.g.
[GR], Appendix II)
$$\eno{
&e_q(x)=\sum_{n\ge 0}{x^n\over (q)_n}={1\over (x)_{\infty}},& (2.53)\cr
&E_q(x)=\sum_{n\ge 0}{q^{n(n-1)\over 2}x^n\over (q)_n}=(-x)_{\infty}.}
$$
Here we use the standard $q$-analysis notation:
$$(x)_{\infty}:=(x;q)_{\infty}=\ds\prod_{n\ge 0}(1-xq^n),~~
(x)_n:={(x;q)_{\infty}\over (xq^n;q)_{\infty}}.
$$

It is clear from the definition (2.53) that
$$(1+x)E_q(qx)=E_q(x),~~e_q(qx)=(1-x)e_q(x),~~e_{q^{-1}}(x)=E_q(-qx).
\eqno (2.54)
$$
\medbreak

{\bf Lemma 8.} {\it We have}

\vskip 0.2cm
$i)$ \ $Li_2(x;q)=\log (e_q(x))$,

\vskip 0.2cm
$ii)$ \ $Li_2(x;q)+Li_2(x;q^{-1})=-\log (1-x)$.
\vskip 0.2cm

Proof. it is clear from (2.54) that
$$Li_2(qx;q)=Li_2(x;q)+\log (1-x).
$$
Hence, if $Li_2(x;q)=\ds\sum_{n\ge 1}a_nx^n$, then $q^na_n=a_n-\ds{1\over
n}$.

\qed

\medbreak
{\bf Proposition H} (Euler-Maclaurin's summation formula). {\it If $f\in
C^{2n}([M,N])$, $M$ and $N$ integers, then}
$$\eno{
\sum_{m=M}^{N}f(m)&=\int_M^Nf(t)dt+{1\over 2}(f(N)+f(M))+
\sum_{k=1}^n{B_{2k}\over (2k)!}\left\{f^{(2k-1)}(N)-
f^{(2k-1)}(M)\right\}\cr
&-\int_M^N{{\ol B}_{2n}(t)\over (2n)!}f^{(2n)}(t)dt.&(2.55)}
$$

Here $B_k$ are the Bernoulli numbers and $B_k(t)$ are the Bernoulli
polynomials which are defined by
$${ze^{tz}\over e^z-1}=\sum_{k\ge 0}{B_k(t)\over k!}z^k,~~B_k:=B_k(0).
$$
${\ol B}_n(t)$ is the so-called modified Bernoulli polynomial: ${\ol
B}_n(t)=B_n(\{ t\})$, where $\{ t\} :=t-[t]$ is the fractional part of
a real number $t$.

As a simple corollary of the Euler-Maclauren summation formula, one can
obtain

\medbreak
{\bf Corollary 10.} {\it If $q=e^{-\epsilon}$ and $\epsilon\to 0$, then
the function $e_q(x)$ has the following asymptotic expansion ~~($0<x<1$)}
$$e_q(x)=(1-x)^{-{1\over 2}}\exp ({Li_2(x)\over\epsilon}+
{x\epsilon\over 12(1-x)})(1+O(\epsilon^3)).
$$

Indeed, it follows from (2.55) that (cf. [Mo], [UN]) if $0<x<1$ and
$q=e^{-\epsilon}$, then as $\epsilon\to 0$,
$$\log ((x;q)_{\infty})\sim -{Li_2(x)\over\epsilon}+{1\over 2}\log (1-x)
-\sum_{k=1}^{\infty}{B_{2k}\over (2k)!}{P_{2k-1}(x)\epsilon^{2k-1}\over
(1-x)^{2k-1}},
$$
where $P_n(x)$ is a polynomial of degree $n$ satisfying
$$P_n(x)=(x-x^2)P_{n-1}'(x)+nxP_{n-1}(x),~~P_0=1,~~n=1,2,3,\ldots .
$$
Moreover, the error in terminating the series is less in absolute value
than that of the first term neglected and has the same sign.

The most important and fundamental property of Euler's dilogarithm is the
five-term relation (in Rogers' form, [Ro])\ \ $Li_2(x)+Li_2(y)-Li_2(xy)=$
$$=Li_2\left({x(1-y)\over 1-xy}\right)
+Li_2\left({y(1-x)\over 1-xy}\right)+\log\left({1-x\over 1-xy}\right)
\log\left({1-y\over 1-xy}\right).\eqno (2.56)
$$

Before discussing the (quantum) analog of the five-term relation for the
quantum dilogarithm $Li_2(x;q)$ (see also [FK]), let us consider in more
detail the properties of $q$-exponent $e_q(x)$.

\medbreak
{\bf Lemma 9.} {\it Assume that variables $a$ and $b$ satisfy the Weyl
relation $ab=qab$. Then we have}

\vskip 0.2cm
$i)$~~~~ $e_q(b)e_q(a)=e_q(a+b)$, \hfill (2.57)

\vskip 0.2cm
$ii)$~~~ $e_q(a)e_q(b)=e_q(b-ba)e_q(a)$ \hfill (2.58)

\vskip 0.2cm
\hskip 2.45cm $=e_q(a+b-ba)$ \hfill (2.59)

\vskip 0.2cm
\hskip 2.45cm $=e_q(b)e_q(-ba)e_q(a)$ \hfill (2.60)
\vskip 0.2cm

\hskip 2.45cm $=e_q(b)e_q(a-ba)$ \hfill (2.61)
\vskip 0.2cm

Proof. The first relation (2.57) is well-known and follows from the
following result due to M.-P. Sch\"utzenberger (the middle of 50's)
$${\rm if}~~ab=qab,~~{\rm
then}~~(a+b)^n=\sum_{k=0}^na^kb^{n-k}\left[\matrix{n\cr k}\right]_q.
$$
As for the identities of part $ii)$, the crucial relation is (2.58). Both
(2.59) and (2.60) follow from (2.58) and (2.57) (hint:
$a(b-ba)=q(b-ba)a$, $bab=qb^2a$). Now let us prove the identity (2.58).
We give two proofs.

{\it First proof.} Let us check that the both sides of (2.58) as the
functions on $b$ satisfy the same functional equation. Indeed, ($e(\cdot
):=e_q(\cdot )$)
$$\eno{
&e(a)e(qb)e^{-1}(b)e^{-1}(a)=e(a)(1-b)e^{-1}(a)=1-e(a)be^{-1}(a)\cr
&=1-be(qa)e^{-1}(a)=1-b(1-a)=1-b+ba;\cr
&e(q(b-ba))e(a)e^{-1}(a)e^{-1}(b-ba)=1-b+ba.}
$$

{\it Second proof} (A. Volkov). We have
$$e(a)e(b)e^{-1}(a)=e(e(a)be(a^{-1}))=e(be(qa)e^{-1}(a))=e(b(1-a)).
$$

In a similar manner, the identity (2.61) may be derived from (2.54).

\qed

It is very surprising that (2.57) and (2.59) have the polynomial analogs.

\medbreak
{\bf Lemma 10.} {\it Under the assumptions of Lemma 9, we have}

\vskip 0.2cm
$i)$ ~ $(a;q)_n\cdot (b;q)_n=(a+b-q^nba;q)_n$, \hfill(2.62)

\vskip 0.2cm
$ii)$ ~ $(b;q)_n\cdot (a;q)_n=(a+b-ba;q)_n$. \hfill(2.63)
\vskip 0.2cm

Proof. Let us prove (2.62) by induction. The case $n=1$ is clear.
Further,
$$\eno{
{\rm LHS}(2.62)&=(1-q^{n-1}a)(a)_{n-1}(1-b)(qb)_{n-1}\cr
&=(1-a)(qa)_{n-1}(qb)_{n-1}-(1-q^{n-1}a)b(qa)_{n-1}(qb)_{n-1}\cr
&=(1-a-b+q^{n-1}ab)(qa)_{n-1}(qb)_{n-1}\cr
&{\buildrel{\rm induction}\over =}~~(1-a-b+q^nba)(qa+qb-q^{n+1}ba)_{n-1}
=(a+b-q^nba)_n.}
$$

Analogously, \ \ LHS(2.63)~=~$(1-b)(qb)_{n-1}(1-q^{n-1}a)(a)_{n-1}$
$$\eno{
&=(1-q^{n-1}b)(b)_{n-1}(a)_{n-1}-q^{n-1}(1-b)(qb)_{n-1}a(a)_{n-1}\cr
&=(1-q^{n-1}b)(b)_{n-1}(a)_{n-1}-q^{n-1}(1-b)a(b)_{n-1}(a)_{n-1}\cr
&=(1-q^{n-1}b-q^{n-1}a+q^{n-1}ba)(b)_{n-1}(a)_{n-1}~~
{\buildrel{\rm induction}\over =}~~(a+b-ba)_n.}
$$

\qed

Now we are ready to give the quantum analog of the five-term relation
(1.4) for the quantum dilogarithm $Li_2(x;q)$.

\medbreak
{\bf Theorem I} (L.D. Faddeev and R.M.  Kashaev, [FK]). {\it The five term
or "pentagon" relation ($ab=qba$)
$$e_q(a)e_q(b)=e_q(b)e_q(-ba)e_q(a)\eqno (2.60)
$$
is a quantum analog of the Rogers five-term relation (1.4) or (2.56),
i.e. an appropriate limit $q\to 1$ in (2.60) reduces the last to the
relation (2.56).}

Let us rewrite the relations (2.57), (2.59) and (2.60) using the quantum
dilogarithm.
For this purpose we need the Baker-Campbell-Hausdorff series. Namely, for
any Lie algebra with Lie bracket $[\cdot ,\cdot ]$, let us denote by
$H(x,y)$ its Campbell-Hausdorff series, i.e.
$$\eno{
&e^x\cdot e^y=e^{H(x,y)},\cr
&H(x,y)=x+y+{1\over 2}[x,y]+{1\over 12}\left\{[x,[x,y]]+[y,[y,x]]\right\}+
{1\over 24}[xyyx]\cr
&+{1\over 144}\left\{ 9[xxyyx]+9[yyxxy]+4[xyyyx]-2[xxxxy]-2[yyyyx]\right\}+
\ldots ,}
$$
where $[a_1a_2\ldots a_n]$ is the multiple commutator $[a_1,[a_2,\ldots ,
[a_{n-1},a_n]\cdots ]$.
\medbreak
{\bf Proposition 2} (Functional equations for quantum dilogarithm). {\it We
have ($ab=qba$)}

\vskip 0.2cm
$i)$ \ $H(Li_2(a;q),Li_2(b;q))=Li_2(a+b-ab;q)$,

\hfill (2.64)

$ii)$ \ $H(Li_2(b;q),Li_2(a;q))=Li_2(a+b;q)$.

\medbreak
{\bf Corollary 11} (Five term relation for quantum dilogarithm).
$$\eno{
&H(Li_2(a;q),Li_2(b;q))=H(Li_2(b;q),H(Li_2(-ba;q),Li_2(a;q)))&(2.65)\cr
&=H(H(Li_2(b;q),Li_2(-ba;q)),Li_2(a;q)).}
$$
\bigbreak

{\bf Exercises to Section 2.5.}
\medbreak

{\bf A} (Weyl algebra and $q$-exponential a function).

{\bf 1.} Let us consider a function $\varphi (a)$
$$\varphi (a):=e_{q^2}(-qa)=\sum_{n\ge 0}{(-qa)^2\over (q^2;q^2)_n}=
{1\over\ds\prod_{j\ge 0}(1+q^{2j+1}a)}.
$$
Show that

\vskip 0.2cm
\hskip -0.5cm $i)$ $\varphi (qa)=(1+a)\varphi (q^{-1}a)$;

\vskip 0.2cm
\hskip -0.5cm $ii)$ If $a$ and $b$ satisfy the Weyl relation $ab=q^2ba$,
then
$$\eno{
\varphi (b)\varphi (a)&=\varphi (a+b),&(1)\cr
\varphi (a)\varphi (b)&=\varphi(b)\varphi (a+qb),&(2a)\cr
&=\varphi (a+b+qba),&(2b)\cr
&=\varphi (b)\varphi (qba)\varphi (a),&(2c)}
$$
Hint: use the relations (2.57)-(2.61).

\medbreak
{\bf 2.} Let us consider the theta-function ~~$\theta (a):=
[\varphi (a)\varphi (a^{-1})]^{-1}$.
Show that

\vskip 0.2cm
\hskip -0.5cm $i)$ $\theta (qa)=a^{-1}\theta (q^{-1}a)$;

\vskip 0.2cm
\hskip -0.5cm $ii)$ $\theta (a)=\ds\sum_{n\in{\bf Z}}q^{n^2}a^n$;

\vskip 0.2cm
\hskip -0.5cm $iii)$ (cf. [FV]) If $ab=q^2ba$, then
$$\theta (a)\tt (b)\tt (a)=\tt (b)\tt (a)\tt (b).\eqno (3)
$$
Hint: use the following relation ~~$\tt (a)\tt (b)a=b\tt (a)\tt (b)$.
\medbreak

{\bf 3.} Let us consider a function ~~$f(a):=f(z,a)=
\varphi (za)/\vp (a)$.
Show that

\vskip 0.3cm
\hskip -0.5cm $i)$ ~$\ds{f(qa)\over f(q^{-1}a)}=\ds{1+za\over 1+a}$, ~~
$\ds{f(qz,a)\over f(q^{-1}z,a)}=1+az$;

\vskip 0.2cm
\hskip -0.5cm $ii)$ ~$f(z,a)=\ds\sum_{n\ge 0}{a^nq^{n^2}(z;q^{-2})_n\over
(q^2;q^2)_n}$;

\vskip 0.2cm
\hskip -0.5cm $iii)$ ~(Cf. [FV]) If $ab=q^2ba$, then $f(b)f(a)=f(a+b+qba)$.

Hint: use the relations
$$\eno{
&\vp (za)\vp (b)=\vp (b)\vp (za+zqba),\cr
&\vp (zb)\vp (za+zqba)=\vp (za+zb+zqba).}
$$

{\bf 4.} Let us consider a function ~~$\pi (a):=\pi (z,a)=
\vp (za)\vp (za^{-1})$.
Show that

\vskip 0.3cm
\hskip -0.5cm $i)$ ~$\pi (z,a)=\ds\sum_{n\ge 0}{(-qz)^nH_n(a)\over
(q^2;q^2)_n}$, ~~where $H_n(a):=H_n(a;q^2)$ is the so-called continuous
$q$-Hermite polynomial. Prove the recurrence relation for the continuous
$q$-Hermite polynomials:
$$H_n(a)=AH_{n-1}(a)-(1-q^{2n-2})H_{n-2}(a),\eqno (4)
$$
where $A=a+a^{-1}$, $H_0(a)=1$, $H_1(a)=a+a^{-1}$.

It follows from (4), that
$$H_n(a;q^2)=\det\left|\matrix{A&1&\ldots &0&0\cr 1-q^2&A&\ldots&0&0\cr
\ldots&&\ldots&&\ldots\cr 0&0&\ldots&A&1\cr 0&0&\ldots&1-q^{2n-2}&A}
\right|_{n\times n}.
$$
Hint: use the functional equation:
$${\pi (qa)\over \pi (q^{-1}a)}={1+za\over 1+za^{-1}}.\eqno (5)
$$

\hskip -0.5cm $ii)$ (Modifyed Yang-Baxter equation). If $ab=q^2ba$, then
$$\pi (z',b)\pi (zz',qba)\pi (z,a)=\pi (z,a)\pi (zz',qab^{-1})\pi
(z',b).\eqno (6)
$$

Proof. Let us remark at first that ($ab=q^2ba$)
$$\eno{
&\pi (z,a)b(1+zq^{-1}a^{-1})=b(1+zqa)\pi (z,a),& (7)\cr
&\pi (z,a)b^{-1}(1+zq^{-1}a)=b^{-1}(1+zqa^{-1})\pi (z,a).}
$$
Indeed, using the functional equation (5), one can find
$$\pi (z,a)b(1+aq^{-1}a^{-1})=b\pi (z,q^2a)(1+zq^{-1}a^{-1})=
b(1+zqa)\pi (z,a).
$$

To go further, let us use the relations (1) and (2). One can check
$$\eno{
&\Phi :=\pi (z',b)\pi (zz',qba)=\vp (z'b)\vp (z'b^{-1})\vp (zz'qba)\vp
(zz'qb^{-1}a^{-1})\cr
&{\buildrel (2c)\over =}~\vp (z'b)\vp (zz'qba)\vp (z(z')^2a)\vp
(z'b^{-1})\vp(zz'qb^{-1}a^{-1})\cr
&{\buildrel (1)\over =}~\vp (z'b(1+qza))\vp (z(z')^2a)\vp
(z'b^{-1}(1+qza^{-1})).}
$$
Consequently (use the relation (7) !),
$$\eno{
\Phi\cdot\pi (z,a)&=\pi (z,a)\vp (z'b(1+zq^{-1}a^{-1}))\vp (z(z')^2a)\vp
(z'b^{-1}(1+zq^{-1}a))\cr
&=\pi (z,a)\pi (zz', qab^{-1})\pi (z', b),}
$$
as it was claimed.

\qed

{\bf 5} (Commutation relations for the continuous $q$-Hermite
polynomials).

Show that ($ab=q^2ba$)

\vskip 0.2cm
\hskip -0.5cm $i)$ ~$q[B_1(a),B_n(b)]=(1-q^{2n})\left\{ B_1(qab^{-1})
B_{n-1}(b)-B_{n-1}(b)B_1(qba)\right\}$,

\vskip 0.2cm
$q[B_n(a),B_1(b)]=(1-q^{2n})\left\{ B_{n-1}(a)
B_{1}(qab^{-1})-B_{1}(qba)B_{n-1}(a)\right\}$.

\vskip 0.3cm
\hskip -0.5cm $ii)$ ~More generally, in the Weyl algebra $\{
a,b~|~ab=q^2ba\}$ there exist the following relations ($m,n=1,2,3,\ldots $)
$$\eno{
&\sum_{k=0}^{\min (n,m)}(-q^{-1})^k\left[\matrix{n\cr k}\right]_{q^2}
\left[\matrix{m\cr
k}\right]_{q^2}(q^2;q^2)_kB_{n-k}(b)B_k(qba)B_{m-k}(a)&(8)\cr \cr
&=\sum_{k=0}^{\min (n,m)}(-q^{-1})^k\left[\matrix{n\cr k}\right]_{q^2}
\left[\matrix{m\cr
k}\right]_{q^2}(q^2;q^2)_kB_{m-k}(a)B_k(qab^{-1})B_{n-k}(b).}
$$
Hint: the relations (8) are equivalent to the modifyed Yang-Baxter
equation (6).

\medbreak
{\bf 6.} Let us consider a function (cf. [FV]) ~~$r(z,a)=
\tt(a)\pi (z,a)$. Show that

\hskip -0.5cm $i)$ (functional equations)
$${r(z,qa)\over r(z,q^{-1}a)}={1+za\over a+z},~~{r(qz,a)\over r(q^{-1}z,a)}
=(1+za)(1+za^{-1}).
$$

\hskip -0.5cm $ii)$ ~$r(z,a)=1+\ds\sum_{n\ge 1}q^{n(n-1)}{(z;q^{-2})_n\over
(zq^2;q^2)_n}(a^n+a^{-n})$.

\vskip 0.2cm
\hskip -0.5cm $iii)$ (Yang-Baxter equation, L.D.~Faddeev and A.Yu.~Volkov
[FV], A.N.~Kirillov).

If $ab=q^2ba$ and $zz'=z'z$, then
$$r(z,a)r(zz',b)r(z',a)=r(z',b)r(zz',a)r(z,b).\eqno (9)
$$
Hint: using the commutation relations
$$\eno{
&\tt (b)^{\mp 1}g(a)\tt (b)^{\pm 1}=g(q^{\pm 1}ab^{\mp 1}),\cr
&\tt (a)^{\mp 1}g(b)\tt (a)^{\pm 1}=g(q^{\mp 1}a^{\pm 1}b),}
$$
to reduce a proof of (9) to that of (6).

\hskip -0.5cm $iv)$ Using the functional equation ~~$(z+a)r(z,qa)=
(1+za)r(z, q^{-1}a)$, ~~check that
$$\eno{
&{\rm LHS}(9)(az^{-1}+qba+z'b)=(z'a+qba+bz^{-1}){\rm LHS}(9),\cr
&{\rm RHS}(9)(az^{-1}+qba+z'b)=(z'a+qba+bz^{-1}){\rm RHS}(9).}
$$
Deduce from these relations that
$${\rm LHS}(9)={\rm RHS}(9).
$$

\medbreak
{\bf B.} (Miscellany).

{\bf 7.} Let $W_q$ be the algebra over the field of rational functions
${\bf C}(q)$ generated by $a$ and $b$ subject to the relation $ab=q^2ba$.

Prove that the maps ~~$\Delta$  $(\wt\Delta ):~W_q\to W_q\otimes W_q$
$$\eno{
&\Delta (a)=1\otimes a+a\otimes b,~~({\wt\Delta}(a)=a\otimes a),\cr
&\Delta (b)=b\otimes b,~~({\wt\Delta}(b)=1\otimes b+b\otimes a)}
$$
define the comultiplications in $W_q$.
\medbreak

{\bf 8.} Consider an element $f(a,b)=\ds\sum_{n,m}c_{n,m}(q)b^na^m$ from
the Weyl algebra $W_{q^{1\over 2}}$ and assume that there exists the limit
$$\lim_{q\to 1}(1-q)f(a,b)={\wt f}({\wt a},{\wt b}),
$$
where $\wt a$ and $\wt b$ commutes.

Prove that

\vskip 0.2cm
\hskip -0.5cm $i)$ ~~$\ds\lim_{q\to 1}(1-q)\left[ L(a),f(a,b)\right]=\log
(1-{\wt a})\left({\wt b}{\partial\over \partial{\wt b}}\right){\wt
f}({\wt a},{\wt b})$,

\vskip 0.3cm
 ~~$\ds\lim_{q\to 1}(1-q)\left[f(a,b),L(b)\right]=\log
(1-{\wt b})\left({\wt a}{\partial\over \partial{\wt a}}\right){\wt
f}({\wt a},{\wt b})$.

\vskip 0.3cm
\hskip -0.5cm $ii)$ ~~$\ds\lim_{q\to 1}(1-q)\sum_{n\ge 0}{1\over n!}{\rm
ad}^n_{L(a)}\left(f(a,b)\right) ={\wt f}({\wt a},{\wt b}-{\wt a}{\wt
b})$,

where ~~ad$_y^0(x)=x$ and ~~ad$_y^{n+1}(x)=[y,{\rm ad}_y^n(x)]$,
$n\ge 0$.

\vskip 0.2cm
Hint: use the Taylor formula ~~($xy=yx$)
$$\exp (\left(yx\ds{\partial\over
\partial x}\right) f(x)=f\left(x\exp (y)\right).
$$
\medbreak

{\bf 9} (Quantum polylogarithm). Let us define the quantum polylogarithm
as the following formal series ($k\ge 1$)
$$Li_k(x;q):=\sum_{n\ge 1}{x^n\over n(1-q^n)^{k-1}}.\eqno (10)
$$

$\bullet$ ~Prove that ($k\ge 2$)
$$\exp\left(Li_k(x;q)\right) =\prod_{m=(m_1,\ldots
,m_{k-2})\in{\bf Z}_+^{k-2}}e_q(q^{|m|}x),\eqno (11)
$$
where $|m|:=m_1+\ldots +m_{k-2}$.

Hint: it is clear that $Li_k(x;q)-Li_k(qx;q)=Li_{k-1}(x;q)$.
Consequently,
$$\eno{
&Li_k(x;q)=\ds\sum_{m\ge 0}Li_{k-1}(q^mx;q),\cr
&\exp\left(Li_k(x;q)\right) =\ds\prod_{m\ge 0}\exp\left(
Li_{k-1}(q^mx;q)\right).}
$$

$\bullet$ ~(Functional equation for quantum trilogarithm). Prove that if
$x$ and $y$ satisfy the Weyl relation $xy=qyx$, then
$$\eno{
\exp \left( Li_3(x;q)\right)\exp\left(Li_3(y;q)\right)
&=\exp\left(Li_3(y;q)\right)\exp\left( Li_3((y)_{\infty}x;q)\right)&(12)\cr
&=\exp\left( Li_3(y(x)_{\infty};q)\right)\exp\left( Li_3(x;q)\right) .}
$$

It is a very intertesting task to study the asymptotic behavior of the
quantum polylogarithm functional
equations (see e.g. (2.64), (2.65) and (9), (11), (12)) when ~$q=\zeta
e^{-\epsilon}\to 1$, ~$\epsilon\to 0$, ~where
{}~$\zeta =\exp\ds\left({2\pi i\over l}\right)$.
We intended to consider this interesting subject in [Kir10].

\vskip 0.4cm
{\bf Appendix.}
\medbreak

Proof of Proposition E, $n=2$, $A=1$. We have the following series of
relations coming from the functional equation (1.4):
$$\eno{
\bullet~& L(\beta_1/\al_2)+L(\al_2x_1)-L(\beta_1x_1)-L\left({\beta_1
(1-\al_2x_1)\over\al_2(1-\beta_1x_1)}\right) -L\left({(\al_2-\beta_1)
x_1\over 1-\beta_1x_1}\right) =0,\cr
\bullet~& L(\beta_1/\al_1)+L(\al_1x_2)-L(\beta_1x_2)-L\left({\beta_1
(1-\al_1x_2)\over\al_1(1-\beta_1x_2)}\right) -L\left({(\al_1-\beta_1)
x_2\over 1-\beta_1x_2}\right) =0,\cr
\bullet~& L(\beta_2/\al_1)+L(\al_1x_1)-L(\beta_2x_1)-L\left({\beta_2
(1-\al_1x_1)\over\al_1(1-\beta_2x_1)}\right) -L\left({(\al_1-\beta_2)
x_1\over 1-\beta_2x_1}\right) =0,\cr
\bullet~& L(\beta_2/\al_2)+L(\al_2x_2)-L(\beta_2x_2) -L\left({\beta_2
(1-\al_2x_2)\over\al_2(1-\beta_2x_2)}\right) -L\left({(\al_2-\beta_2)
x_2\over 1-\beta_2x_2}\right) =0,\cr
\bullet~& L\left({(\al_1-\beta_1)x_2\over 1-\beta_1x_2}\right) +
L\left({(\al_2-\beta_1)x_1\over 1-\beta_1x_1}\right) -L(t) -L\left(
{x_1(\al_2-\beta_1)(1-\beta_2x_2)\over (1-\beta_1x_1)(1-\al_2x_2)}
\right)\cr
&-L\left({x_2(\al_1-\beta_1)(1-\beta_2x_1)\over (1-\beta_1x_2)
(1-\al_1x_1)}\right) =0,\cr
\bullet~&L\left({(\al_1-\beta_2)x_1\over 1-\beta_2x_1}\right) +
L\left({(\al_2-\beta_2)x_2\over 1-\beta_2x_2}\right) -L(t)-L\left(
{x_1(\al_1-\beta_2)(1-\beta_1x_2)\over (1-\beta_2x_1)(1-\al_1x_2)}
\right)\cr
&-L\left({x_2(\al_2-\beta_2)(1-\beta_1x_1)\over (1-\beta_2x_2)
(1-\al_2x_1)}\right) =0,\cr
\bullet~&L\left({\beta_1(1-\al_2x_1)\over\al_2(1-\beta_1x_1)}\right) +
L\left({\beta_2(1-\al_1x_1)\over\al_1(1-\beta_2x_1)}\right) -
L\left({\beta_1\beta_2(1-t)\over \al_1\al_2}\right)\cr
& -L\left(
{\beta_1(1-\al_2x_1)(1-\beta_2x_2)\over (1-\beta_1x_1)(\al_2-\beta_2)}
\right)-L\left({\beta_2(1-\al_1x_1)(1-\beta_1x_2)\over (1-\beta_2x_1)
(\al_1-\beta_1)}\right) =0,\cr
\bullet~&L\left({\beta_1(1-\al_1x_2)\over\al_1(1-\beta_1x_2)}\right) +
L\left({\beta_2(1-\al_2x_2)\over\al_2(1-\beta_2x_2)}\right) -
L\left({\beta_1\beta_2(1-t)\over \al_1\al_2}\right)\cr
&-L\left(
{\beta_1(1-\al_1x_2)(1-\beta_2x_1)\over (1-\beta_1x_2)(\al_1-\beta_2)}
\right) -L\left({\beta_2(1-\al_2x_2)(1-\beta_1x_1)\over (1-\beta_2x_2)
(\al_2-\beta_1)}\right) =0,\cr
\bullet~&L\left({(\al_2-\beta_2)x_2(1-\beta_1x_1)\over (1-\al_2x_1)
(1-\beta_2x_2)}\right) +
L\left({\beta_1(1-\al_2x_1)(1-\beta_2x_2)\over (\al_2-\beta_2)
(1-\beta_1x_1)}\right) -L\left( x_2\beta_1\right)\cr
& -L\left({-\beta_1/\al_1\over 1-\beta_1/\al_1}\right)
-L\left({-\al_1x_2\over 1-\al_1x_2}\right) =0,\cr
\bullet~&L\left({(\al_1-\beta_1)x_2(1-\beta_2x_1)\over (1-\al_1x_1)
(1-\beta_1x_2)}\right) +
L\left({\beta_2(1-\al_1x_1)(1-\beta_1x_2)\over (\al_1-\beta_1)
(1-\beta_2x_1)}\right) -L\left( x_2\beta_2\right)\cr
& -L\left({-\beta_2/\al_2\over 1-\beta_2/\al_2}\right)
-L\left({-\al_2x_2\over 1-\al_2x_2}\right) =0,\cr
\bullet~&L\left({(\al_2-\beta_1)x_1(1-\beta_2x_2)\over (1-\al_2x_2)
(1-\beta_1x_1)}\right) +
L\left({\beta_2(1-\al_2x_2)(1-\beta_1x_1)\over (\al_2-\beta_1)
(1-\beta_2x_2)}\right) -L\left( x_1\beta_2\right)\cr
& -L\left({-\beta_2/\al_1\over 1-\beta_2/\al_1}\right)
-L\left({-\al_1x_1\over 1-\al_1x_1}\right) =0,\cr
\bullet~&L\left({(\al_1-\beta_2)x_1(1-\beta_1x_2)\over (1-\al_1x_2)
(1-\beta_2x_1)}\right) +
L\left({\beta_1(1-\al_1x_2)(1-\beta_2x_1)\over (\al_1-\beta_2)
(1-\beta_1x_2)}\right) -L\left( x_1\beta_1\right)\cr
& -L\left({-\beta_1/\al_2\over 1-\beta_1/\al_2}
\right) -L\left({-\al_2x_1\over 1-\al_2x_1}\right) =0.}
$$
Let us denote by $S$ the LHS of the sum of all of these relations. Let
$D$ be a difference between the LHS and the RHS of the identity from
Proposition E. Then one can check
$$S=-2D-\sum_{i,j}\left\{ L(\al_ix_j)+L\left({-\al_ix_j\over
1-\al_ix_j}\right) -2L(\al_i /\al_j)\right\} +2L(1)=0.
$$
But it follows from the integral representation (1.2) that a function
$$l(t):=\sum_{i,j}\left\{L(\al_ix_j(t))+L\left({-\al_ix_j(t)\over
1-\al_ix_j(t)}\right)\right\}
$$
does not depend on $t$. Taking $t=1$ one can find this constant. The result
is
$$l(t)=2\sum L(\al_i/\al_j)+2L(1),$$
consequently, $D=0$.

Note that the next Lemma and identities (A1)-(A2) had played the key role
in the proving  of Proposition E.
\medbreak

{\bf Lemma A1.} {\it Let $r(x,t)$ be a polynomial as in Proposition E.
Then we have

$i)$ $\left(1-\ds{A\beta_1\ldots \beta_n\over\al_1\ldots\al_n}(1-t)\right)
\ds\prod_{l=1}^n(1-\beta x_l)=\prod_{l=1}^n\left(1-{\beta\over\al_l}\right)$,
where $\beta =\beta_k$ for some $k$, $1\le k\le n$,

$ii)$ $\left(1-\ds{A\beta_1\ldots \beta_n\over\al_1\ldots\al_n}(1-t)\right)
\ds\prod_{l=1}^n(1-\al
x_l)=-(1-t)\prod_{l=1}^n\left(1-{\beta_l\over\al}\right)$,
where $\al =\al_k$ for some $k$, $1\le k\le n$.}
\medbreak

{\bf Corollary} (of Lemma A1, with $n=2$). {\it We have}
$$1-{(\al_1-\beta_1)x_2(1-\beta_2x_1)\over (1-\beta_1x_2)(1-\al_1x_1)}
=-{(\al_1-\beta_1)(\al_1-\beta_2)\over
(1-\beta_1x_2)(1-\al_1x_1)(\al_1\al_2-\beta_1\beta_2(1-t))},\eqno(A1)
$$
$$1-{\beta_1(1-\beta_2x_2)(1-\al_2x_1)\over (\al_2-\beta_2)(1-\beta_1x_1)}
={\al_1(\al_2-\beta_1)\over
(1-\beta_1x_1)(\al_1\al_2-\beta_1\beta_2(1-t))}.\eqno(A2)
$$

\qed
\vfil\eject
\vskip 1.0cm

{{\bf References}\footnote*{Most of the following are referred in our
lectures. All are relevant in one way or the other to the topics presented.}}

\medbreak
\item{[A]} K. Aomoto. Addition theorem of Abel type for hyper-logarithms.
{\it Nagoya Math. J.}, 1982, v.88, 55-71.
\item{[AL]} D. Acreman and J.H. Loxton. Asymptotic analysis of Ramanujan
pairs. {\it Aequat. Math.}, 1986, v.30, 106-117.
\item{[An1]} G. Andrews. The theory of partitions. {\it Addison-Wesley
Publishing Company}, 1976.
\item{[An2]} G. Andrews. Multiple series Rogers-Ramanujan type
identities. {\it Pacific J. Math.}, 1984, v.114, 267-283.
\item{[An3]} G. Andrews. $q$-series: their development and application in
analysis, number theory, combinatorics, physics, and computer algebra. {\it
Regional conference series in mathematics, ISSN 0160-7642}, 1985, no.66.
\item{[ABBFV]} G.~Andrews, R.~Baxter, D. Bressoud, W. Burge, P. Forrester
and G.~Viennot. Partitions with prescribed hook differences. {\it Europ.
J. Combinatorics}, 1987, v.8, 341-350.
\item{[ABF]} G.~Andrews, R.~Baxter and P.~Forrester.
Eight-vertex ${SOS}$-model and generalized {R}ogers-{R}amanujan-type
identities. {\it J. Stat. Phys.}, 1984, v.35, 193-266.
\item{[AB]} G.~Andrews and D. Bressoud. On the Burge correspondence
between partitions and binary words. {\it Rocky Mtn. J. Math.}, 1985,
v.15, 225-233.
\item{[AAB]} A. Agarwal, G.~Andrews and D. Bressoud. The Bailey lattice.
{\it J. Indian Math. Soc.}, 1987, v.51, 57-73.
\item{[B]} W. Bailey. Identities of the Rogers-Ramanujan type. {\it Proc.
London Math. Soc. (2)}, 1949, v.50, 1-10.
\item{[BR]} V. Bazhanov and N. Reshetikhin. Restricted solid-on-solid
models connected with simply laced algebras and conformal field theory.
{\it J. Phys. A}, 1990, v.23, 1477-1492.
\item{[BGSV]} A. Beilinson, A. Goncharov, V. Schechtman and A. Varchenko.
Aomoto dilogarithm, mixed Hodge structures, and motivic cohomology of
pairs of triangles on the plane. {\it The Grothendieck Festschrift,
vol.1, Birkhauser, Progress in Math. Series, Boston}, 1990, v.86, 135-172.
\item{[Ber]} A. Berkovich. Fermionic counting of RSOS-states and Virasoro
character formulas for the unitarity minimal series $M(\nu ,\nu +1)$.
Exact results. {\it Preprint Univ. Bonn}, 1994, 31p.; hep-th/9403073.
\item{[Bl]}S.~Bloch. Application of the dilogarithm function in algebraic
$K$-theory  and algebraic geometry. {\it Proc. of the International Symp.
on Alg. Geom.(Kyoto Univ., Kyoto,1977)},\ 103-114, Kinokuniya Book Store,
Tokyo, 1978.
\item{[BS]} P. Bouwknegt and K. Schoutens. $\cal W$ symmetry in Conformal
Field Theory. {\it Phys. Rep.}, 1993, v.223, 183-276.
\item{[Bre1]} D. Bressoud. Analytic and combinatorial generalization of the
Rogers-Ramanujan identities. {\it Mem. Amer. Math. Soc.}, 1980, v.224, 54p.
\item{[Bre2]} D. Bressoud. An easy proof of the Rogers-Ramanujan
identities. {\it J. Number Theory}, 1983, v.16, 235-241.
\item{[Bre3]} D. Bressoud. Problems on the $Z$-statistic. {\it Discrete
Math.}, 1988/89, v.73, 37-48.
\item{[Bre4]} D. Bressoud. Lattice paths and the Rogers-Ramanujan
identities. {\it Lect, Notes in Math.}, 1987, v.1395, 140-172.
\item{[Bre5]} D. Bressoud. In the land of OZ. {\it in "$q$-series and
partitions"} (ed. D.~Stanton), IMA Volumes in Math. and Applic., v.18,
Springer-Verlag, New-York, 1989, 45-55.
\item{[Bre6]} D. Bressoud. The Bailey lattice: an introduction. In {\it
Ramanujan Revisited (G.~Andrews et al. eds.), Academic Press, Inc.}, 1988,
57-67.
\item{[Bro1]} J. Browkin. Conjectures on the dilogarithm. In {\it
$K$-theory}, 1989, v.3, 29-56.
\item{[Bro2]} J. Browkin. $K$-theory, cyclotomic equations and Clansen's
functions. {\it In [Le4], Chapter 11}, 233-273.
\item{[Bry]} R.K. Brylinski. Limits of weight spaces, Lusztig's $q$-analog
and fiberings of adjoint orbits. {\it J. Amer. Math. Soc.}, 1989, v.2,
517-533.
\item{[Bur]} W. Burger, A three-way correspondence between partitions.
{\it Eorop. J. Comb.}, 1982, v.3, 195-213.
\item{[Ca]} J.L. Cathelineau. Homologie du groupe lineaire et
polylogarithmes. {\it Seminaire BOURBAKI}, 1992-93, $n^0$ 772, 23p.
\item{[Co]} H.S.M. Coxeter. The functions of Schl\"afli and Lobachevsky.
{\it Quart. J. Math. Oxford Ser.}, 1935, v.6, 13-29.
\item{[D1]} S. Dasmahapatra. String hypothesis and characters of coset CFTs.
{\it Preprint IC/93/91}, 1993, 13p.
\item{[D2]} S. Dasmahapatra. On state counting and characters. {\it City
Univ. Preprint}, 1994, 22p.
\item{[DJKMO1]} E. Date, M. Jimbo, A. Kuniba, T. Miwa and M. Okado.
Exactly solvable SOS models II: Proof of the star-triangle relation and
combinatorial identities. {\it Adv. Studies in Pure Math.}, 1988, v.16,
17-122.
\item{[DJKMO2]} E. Date, M. Jimbo, A. Kuniba, T. Miwa and M. Okado. Path,
Maya Diagrams and representations of $\wh{\hbox{\germ s}l}(r,{\bf C})$.
{\it Adv. Studies in Pure Math.}, 19, v., 149-191.
\item{[DKKMM]} S. Dasmahapatra, R. Kedem, T.R. Klassen, B.M. McCoy and
E. Melzer. Quasi-particles, Conformal Field Theory, and $q$-series. {\it
Int. J. Modern Phys.}, 1993, v.7, 3617-3648.
\item{[DKMM]} S. Dasmahapatra, R. Kedem, B.M. McCoy and E. Melzer.
Virasoro characters from Bethe equations for the critical ferromagnetic
three-state Potts model. {\it J. Stat. Phys.}, 1994, v.74, 239-274.
\item{[DS]} J.L. Dupont and C.-H. Sah. Dilogarithm identities in
Conformal Field Theory and group homology. {\it Preprint}, 1993.
\item{[EKK]} F. Essler, V.E. Korepin and K. Schoutens. Fine structure of
the Bethe ansatz for the spin-${1\over 2}$ Heisenberg $XXX$ model. {\it
J. Phys. A: Math. Gen.}, 1992, v.25, 4115-4126.
\item{[FK]} L.D. Faddeev and R.M. Kashaev. Quantum dilogarithm. {\it
Preprint}, 1993, 8p.
\item{[FV]} L.D. Faddeev and A.Yu. Volkov. Abelian current algebra and
the Virasoro algebra on the lattice. {\it Preprint HU-TFT-93-29}, 1993.
\item{[FeFr]} B. Feigin and E. Frenkel. Coinvariants of nilpotent
subalgebras of the Virasoro algebra and partition identities. {\it
Advances in Soviet Mathematics}, 1993, v.9.
\item{[FF]} B. Feigin and D. Fuchs. Representations of the Virasoro
algebra. {\it in "Representations of infinite-dimensional Lie groups and
Lie algebras"}, eds. A.M. Vershik and D.P. Zhelobenko, Gordon and Breach,
1990, 465-554.
\item{[FeNO]} B. Feigin, T. Nakanishi and H. Ooguri. The annihilating
ideals of minimal models. {\it Int. J. Mod. Phys.}, 1992, v.1A, 217-238.
\item{[FeSt]} B.L. Feigin and A.V. Stoyanovsky. Quasi-particles models for
the representations of Lie algebras and geometry of flag manifold. {\it
Preprint}, 1993, 35p.
\item{[Fi]} S. Fishel. The combinatorics of certain entries in the
two-parameter Kostka matrix. {\it Preprint}, 1993, 9p.
\item{[Fo]} D. Foata. Distribution eul\'eriennes et manoniennes sur le
group des permutations. {\it in Higher Combinatorics}, [M.~Aigner, ed.
Berlin, 1976], p.27-49, Amsterdam, D.~Reidel, 1977 (Proc. NATO Adv. Study
Inst.)
\item{[FQ]} O. Foda and Y.-H. Quano. Virasoro character identities from
the Andrews-Bailey construction. {\it Univ. Melbourne Preprint No.26},
1994, 18p.
\item{[FB]} P. Forrester and R. Baxter. Further exact solutions of the
eight-vertex $SOS$ model and generalizations of the Rogeres-Ramanujan
identities. {\it J. Stat. Phys.}, 1985, v.38, 435-472.
\item{[FrSz1]} E. Frenkel and A. Szenes. Dilogarithm identities,
$q$-difference equations and the Virasoro algebra. {\it Duke Math. J.,
Int. Math. Res. Notices}, 1993, v.2, 53-60.
\item{[FrSz2]} E. Frenkel and A. Szenes. Crystal bases, dilogarithm
identities and torsion in algebraic $K$-groups. {\it Preprint}, 1993, 33p.
\item{[Ga]} A. Garsia. Orthogonality of Milne's polynomials and raising
operators. {\it Discrete Math.}, 1992, v.99, 247-264.
\item{[GM]} A.~Garsia and S.~Milne. A {R}ogers-{R}amanujan bijection.
{\it J. Comb. Theory, Ser. A}, 1981, v.31, 289-339.
\item{[GR]} G. Gasper and M. Rahman. Basic hypergeometric series. {\it
Cambridge Univ. Press}, 1990.
\item{[GS]} I. Gessel and D. Stanton. Applications of $q$-Lagrange
inversion to basic hypergeometric series. {\it Trans. Amer. Math. Soc.},
1983, v.277, 173-201.
\item{[Gre]} J. Greene. Bijections related to statistics on words. {\it
Discrete Math.}, 1988, v.68, 15-29.
\item{[GKO]} P.~Goddard, A.~Kent, and D.~Olive. Virasoro algebras
and coset space models. {\it Phys. Lett.}, 1985, v.B152, 88-93.
\item{[GW]} F. Goodman and H. Wenzl. Littlewood-Richardson coefficients
for Hecke algebras at roots of unity. {\it Adv. Math.}, 1990, v.82,
244-265.
\item{[Go]} A, Goncharov. Hyperlogarithms, mixed Tate motives and
multiple $\zeta$-numbers. {\it Preprint MSRI-058-93}, 1993, 35p.
\item{[Han]} G.-N. Han. Une courte d\'emonstration d'un resultat sur la
$Z$-statistique. {\it C.R. Acad. Sci. Paris, Serie I}, 1992, t.314,
969-971.
\item{[JKS]} U. Jannsen, S. Kleiman and J.-P. Serre (eds). Motives. {\it
Proc. Symp. in Pure Math.}, AMS, Providence, 1994, v.55, part 2.
\item{[JMMO]} M. Jimbo, K.C. Misra, T. Miwa and M. Okado. Combinatorics of
representations of $U_q(\wh{\hbox{\germ s}l}(n))$ at $q=0$. {\it
Commun. in Math. Phys.}, 1991, v.136, 543-566.
\item{[Kac]} V.G. Kac. Infinite dimensional Lie algebras. {\it Cambridge Univ.
Press}, 1990.
\item{[KP]} V. Kac and D. Peterson. Infinite-dimensional Lie algebras,
theta functions and modular forms. {\it Adv. Math.}, 1984, v.53, 125-264.
\item{[KW1]} V. Kac and M. Wakimoto. Modular invariant representations of
infinite-dimensional Lie algebras and superalgebras. {\it Proc. Natl.
Acad. Sci. USA}, 1988, v.85, 4956-4960.
\item{[KW2]} V. Kac and M. Wakimoto. Modular and conformal invariance
constraints in representation theory of affine algebras. {\it Adv. Math.},
1988, v.70, 156-236.
\item{[KW3]} V. Kac and M. Wakimoto. Integrable highest weight modules
over affine superalgebras and number theory. {\it Preprint}, 1994, 44p.
\item{[KKMMNN]} S.-J. Kang, M. Kashiwara, K.C. Misra, T. Miwa, T.
Nakashima and A. Nakayashiki. Affine crystals and vertex models. {\it
Int. J. Mod. Phys.}, 1992, v.A7, 449-484.
\item{[Ka]} M. Kashiwara. On the crystal bases of the $q$-analog of
universal enveloping algebras. {\it Duke Math. J.}, 1991, v.63, 465-516.
\item{[K]} R. Kaufmann. Pathspace decompositions for the Virasoro algebra
and its Verma modules. {\it Preprint Bonn-TH-94-05}, 1994, 19p. and
hep-th/9405041.
\item{[KKMM1]} R. Kedem, T.R. Klassen, B.M. McCoy and E. Melzer.
Fermionic quasi-particle representations for characters  of
$(G^{(1)})_1\times (G^{(1)})_1/(G^{(1)})_2$. {\it Phys. Lett.}, 1993,
v.B304, 263-270.
\item{[KKMM2]} R. Kedem, T.R. Klassen, B.M. McCoy and E. Melzer.
Fermionic sum representations for conformal field theory characters. {\it Phys.
Lett.}, 1993, v.B307, 68-76.
\item{[KeMa]} R. Kedem and B.M. McCoy. Construction of modular branching
functions from Bethe's equations in the 3-state Potts chain. {\it J.
Stat. Phys.}, 1993, v.71, 865-901.
\item{[KMM]} R. Kedem, B.M. McCoy and E. Melzer. The sums of Rogers,
Schur and Ramanujan and Bose-Fermi correspondence in 1+1-dimensional
quantum field theory. {\it Preprint ITP-SB-93-19}, 1993, 27p.
\item{[Kir1]} A.N. Kirillov. Combinatorial identities and completeness of
states for the generalized Heisenberg magnet. {\it Zap. Nauch. Sem. LOMI
(in Russian)},
1984, v. 131, 88-105.
\item{[Kir2]} A.N. Kirillov. Completeness of the Bethe vectors for
generalized Heisenberg magnet. {\it Zap. Nauch. Sem. LOMI (in Russian)},
1984, 134, 169-189.
\item{[Kir3]} A.N. Kirillov. On the Kostka-Green-Foulkes polynomials and
Clebsch-Gordan numbers. {\it Journ. Geom. and Phys.},
1988, v.5, p.365-389.
\item{[Kir4]} A.N. Kirillov. On identities for the Rogers dilogarithm
function related to simple Lie algebras. {\it Zap. Nauch. Sem. LOMI (in
Russian)}, 1987, v.164, 121-133 and {\it J. Soviet Math.}, 1989,
v.47, 2450-2459.
\item{[Kir5]} A.N. Kirillov. Unimodality of generalized Gaussian coefficients.
{\it  C.R.Acad.Sci.Paris}, 1992, t.315, Serie I, 497-501.
\item{[Kir6]} A.N. Kirillov. Fusion algebra and Verlide's formula. {\it
Preprint}, 1992, 9p.
\item{[Kir7]} A.N. Kirillov. Dilogarithm identities, partitions and spectra in
conformal field theory, Part I. {\it Submitted to Algebra and Analysis},
1992, 25p.
\item{[Kir8]} A.N. Kirillov. Generalization of the Gale-Ryser theorem.
{\it Preprint LITP-93-22}, 1993, 10p.
\item{[Kir9]} A.N. Kirillov. Decomposition of symmetric and exterior powers
of the adjoint representation of $\hbox{\germ g}l(N)$. 1.Unimodality of
principal specialization of the internal product of the Schur functions.
 {\it Int. J. Mod. Phys.}, 1992, v.7, 545-579.
\item{[Kir10]} A.N. Kirillov. Quantum polylogarithms. {\it Preprint}, 1994.
\item{[KK]} A.N. Kirillov and S.V. Kerov. Combinatorial algorithms related
with Bete's ansatz for $GL(n,C)$. {\it submitted to Algebra and Analysis},
1992.
\item{[KL]} A.N. Kirillov and N.A. Liskova. Completeness of Bethe's
states for generalized $XXZ$ model. {\it Preprint UTMS-94-47, Tokyo Univ.},
1994, 20p.
\item{[KlMe]} T. Klassen and E. Melzer. Purely elastic scattering
theories and their ultra-violet limit. {\it Nucl. Phys.}, 1990, v.B338,
485-528.
\item{[KR1]} A.N. Kirillov and N.Yu. Reshetikhin. Exact solution of the
$XXZ$ Heisenberg model of spin $S$. {\it J. Soviet Math.}, 1986, v.35,
2627-2643.
\item{[KR2]} A.N. Kirillov and N.Yu. Reshetikhin.
The Bethe ansatz and the combinatorics of Young tableaux.
{\it J. Soviet Math.}, 1988, v.41, p.925-955.
\item{[Ku]} A. Kuniba. Thermodynamics of the $U_q(X_r^{(1)})$ Bethe ansatz
system with $q$ a root of unity. {\it Nucl. Phys.}, 1993, v.B389,
209-244.
\item{[KN]} A. Kuniba and T. Nakanishi. Spectra in
Conformal Field Theories from the Rogers dilogarithm. {\it Modern Phys.
Lett.}, 1992, v.A7, 3487-3494.
\item{[KNS]} A. Kuniba, T. Nakanishi and J. Suzuki. Characters in
Conformal Field Theories from thermodynamic Bethe ansatz. {\it Modern
Phys. Lett.}, 1993, v.A7, 1649-1659.
\item{[LS]} A.~Lascoux and M.-P. Sch\"utzenberger.
Sur une conjecture de H.O. Foulkes.
{\it C. R. Acad. Sci. Paris}, 1978, v.286A, p.323--324.
\item{[LP]} J. Lepowsky and M. Primc. Structure of the standard modules
for the affine Lie algebra $A_1^{(1)}$. {\it Contemporary Math., AMS,
Providence}, 1985, v.46.
\item{[Lv]} M. Levine. The indecomposable $K_3$ of fields. {\it Ann. Ec.
Norm. Sup.}, 1989, v.22, 255-344.
\item{[Le1]} L. Lewin. Dilogarithm and associated functions. {\it North
Holland, Amsterdam}, 1981.
\item{[Le2]} L. Lewin. The dilogarithm in algebraic fields. {\it J.
Austral. Math. Soc. Series A}, 1982, v.33, 302-330.
\item{[Le3]} L. Lewin. The inner structure of the dilogarithm  in
algebraic fields. {\it J. Number Theory}, 1984, v.19, 345-373.
\item{[Le4]} L. Lewin. The order-independence of the polylogarithmic
ladder structure--implications for a new category of functional equations.
{\it Aequat. Math.}, 1986, v.30, 1-20.
\item{[Le5]} L. Lewin (ed). Structural properties of polylogarithms. {\it
Math. Surveys and Monographs}, 1991, v.37.
\item{[Li]} S. Lichtenbaum. Groups related to scissors-congruence groups.
{\it Contemporary Math.}, 1989, v.83, 151-157.
\item{[Lin]} X.-S. Lin. Knot invariants and iterated integrals. {\it
Prepr. IAS}, 1994, 10p.
\item{[Lit]} D.E. Littlewood.
The theory of group characters.
{\it Oxford University Press}, 1950.
\item{[Lo1]} J.H. Loxton. Special values of the dilogarithm function.
{\it Acta Mathematica}, 1984, v.43, 155-166.
\item{[Lo2]} J.H. Loxton. Partition identities and the dilogarithm, in
[Le4], Chapter 13, pp.287-299.
\item{[Lu1]} G.~Lusztig.
Green polynomials and singularities of unipotent classes.
{\it Adv. Math.}, 1981, v.42, p.169-178.
\item{[Lu2]} G.~Lusztig. Singularities, character formulas, and a $q$-analog
of weight  multiplicities. {\it Ast\'erisque}, 1983, v.101-102, p.208-227.
\item{[Lu3]} G.~Lusztig. Modular representations and quantum groups. {\it
Contemporary Math.}, 1989, v.82, 59-77.
\item{[Ma]} I.G. Macdonald.
Symmetric Functions and Hall Polynomials.
{\it Oxford University Press}, 1979.
\item{[Me1]} E. Melzer. The many faces of a character. {\it Lett. Math.
Phys.}, 1994, v.31, 233-246.
\item{[Me2]} E. Melzer. Fermionic character sums and the corner transfer
matrix. {\it Int. J. Mod. Phys.}, 1994, v.A9, 1115-1136.
\item{[Ml1]} S. Milne. Classical partition functions and the $U(n+1)$
Rogers-Selberg identity. {\it Discrete Math.}, 1992, v.99, 199-246.
\item{[Ml2]} S. Milne. The $C_l$ Rogers-Selberg identity. {\it SIAM J.
Math. Anal.}, 1994, v.25, 571-595.
\item{[Ml3]} S. Milne. The multidimensional $_1\psi_1$ sum and Macdonald
identities for $A_l^{(1)}$. {\it Proc Sympos. Pure Math.}, 1989, v.49
(part 2), 323-359.
\item{[ML]} S. Milne and G. Lilly. Consequences of the $A_l$ and $C_l$
Bailey transform and Bailey  lemma. {\it Preprint}, 1992, 28p.
\item{[Mi1]} J. Milnor. Introduction to algebraic $K$-theory. {\it Annals
of Math. Studies}, v.72, Princeton Univ. Press, 1971.
\item{[Mi2]} J.~Milnor. Hyperbolic geometry : the first 150 years.
{\it Bull A.M.S.}, 1982, v.6, 9-24.
\item{[Mi3]} J. Milnor. On polylogarithms, Hurwitz zeta-function, and the
Kubert identities. {\it Enseign. Math.}, 1983, v.29, 281-322.
\item{[MM]} K.C. Misra and T. Miwa. Crystal base for the basic
representation of $U_q(\wh{\hbox{\germ s}l(n)})$. {\it Commun. in
Math. Phys.}, 1990, v.134, 79-88.
\item{[MS]} A.S. Merkuriev and A.A. Suslin. On the $K_3$ of a field. {\it
Math. USSR Izv.}, 1991, v.36, 541-565.
\item{[Mo]} D. Moak. The $q$-analogue of Stirling's formula. {\it Rocky
Mtn. J. Math.}, 1984, v.14, 403-413.
\item{[NRT]} W. Nahm, A. Recknagel and M. Terhoeven. Dilogarithm
identities in Conformal Field Theory. {\it Mod. Phys. Lett.}, 1993, v.A8,
1835-1848.
\item{[O]} J. Oesterl\'{e}. Polylogarithmes. {\it Seminaire BOURBAKI},
1992-93, $n^0$762, 16p.
\item{[P1]} P. Paule. On identities of the Rogers-Ramanujan type. {\it J.
Math. Analysis and Appl.}, 1985, v.107, 255-284.
\item{[P2]} P. Paule. The concept of Bailey chains. {\it Publ. I.R.M.A.,
Strasbourg}, 1988, Actes $18^e$ S\'eminaire Lotharingien, 53-76.
\item{[Q1]} Y.-H. Quano. Polynomial identities of the Rogers-Ramanujan
type. {\it Univ. Melbourne Preprint, No. 25}, 1994, 25p.
\item{[Q2]} Y.-H. Quano. On equivalence between Bose/Fermi Virasoro
characters. {\it Univ.  Melbourne Preprint, No. 27}, 1994, 9p.
\item{[Ray]} G. Ray. Multivariable polylogarithm identities, in [Le5],
Chapter 7, pp. 123-169.
\item{[R]} N.Yu. Reshetikhin. Quasitriangularity of quantum groups at
root of unity. {\it Preprint}, hep-th/9403105, 1994, 17p.
\item{[RS]} B. Richmond and G. Szekeres. Some formulas related to
dilogarithm, the zeta function and the Andrews-Gordon identities. {\it
J. Austral. Math. Soc.}, 1981, v.A31, 362-373.
\item{[R-C]} A. Rocha-Caridi. Vacuum vector representations of the
Virasoro algebra. {\it in "Vertex Operator in Mathematical Physics"}
(eds. J. Lepowsky, S. Mandelstam and I.M. Singer), Publications of MSRI,
Springer-Verlag, 1984, v.3, 451-473.
\item{[Ro]} L. Rogers. On function sum theorems connected with the series
$\ds\sum_{n=1}^{\infty}{x^n\over n^2}$. {\it Proc. London Math. Soc.},
1907, v.4, 169-189.
\item{[Ros]} M. Rosso. Finite dimensional representations of the quantum
analog of the enveloping algebra of a complex simple Lie algebra. {\it
Commun. Math. Phys.}, 1988, v.117, 581-593.
\item{[Sa]} B. Sagan. The symmetric group. {\it Wadsworth} \& {\it Brooks/Cole,
Pacific Grove, CA}, 1991.
\item{[Se]} A. Sergeev. Tensor algebra of the defining representation
as the module over Lie superalgebras $\g l(m,n)$ and $Q(n)$. {\it Math.
USSR, Sbornik}, 1985, v.51, 419-424.
\item{[Ser]} V. Serganova. Kazhdan-Lusztig polynomials for the Lie
superalgebras $\g l(m/n)$. {\it Yale Univ. Preprint}, 1992, 8p.
\item{[Sl]} L.J. Slater. Further identities of the Rogers-Ramanujan type.
{\it Proc. London Math. Soc.}, 1951, v.54, 147-167.
\item{[Stn1]} R.P. Stanley. $GL(n,{\bf C})$ for combinatorialists.
{\it in ``Surveys in Combinatorics'' (ed. E.K.Lloyd), London Math. Soc.,
Lecture Note Series, No.82, Cambridge University Press, Cambridge}, 1983,
187-199.
\item{[Stn2]} R.P. Stanley. The stable behavior of some characters of
$SL(n,{\bf C})$. {\it Linear and Multil. Algebra}, 1984, v.16, 3-27.
\item{[Stm]} J.R. Stembridge. Hall-Littlewood polynomials, plane partitions
and the Rogers-Ra\-ma\-nu\-jan identities. {\it Trans Amer. Math. Soc.},
1990, v.319, 469-498.
\item{[Su]} A.A. Suslin. $K_3$ of a field and the Bloch group. {\it Proc.
of the Steklov Inst. of Math.}, 1991, v.4, 217-239.
\item{[Te]} I. Terada. A generalization of the length symmetry and
variety of $N$-stable flags. {\it Preprint}, 1993, 26p.
\item{[Tr1]} M. Terhoeven. Lift of dilogarithm to partition identities.
{\it Preprint Bonn-HE-92-36}, 1992 and hep-th/9211120.
\item{[Tr2]} M. Terhoeven. Dilogarithm identities, fusion rules and
structure constants of CFTs. {\it Preprint Bonn-HE-93-24}, 1993, 11p.
\item{[TV]} V. Tarasov and A. Varchenko. Asymptotic solutions to the
quantized Knizhnik-Za\-mo\-lod\-chi\-kov equation and Bethe vectors.
{\it Prepr. UTMS-94-46, Tokyo Univ.}, 1994.
\item{[UN]} K. Ueno and M. Nishizawa. Quantum groups and zeta-functions.
{\it Preprint}, 1994, 12p.
\item{[V]} A. Varchenko. Multidimensional hypergeometric functions in
conformal field theory, algebraic $K$-theory, algebraic geometry. {\it
ICM, Kyoto}, 1990, v.1, 281-300.
\item{[W1]} G.N. Watson. A new proof of the Rogers-Ramanujan identities.
{\it J. London Math. Soc.}, 1929, v.4, 4-9.
\item{[W2]} G.N. Watson. A note on Spence's logarithmic transcendent. {\it
Quart. J. Math. Oxford}, 1937, Ser. 8, 39-42.
\item{[Za1]} D. Zagier. Hyperbolic manifolds and special values of
Dedekind zeta-functions. {\it Invent. Math.}, 1986, v.83, 285-301.
\item{[Za2]} D. Zagier. The remarkable dilogarithm. {\it J. Math. Phys.
Sci.}, 1988, v.22, 131-145.
\item{[Za3]} D.~Zagier.
The remarkable dilogarithm (number theory and related topics).
{\it Papers presented at the Ramanujan colloquium, Bombay}, 1988,
  TIFR.
\item{[Za4]} D. Zagier. Polylogarithms, Dedekind zeta-functions and the
algebraic $K$-theory of fields. {\it in Arithmetic Algebraic Geometry,
Birkhauser}, G. van der Geer e.a. (eds), 1991, v.89, 391-430.
\item{[Zm]} Al. Zamolodchikov. Thermodynamic Bethe ansatz in relativistic
models: scaling 3-state Potts and Lee-Yang models. {\it Nucl. Phys.},
1990, v.B342, 695-720.
\item{[ZB]} D. Zeilberger and D. Bressoud. A proof of Andrew's $q$-Dyson
conjecture. {\it Discrete Math.}, 1985, v.54, 201-224.

\end